\documentclass[structabstract]{aa}  
\usepackage{lscape}
\usepackage{graphicx}
\usepackage{supertabular}
\usepackage{longtable}
\usepackage{txfonts}
\def\gtsima{$\; \buildrel > \over \sim \;$}
\def\ltsima{$\; \buildrel < \over \sim \;$}
\def\gsim{\lower.5ex\hbox{\gtsima}}
\def\lsim{\lower.5ex\hbox{\ltsima}}

\begin{document}
    \title{Probing variability patterns of the Fe K line complex in bright nearby AGNs}

   \subtitle{}

   \author{B. De Marco
          \inst{1}
          \and K. Iwasawa
          \inst{2}
           \and M. Cappi
         \inst{3}
           \and M. Dadina
          \inst{3}
          \and F. Tombesi 
          \inst{3, 4, 5, 6} 
          \and G. Ponti 
          \inst{7}
          \and A. Celotti
          \inst{1}
          \and G. Miniutti
          \inst{8}
          }

   \offprints{demarco@sissa.it}

   \institute{SISSA International School for Advanced Studies, Via Beirut 2-4, I-34151 Trieste, Italy
         \and 
    INAF-Osservatorio Astronomico di Bologna, Via Ranzani 1, I-40127 Bologna, Italy
          \and
    INAF-IASF Bologna, Via Gobetti 101, I-40129 Bologna, Italy 
          \and 
    Dipartimento di Astronomia, Universit\`{a} degli Studi di Bologna, Via Ranzani 1, I-40127 Bologna, Italy
          \and
    Department of Physics and Astronomy, Johns Hopkins University, 3400, Baltimore, MD 21218, USA
    \and
    Laboratory for High Energy Astrophysics, NASA/Goddard Space Flight Center, Greenbelt, MD 20771, USA
    \and 
   APC Universit\'e Paris 7 Denis Diderot, 75205, Paris, France 
    \and
   LAEX, Centro de Astrobiologia (CSIC--INTA); LAEFF, P.O: Box 78, E-28691, Villanueva de la Ca\~nada, Madrid, Spain
             }


 
  \abstract
   {The unprecedented sensitivity of current X-ray telescopes allows for the
  first time to address the issue of the Fe K line complex variability
  patterns in bright, nearby AGNs. These kind of studies have the potential to map the accretion flow in the strong gravity regime of supermassive black holes.}
   {We examine XMM-{\it Newton} observations of the brightest sources of the FERO sample of radio--quiet type 1 AGNs (for a total of 72 observations) with the aim of characterizing the temporal behaviour of Fe K complex features.}
   {A systematic mapping of residual flux above and below the continuum 
in the 4--9 keV range is performed in the time vs energy domain, with the purpose of identifying
  interesting spectral features in the three energy bands: 5.4--6.1 keV, 6.1--6.8 keV and 6.8--7.2 keV, corresponding respectively to the redshifted, rest frame and blueshifted or highly ionized Fe K$\alpha$ line bands. The variability significance of rest frame and energy shifted Fe K lines is assessed by extracting light curves and comparing them with Monte Carlo simulations.}
  {The time-averaged profile of the Fe K complex revealed spectral complexity in several observations. Red- and blue- shifted components (either in emission or absorption) were observed in 30 out of 72 observations, with an average $\langle EW\rangle\sim$ 90 eV for emission and $\langle EW\rangle\sim$ -30 eV for absorption features. We detected significant line variability (with confidence levels ranging between 90\% and 99.7\%) within at least one of the above energy bands in 26 out of 72 observations on time scales of $\Delta t$$\sim$6--30 ks. Reliability of these features has been carefully calculated using this sample and has been assessed at $\sim$3$\sigma$ confidence level.}
   {This work increases the currently scanty number of detections of variable, energy shifted, Fe lines and confirms the reliability of the claimed detections. We found that the distribution of detected features is peaked at high variability significances in the red- and blue- shifted energy bands rather than at rest-frame energies, suggesting an origin in a relativistically modified accretion flow.}


   \maketitle

\section{Introduction}

Reflection features observed in the X-ray spectra of many Active Galactic
Nuclei (AGNs) represent a powerful tool to study the properties of the
matter flow accreting onto Supermassive Black Holes (SMBH). In particular the fluorescent Fe K line can be used as
a probe of the innermost regions of AGNs (e.g. Fabian et al 1989; Fabian et al. 2000; Reynolds \& Nowak 2003; Miniutti \& Fabian 2005), where the effects of
general and special relativity become significant. Indeed, the relativistically skewed,
redshifted and broadened profile of the line and its variability pattern
strongly depend on the physical conditions of the accreting matter and of the
SMBH (e.g its spin). By now about 25\% of all AGNs observed by XMM-{\it Newton}
appear to show the
presence of a relativistic Fe line in their X--ray spectra (Guainazzi et al. 2006; hereafter GBD06). 
Moreover, from a systematic study of 37 very high S/N XMM-{\it Newton} observations of nearby ($z<0.05$) radio quiet AGN, Nandra et al. 2007 (hereafter NOGR07) demonstrated that Fe K line relativistic broadening is quite common, being effective in $\sim$45\% of the cases.\\
In recent years, more diversity has been added to the general picture by the
discovery of narrow and (apparently) transient features in the E$\sim$4--6 keV
energy range (e.g. Turner et al. 2002; Longinotti et al. 2007a and references therein; Vaughan \& Uttley 2008), characterized by
relatively short-time scale variability, of the order of tens of ks (e.g. Yaqoob et al. 2003; Iwasawa et al. 2004, hereafter IMF04; Tombesi et al. 2007, Petrucci et al. 2007). Their nature is not yet
understood. Vaughan \& Uttley (2008) casted
major doubts about the reliability of such shifted features, pointing out that 
many of the detections reported in the literature are   
not statistically significant and most probably drawn from random fluctuations in the bulk of analysed X-ray data.
 As suggested by these authors,
the detailed and deep investigation of a well defined sample of objects, as the one presented here,
is necessary in order to achieve a firm conclusion on this issue.\\
The importance of such features relies on the fact that their energy is similar to that typical of relativistic Fe K line components, suggesting that we are observing the peak of a variable
relativistic line profile produced in an extended region of the disk (in such
scenario, the red tail would be too faint to be detectable above the
continuum). Alternatively, intrinsically narrow emission lines could be
produced in discrete regions of inflows or outflows and redshifted as a consequence of relativistic motions. Moreover, recent results point towards an
origin in small emitting regions within the inner accretion disk (e.g. IMF04, Tombesi et al. 2007, Petrucci et al. 2007),
as predicted by several models, as the ``hot'' orbiting spot model (Nayakshin \& Kazanas 2001; Dov\v{c}iak et al. 2004) or the lamp-post model (Martocchia \& Matt 1996; Dabrowski \& Lasenby 2001; Miniutti \& Fabian 2004).
Unfortunately, despite the unprecedented sensitivity of current X-ray telescopes, attempts to test these models are hampered by the too fine spectral features predicted (Goosmann et
al. 2007). In this context, however, variability
  studies could be a key tool in providing insight onto their
real nature (e.g. Ponti et al. 2004; IMF04; Miller et al. 2006; Tombesi et al. 2007).\\
In this work we present the results obtained from the time-resolved X-ray
spectral analysis of a well-defined sample (94\% complete) of radio--quiet AGNs selected from the
XMM-{\it Newton} archive. The sample is a subset of the FERO sample (FERO stands for ``{\it Finding Extreme Relativistic Objects}'') and was originally selected by GBD06 to study the general properties of relativistically broadened Fe
K$\alpha$ lines. We consider the same dataset with the aim of looking for transient features in the Fe K band and
analyse their variability patterns by exploiting a uniform analysis of their highest quality observations. Several time resolved studies have so far highlighted the variable
nature of the Fe K line complex (e.g. Ponti et al. 2004; IMF04; De Marco et al. 2006; Miller et al. 2006; Petrucci et al. 2007; Tombesi et al. 2007). Up to now, there is, however, no
  statistically solid characterization of its temporal
behaviour.
 For the time-resolved analysis we adopted the method based on the
mapping of the
excess residuals above and below the continuum in the time vs energy plane; 
this technique has been already widely used (Ballantyne et al. 2004; IMF04; Turner et al. 2006; Miller et al. 2006; Porquet et al. 2007;
Tombesi et al. 2007; Petrucci et al. 2007) and proved to be very useful to characterize such variable X-ray
spectral features.

\section{Selection of the sample}
\label{sec:select}

The sample analysed in this paper was first presented in GBD06 and comprises 33 poorly absorbed (namely with intrinsic N$_{H}<2-3\times 10^{22}$ cm$^{-2}$) radio quiet AGNs
 selected from the RXTE Slew Survey (XSS, Revnivtsev et al. 2004) to have F$_{2-10 keV}\geq
 1.5\times 10^{-11}$erg cm$^{-2}$s$^{-1}$.
To carry out a meaningful timing analysis, all the XMM-{\it Newton} observations with EPIC pn exposure $\geq$10 ks were considered.\\
This selection excluded the sources: H1846-786 (being observed two times for a duration of $<$10 ks) and UGC 10683 (because no XMM-{\it Newton} observations were available as of 2009 February 1).
Furthermore, the source H0557-385 was discarded as its hard X-ray emission has been shown to be highly absorbed during the two $\geq$10 ks XMM-{\it Newton} observations (the spectral curvature below 6 keV implies the presence of an intervening neutral gas cloud, partially covering the X-ray source, with a column density N$_{H}\sim$8$\times$10$^{23}$cm$^{-2}$, Longinotti et al. 2009).
Four observations (MCG-5-23-16, ID 0112830301; MRK 279, ID 0083960101; AKN 564, ID 0006810301; NGC 4593, ID 0109970101) were neglected after Good Time Interval (GTI) filtering, as they result badly affected by proton flares, reducing the EPIC pn effective exposure to $<$10 ks. Finally a 15 ks observations of AKN 564 (ID 0006810101) was rejected, as the effective exposure, although $>$10 ks, is too short with respect to the time resolution needed for the timing analysis.\\
Our final sample consists of 30 sources and 72 observations. Information about the sample is given in Table \ref{tab:info}.\\
In the following each observation is denoted by the corresponding XMM-{\it Newton} revolution number. Whenever more than one observation is carried out during the same revolution, a letter index indicates the chronological order.

\section{Analysis}

In order to characterize the variability pattern
of the Fe K features using the so-called excess map technique
(IMF04), it is important to identify a-priori the strongest lines present in the spectrum. The first step is thus to define the properties of the average spectrum, and then to analyse its temporal behaviour.

\subsection{Data reduction and time-averaged spectral analysis}

The analysis has been carried out using only the EPIC pn camera data because
of its higher sensitivity (about a factor of two, e.g. Watson et al. 2001) in the Fe K band, with respect 
to the EPIC MOS. We used the XMM-SAS v. 8.0.0 software with CCF release as of October 2006
for data reduction, and the \emph{lheasoft} v. 6.0.3 package for
data analysis.
Time intervals free from high background events were
selected by calculating the time series average count rate $\mu$ and
standard deviation $\sigma_c$ during strong flares-free periods, in the E $\geq$ 10 keV energy band and filtering
out all the events with count rate exceeding the (arbitrarily fixed) threshold
of $\mu + 3\sigma_c$. This choice allows to avoid large data gaps in the middle of the observation, which is a necessary condition for a timing analysis to be properly carried out.
For each object,
the source photons were extracted from a circular region of 45 arcsec radius,
while the background ones were collected from adjacent source-free rectangular
regions. Using the SAS task \emph{epatplot} we verified the presence of significant pile-up; whenever some degree of pile-up was
  found it was minimized adopting an annular region for source counts extraction\footnote{Pile-up was revealed in the following observations: IC 4329a rev. 210 (carried out in ``FullWindow'' mode), MCG-5-23-16 rev. 363 (carried out in ``FullWindow'' mode) and 1099 (carried out in ``LargeWindow'' mode), NGC 3227 rev. 1279 (carried out in ``LargeWindow'' mode) and ESO 141-G055 rev. 1435, 1436 and 1445 (carried out in ``FullWindow'' mode).}.\\
Average spectra were produced for each source and analysed using the XSPEC
v. 11.2.3 software package. The spectral fitting was performed in the 4--9
keV energy band, where Fe spectral signatures are expected, using simple
models (i.e. a power law for the continuum and Gaussian
components for emission and absorption lines). 
Some
residual curvature of the spectrum produced by \emph{warm absorber} systems
can yet be detected in this energy band; in order to correct the shape of the
continuum for it, a cold absorption
component was added to the model (clearly the derived column
densities, listed in Table \ref{tab:info}, do not have any physical meaning). It is worth noticing that the presence of narrow red- and blue- shifted absorption
lines (Dadina et al. 2005; Reeves et al. 2004; Risaliti et al 2005; Cappi 2006 and references
therein; Braito et al. 2007; Tombesi et al. 2009; Cappi et al. 2009) in the chosen energy band, did not influence
significantly the determination of the continuum spectral shape.\\
We modelled emission and absorption lines in the time-averaged spectra with simple Gaussian
templates: a Gaussian was added to the model whenever we found evidence for it
from the 99\% confidence contours for intensity vs energy. The best fit models
are reported in Table \ref{tab:models}.

\subsection{Results from time-averaged spectra}

The ``core'' of the Fe K$\alpha$ emission line is detected in all the observations (see Table \ref{tab:models}). This line results usually relatively narrow with 1$\sigma$ width of the order of the EPIC pn spectral resolution at these energies (i.e. $\sim$0.13 keV) or less.
A broader line at E$\sim$6.4 keV  ($\sigma \sim$0.2--0.4 keV) is required in the spectral fitting of: NGC 3516 (rev. 1251), MCG -6-30-15 (rev. 108A, 301, 303), ARK 120 (rev. 679), MKN 766 (rev. 1000, 1001, 1003, 1004), MCG -2-58-22 (rev. 180), ESO 141-G055 (rev. 1435A, 1435B, 1436), ESO 198-G024 (rev. 201), AKN 564 (rev. 930).
The line energies are overall consistent with emission from neutral iron, although in some cases (NGC 3516 rev. 1251, 1253; MCG -6-30-15 (rev. 108A); MKN 766 rev. 999, 1003; ESO 141-G055 rev. 1435B; ESO 198-G024 rev. 207; AKN 564 rev. 930) the implied energies might be indicative of mildly ionized matter (E$\geq$6.45 keV) or more complex line profiles.\\
Emission and/or absorption features at red- and blue- shifted energies were observed in 30 out of 72 observations.
We detected significant redshifted emission in 21 out of 72 observations (see Table \ref{tab:detect}), with 1$\sigma$ width in the range $<$0.7 keV and $\langle EW\rangle\sim$ 90 eV. It must be stressed that this is to be considered as a conservative number of detections because transient redshifted features might result smoothed out in the time averaged spectra analysis, and hence not significantly detectable. Time resolved surveys are the most suitable techniques for the study of this kind of features.\\
Signatures of significant absorption lines blue ward the narrow ``core'' were instead detected in 15 out of 72 observations (see Table \ref{tab:detect}), at energies in the range E$\sim$6.6--8.5 keV, with $\langle EW\rangle\sim$ -30 eV.\\
Finally, lines consistent with emission from highly ionized Fe K$\alpha$ and/or Fe K$\beta$ were revealed in 28 out of 72 observations (see Table \ref{tab:detect}). The 1$\sigma$ width of these lines were overall of the order of the energy resolution of the detector except for 12 observations (MRK 509 rev. 250, 1168; IRAS 05078+1626 rev. 1410; MKN 766 rev. 82, 265, 999, 1001, 1002; NGC 526A rev. 647; NGC 4051 rev. 541; ESO 141-G055 rev. 1445, NGC 7213 rev. 269), where a broader component was detected ($\sigma$$\sim$0.2--0.7 keV, EW$\sim$80--160 eV), which might be indicative of a blending, either between the ionized Fe K$\alpha$ and neutral Fe K$\beta$ or with the blue peak of a relativistic component.

\begin{table*}
\caption{Time-averaged spectral properties of the observations. The revolution number of the observations where the most relevant features in the 4--9 keV spectra were detected is reported, including a flag indicating significances obtained from intensity vs energy contour plots. Multiple flags refer to the corresponding features in order of increasing energy.}
\label{tab:detect}
\centering
\vspace{0.2cm}
\begin{scriptsize}
\begin{tabular}{l l l l l}
\hline\hline             
Source & Redshifted Emission features & Broad Fe K$\alpha$ & Highly Ionized Fe K$\alpha$/Fe K$\beta$ & Absorption features \\
\hline

IC 4329a & 670$^{*}$ &      & 670$^{*}$ &   670$^{**}$ \\
MCG -5-23-16 &   363$^{**}$, 1099$^{*}$  &      &   1099$^{*}$ &   \\
NGC 3783     &   193$^{**}$, 371$^{**}$, 372$^{*}$ &          &   193$^{*}$, 371$^{*}$, 372$^{*}$ &   193$^{*,**}$, 372$^{*}$ \\ 
NGC 3516     &   245$^{*}$, 352$^{*}$, 1250$^{**}$ &   1251$^{*}$ &     &   1250$^{*}$, 1251$^{*}$, 1252$^{*}$, 1253$^{*}$ \\
MRK 509      &     &       &   250$^{**}$, 1073$^{**}$, 1074$^{**}$, 1168$^{*}$ &   1073$^{**}$, 1168$^{*,**}$ \\
MCG -6-30-15 &   108A$^{**}$, 302$^{**}$, 303$^{*}$ &   108A$^{**}$, 301$^{*}$, 303$^{*}$ &   303$^{**}$ &   108A$^{*}$ \\
MCG +8-11-11 &     &       &    794$^{**}$ &     \\
NGC 7314     &     &    &   256$^{*}$ &     \\
ARK 120      &     &   679$^{*}$ &   679$^{*}$ &     \\
MRK 279      &     &     &       &   1088$^{**,*}$ \\
NGC 3227     &   1279$^{*}$ &     &     &           \\
IRAS 05078+1626 &  &   &    1410$^{**}$ &     \\
MRK 590      &   &   &   837$^{*}$  &      \\
MRK 766      &   265$^{**}$, 999$^{*}$, 1000$^{**}$, 1001$^{**}$, 1002$^{**}$, 1003$^{**}$ &   1000$^{*}$, 1001$^{*}$, 1003$^{*}$, 1004$^{**}$ &   82$^{*}$, 265$^{*}$, 999$^{**}$, 1000$^{*}$, 1001$^{*}$, 1002$^{*}$ &    \\
NGC 7469     &   &   &    912$^{**}$ &      \\
MCG -2-58-22 &   &   180$^{**}$ &    &    \\
NGC 526A     &   &      &   647$^{*}$ &   647$^{**}$ \\
NGC 4051     &   263$^{*}$, 541$^{*}$ &     &   541$^{**}$ &   263$^{**}$ \\
ESO 141-G055 &   &   1435A$^{*}$, 1435B$^{*}$, 1436$^{*}$  &    1436$^{**}$, 1445$^{*}$ &   1445$^{*}$ \\
ESO 198-G024 &   &  207$^{**}$  &      &   1128$^{**}$ \\
AKN 564      &   &    930$^{*}$ &     &      \\
NGC 7213     &   &    &   269$^{**}$ &       \\
NGC 4593     &   &    &    465$^{*}$ &     \\
\hline
\hline
\end{tabular}
\begin{list}{}{}
\item[($^{*}$):] detection significance $>4\sigma$;
\item[($^{**}$):] detection significance between $\sim 3\sigma$--4$\sigma$;
\end{list}
\end{scriptsize}
\end{table*}

\subsection{Spectral variability}
\label{sec:proced}

The spectral variability has been investigated by systematically mapping residuals to the continuum emission in the time vs energy domain over the 4--9 keV band. 
As most of the features observed in the time-averaged spectra are
unresolved in their fine structure, the chosen energy resolution for the time-energy `mapping'
is 0.1 keV for nearly all the sources\footnote{In the case of NGC 4051 (rev. 541), AKN 564 (rev. 930) and MRK 704 (rev. 1074) we opted for a lower energy resolution (i.e. 0.2 keV, see Table \ref{tab:info}) in order to keep a high enough time resolution.} (i.e. approximately equal to the detector resolution
in the analysed band). Though not producing a good oversampling for most of the features, this energy resolution represents the only viable choice for the available data.
The time resolution of the map has to be chosen in order to garantee a sufficient number of counts into each energy bin. Hence, fixing the energy resolution, we extracted spectra at different time resolutions during the period of minimum flux of the source, and required a minimum of 20 counts per bin (for the $\chi^{2}$ statistics to be applicable) in the Fe K complex energy band (i.e. at least for E$<$7.5 keV).
The adopted time resolutions are listed in
Table \ref{tab:info}\\
It is useful to compare the time resolutions used in the analysis with a typical time scale of the system.
As indicative time-scale we can consider the estimated orbital period at 10 r$_{g}$, given by (e.g. Bardeen et al. 1972):
$$
T_{orb}=310\ [a+(r/r_{g})^{3/2}]\ M_{7}\ \ (sec),
$$
r$_{g}$ being the gravitational radius, $M_{7}$ the mass of the black hole in units of 10$^{7}M_{\odot}$ and $a$ the adimensional angular momentum per unit mass.
 Following the Nyquist-Shannon sampling theorem, the condition $\Delta t=T_{orb}/2\beta$ (being $\beta \geq$1 the sampling factor and $\Delta t$ the chosen time resolution) must be satisfied for the chosen time resolution to oversample this characteristic time-scale.
This is thus generally the case here that the estimated orbital time-scale at 10r$_{g}$ is oversampled by the chosen time resolution. Only in four cases, being the orbital period very short (i.e. $\leq$2.5 ks), the time resolution cannot adequately oversample it, as the minimum number of counts per energy bin must be preserved. The sources for which we were forced to undersample the orbital time-scale at 10r$_{g}$ are: MCG-6-30-15, MRK 766, NGC 4051 and AKN 564.

\begin{figure*}
\caption{\emph{Left panels}: data-to-model (power law plus cold absorption) ratios of the 4--9 keV time-averaged spectra; \emph{middle panels}: S/N map of excess residuals in the time-energy plane (at a time resolution of 2500 s); \emph{right panels}: 0.3--10 keV background subtracted continuum and residuals light curves (in bands A, B and C). The light curves are renormalized for the corresponding average flux.}
\label{fig:maps1}

\centering
\vspace{1.5cm}
\begin{tabular}{p{4cm}p{4cm}p{4cm}}

\includegraphics[height=4.0cm,width=4.0cm]{figure/spectra/IC4329a_10ks_spec.ps} & \includegraphics[height=4.2cm,width=4.0cm]{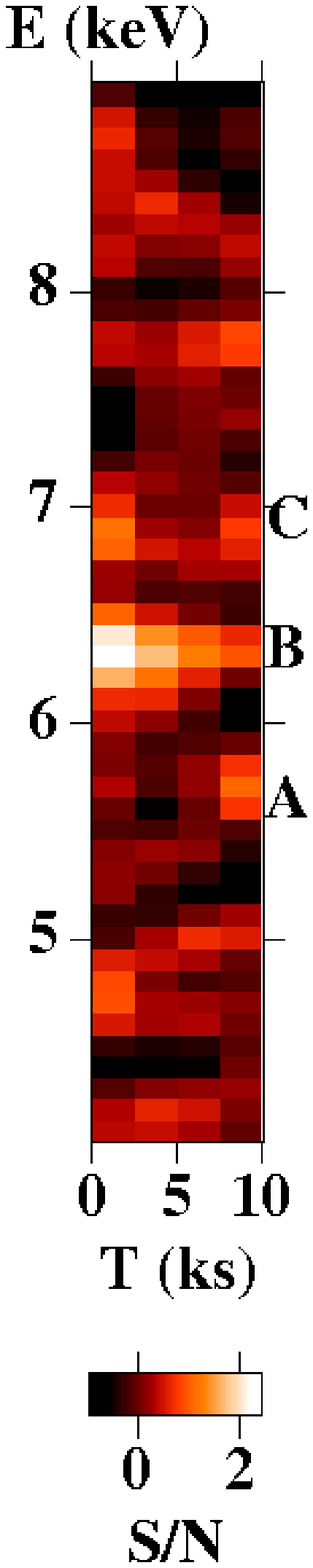} & \includegraphics[width=4.8cm,height=4.0cm]{figure/light_curves/IC4329a_10ks_lc.ps}\\
 & & \\
\includegraphics[height=4.0cm,width=4.0cm]{figure/spectra/IC4329a_133ks_spec.ps} & \includegraphics[height=4.2cm,width=4.0cm]{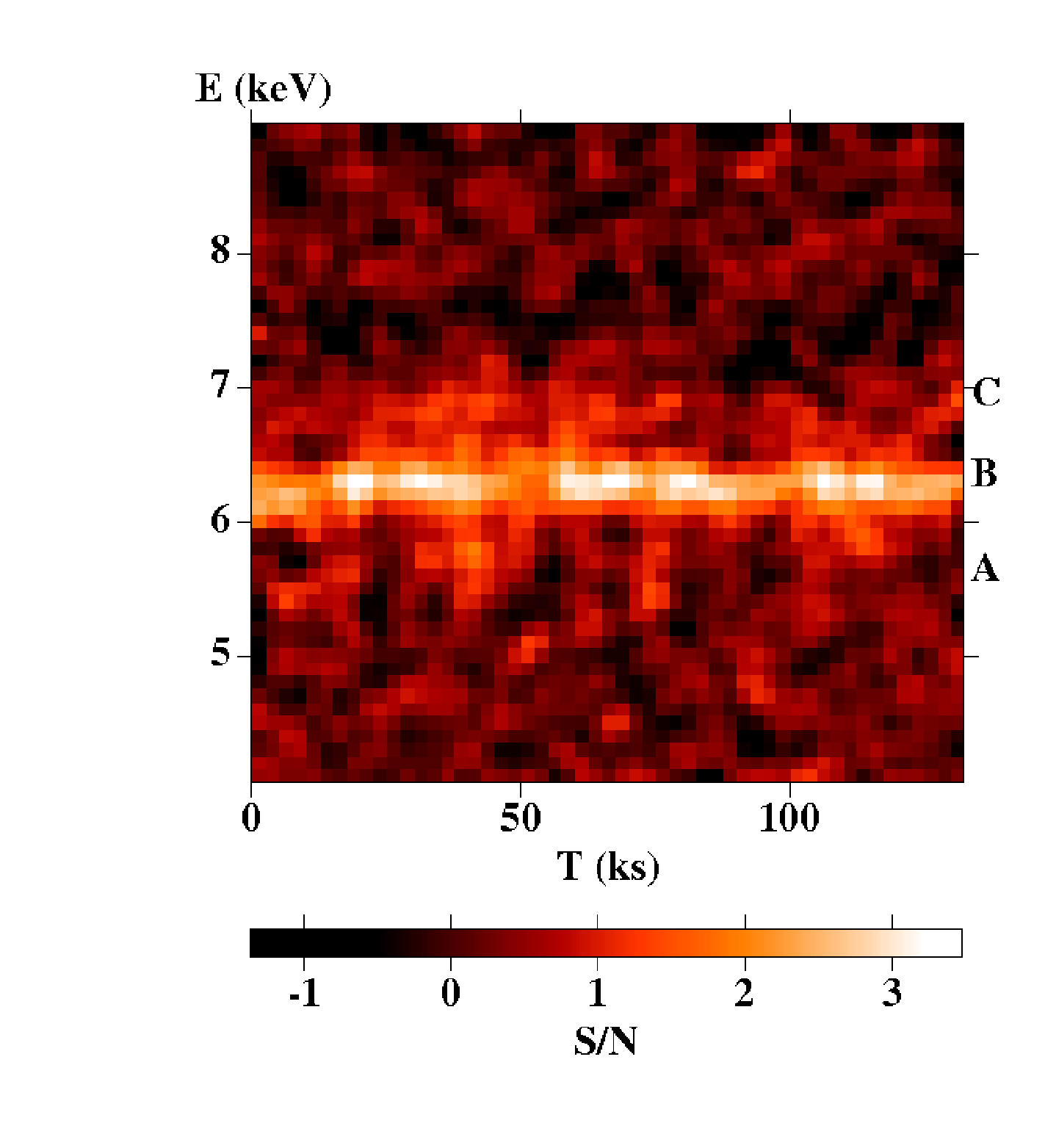} & \includegraphics[width=4.8cm,height=4.0cm]{figure/light_curves/IC4329a_133ks_lc.ps}\\
 & & \\
\includegraphics[height=4.0cm,width=4.0cm]{figure/spectra/NGC3783_138ksB_spec.ps} & \includegraphics[height=4.2cm,width=4.0cm]{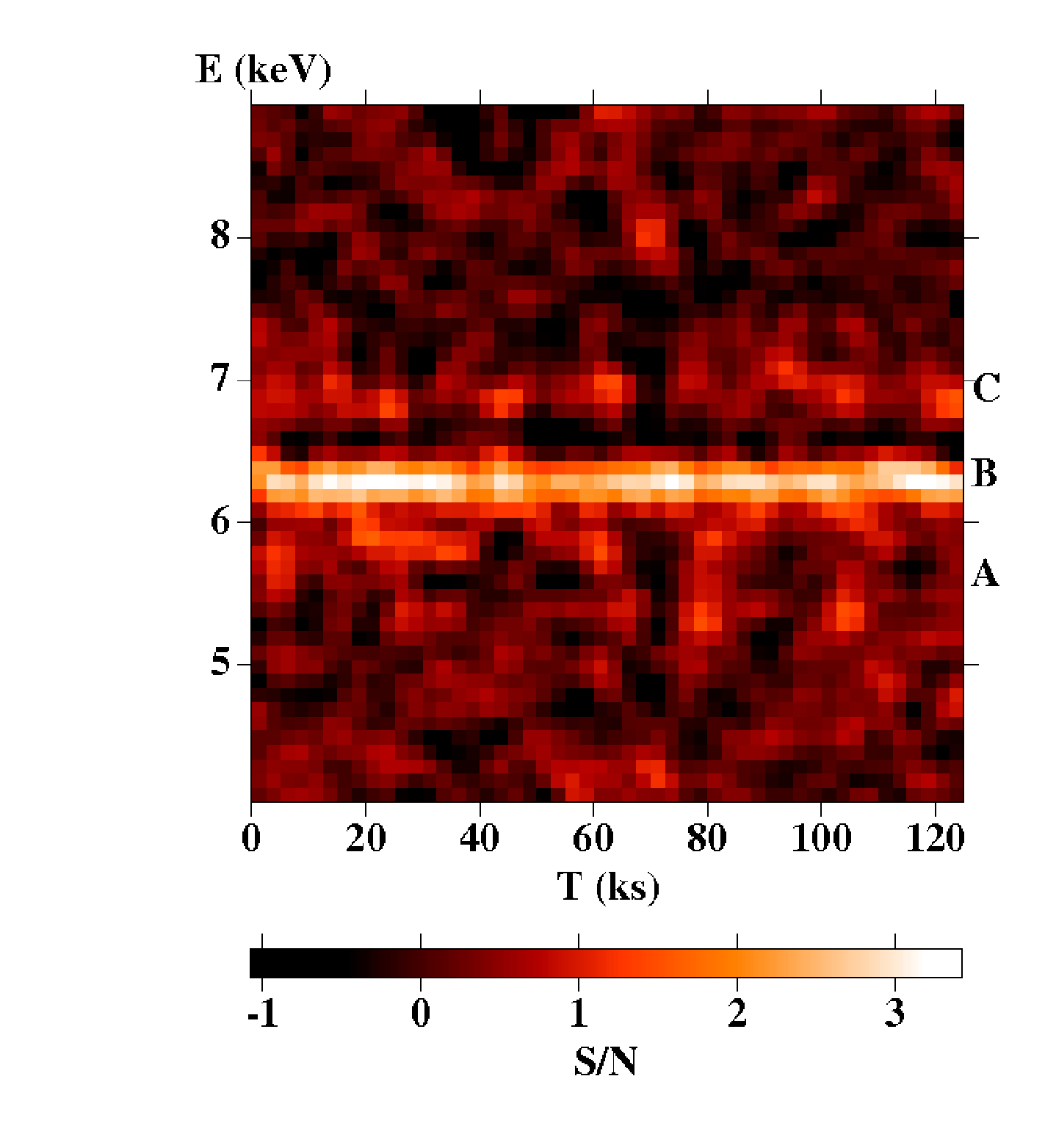} & \includegraphics[width=4.8cm,height=4.0cm]{figure/light_curves/NGC3783_138ksB_lc.ps}\\
 & & \\
\includegraphics[height=4.0cm,width=4.0cm]{figure/spectra/NGC5548_26ks_spec.ps} & \includegraphics[height=4.2cm,width=4.0cm]{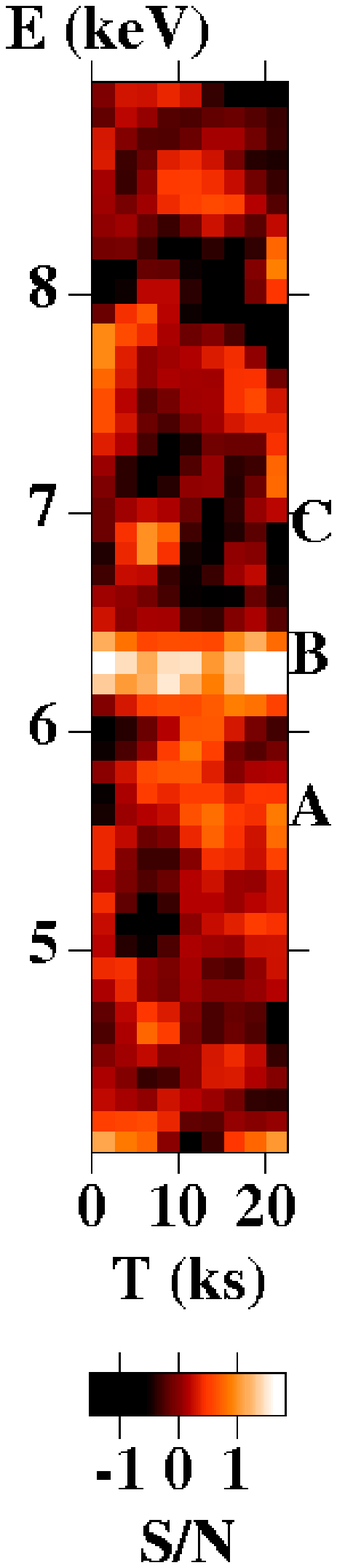} & \includegraphics[width=4.8cm,height=4.0cm]{figure/light_curves/NGC5548_26ks_lc.ps}\\ 

\includegraphics[height=4.0cm,width=4.0cm]{figure/spectra/NGC3516_130ks_spec.ps} & \includegraphics[height=4.2cm,width=4.0cm]{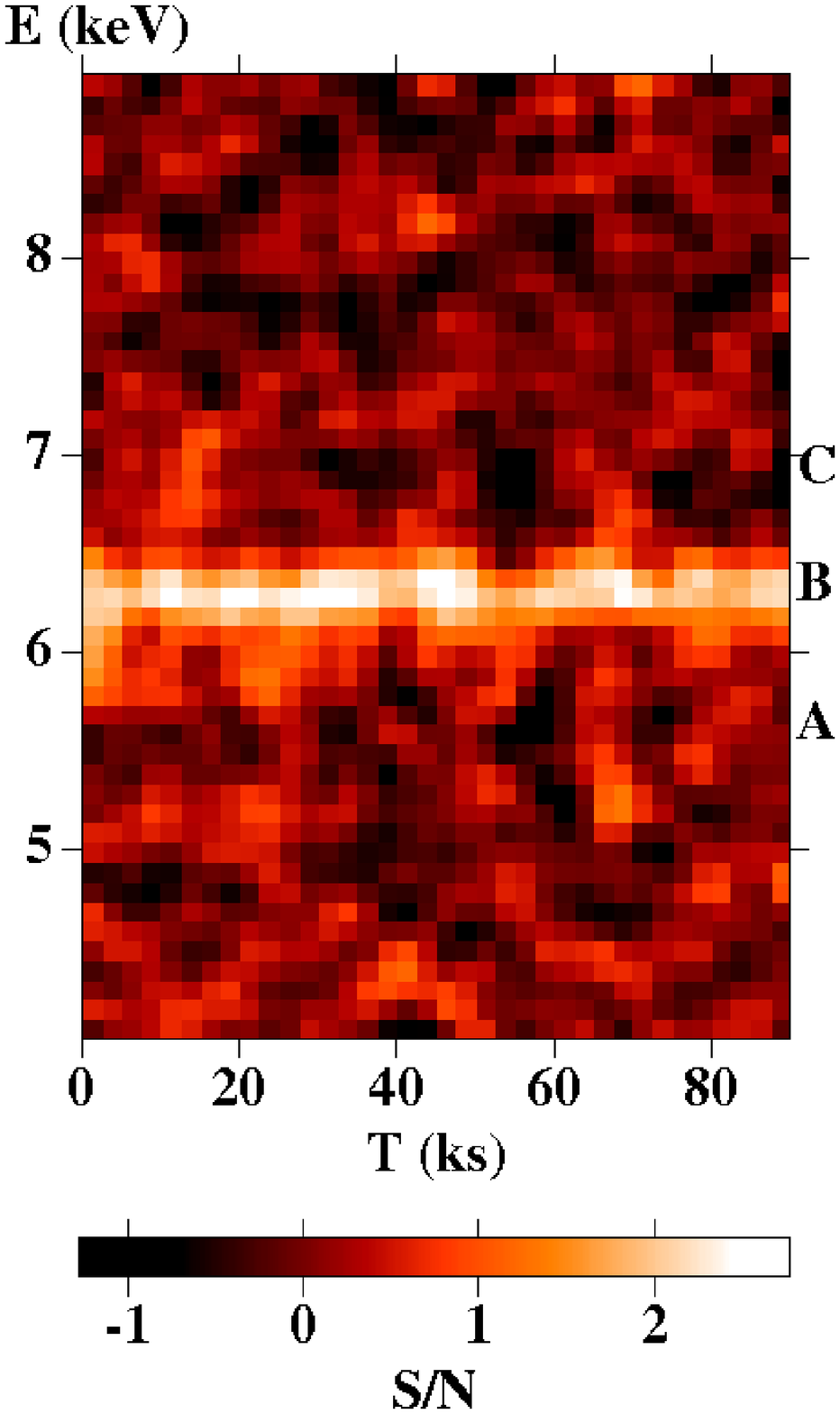} & \includegraphics[width=4.8cm,height=4.0cm]{figure/light_curves/NGC3516_130ks_lc.ps}\\
 & & \\
\end{tabular}
\end{figure*}

\begin{figure*}
\caption{\emph{Left panels}: data-to-model (power law plus cold absorption) ratios of the 4--9 keV time-averaged spectra; \emph{middle panels}: S/N map of excess residuals in the time-energy plane (at a time resolution of 2500 s); \emph{right panels}: 0.3--10 keV background subtracted continuum and residuals light curves (in bands A, B and C). The light curves are renormalized for the corresponding average flux.}
\label{fig:maps2}
\centering
\vspace{1.5cm}
\begin{tabular}{p{4cm}p{4cm}p{4cm}}
\includegraphics[height=4.0cm,width=4.0cm]{figure/spectra/NGC3516_52ks_spec.ps} & \includegraphics[height=4.2cm,width=4.0cm]{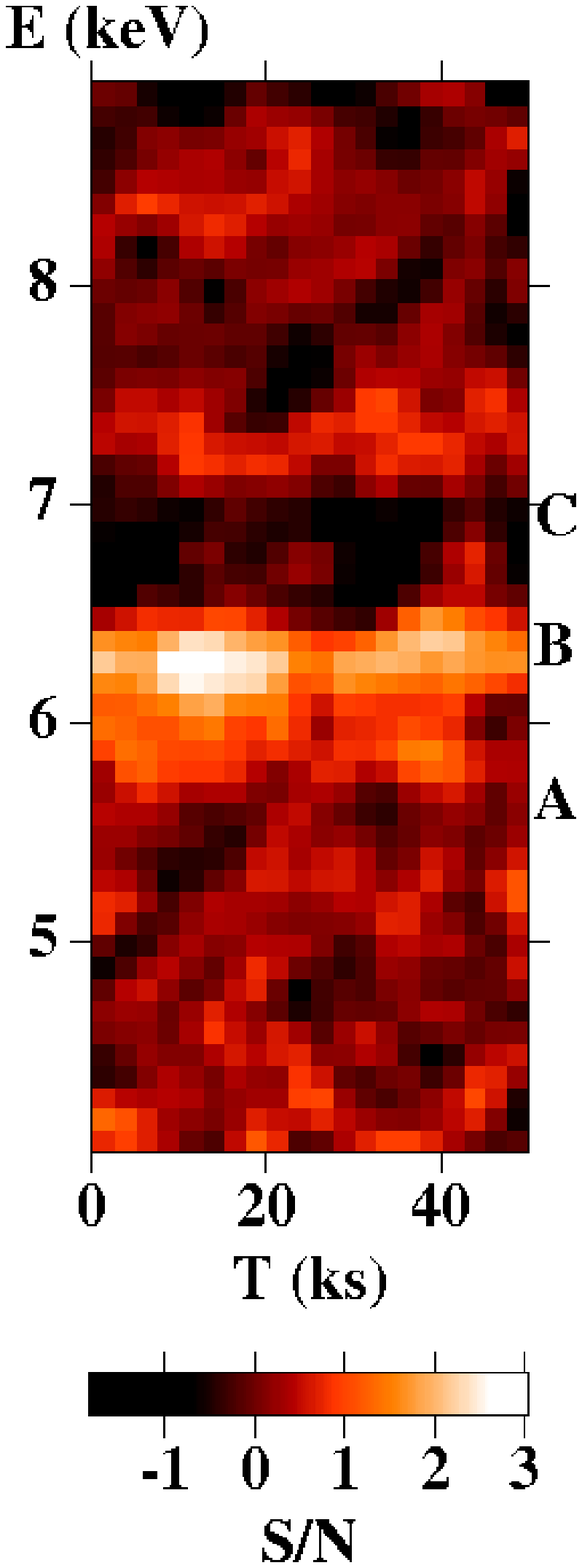} & \includegraphics[width=4.8cm,height=4.0cm]{figure/light_curves/NGC3516_52ks_lc.ps}\\
 & & \\
\includegraphics[height=4.0cm,width=4.0cm]{figure/spectra/NGC3516_69ks_spec.ps} & \includegraphics[height=4.2cm,width=4.0cm]{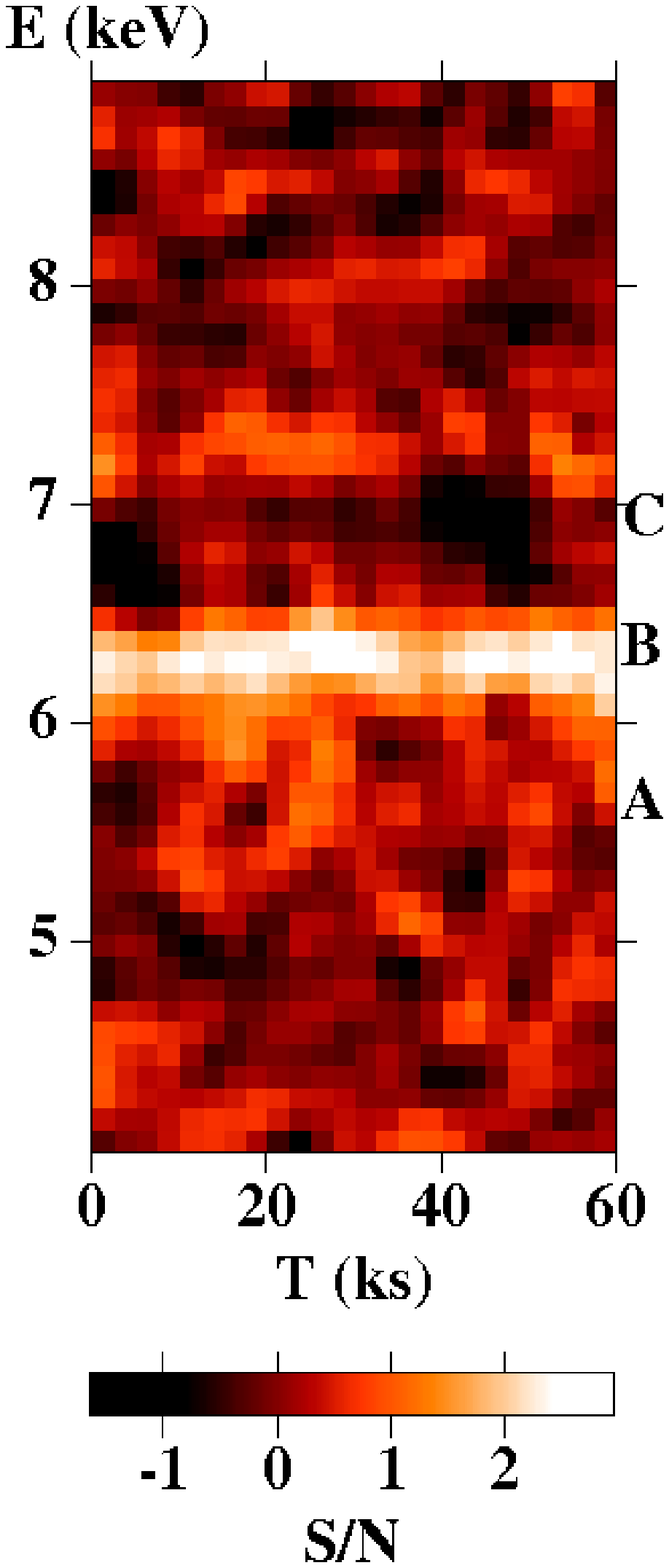} & \includegraphics[width=4.8cm,height=4.0cm]{figure/light_curves/NGC3516_69ks_lc.ps}\\
 & & \\
\includegraphics[height=4.0cm,width=4.0cm]{figure/spectra/NGC3516_68ksA_spec.ps} & \includegraphics[height=4.2cm,width=4.0cm]{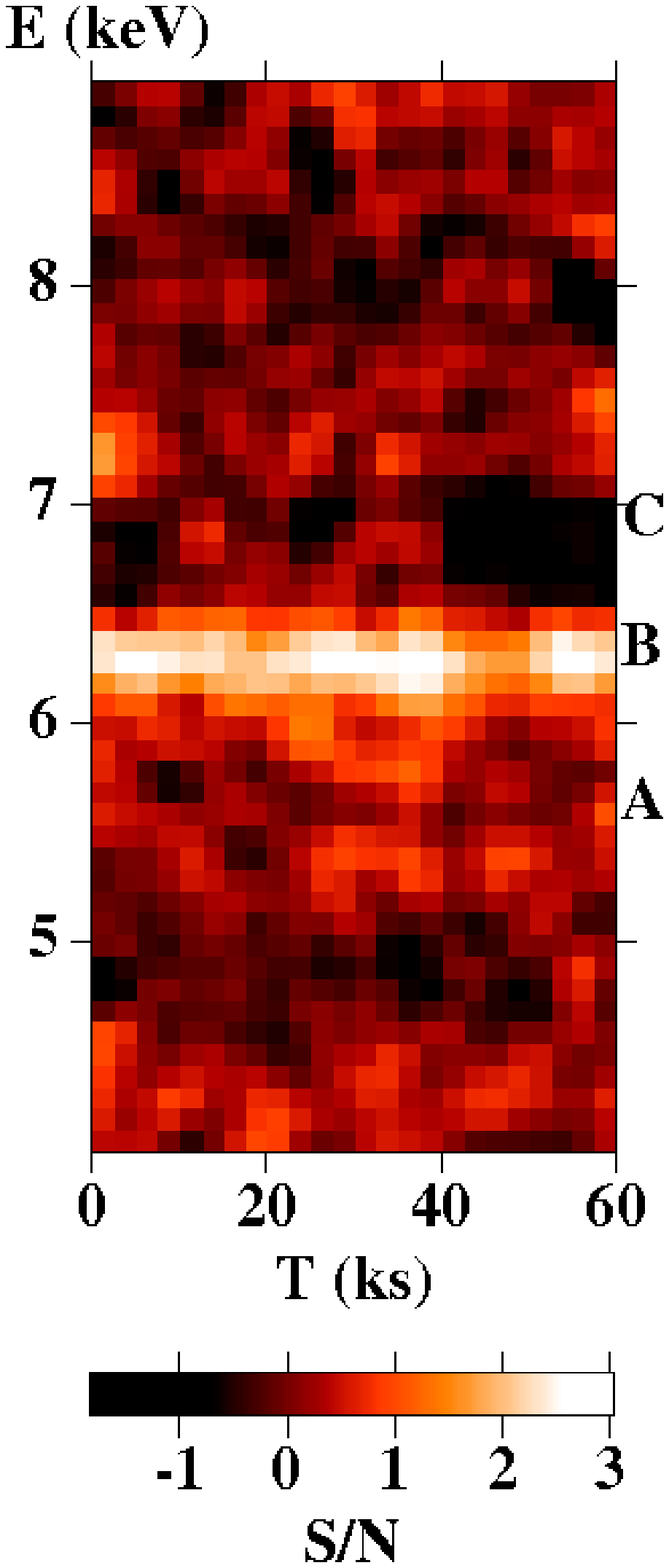} & \includegraphics[width=4.8cm,height=4.0cm]{figure/light_curves/NGC3516_68ksA_lc.ps}\\
 & & \\
\includegraphics[height=4.0cm,width=4.0cm]{figure/spectra/MRK509_31ks_spec.ps} & \includegraphics[height=4.2cm,width=4.0cm]{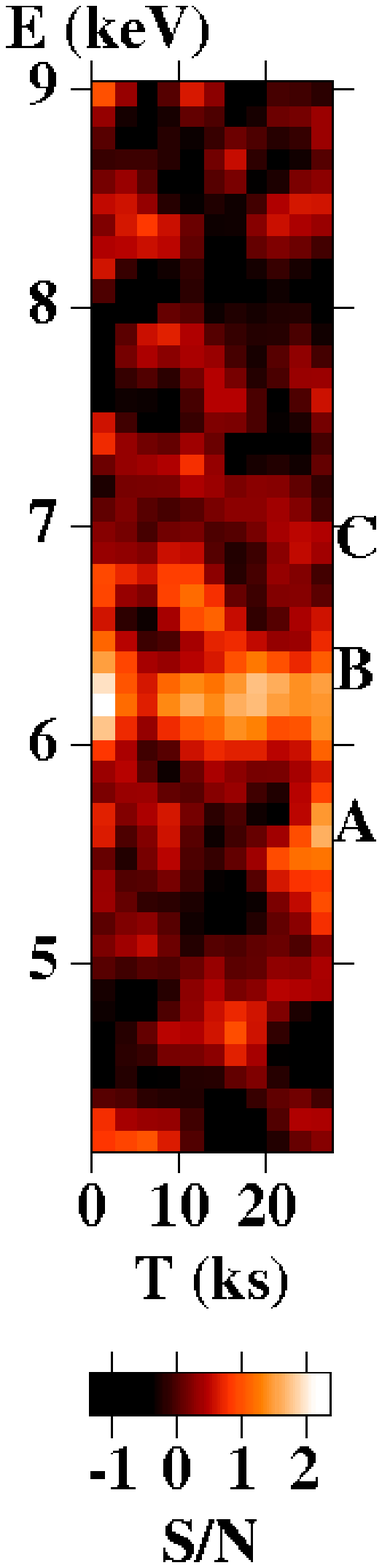} & \includegraphics[width=4.8cm,height=4.0cm]{figure/light_curves/MRK509_31ks_lc.ps}\\
 & & \\
\includegraphics[height=4.0cm,width=4.0cm]{figure/spectra/MRK509_86ks_spec.ps} & \includegraphics[height=4.2cm,width=4.0cm]{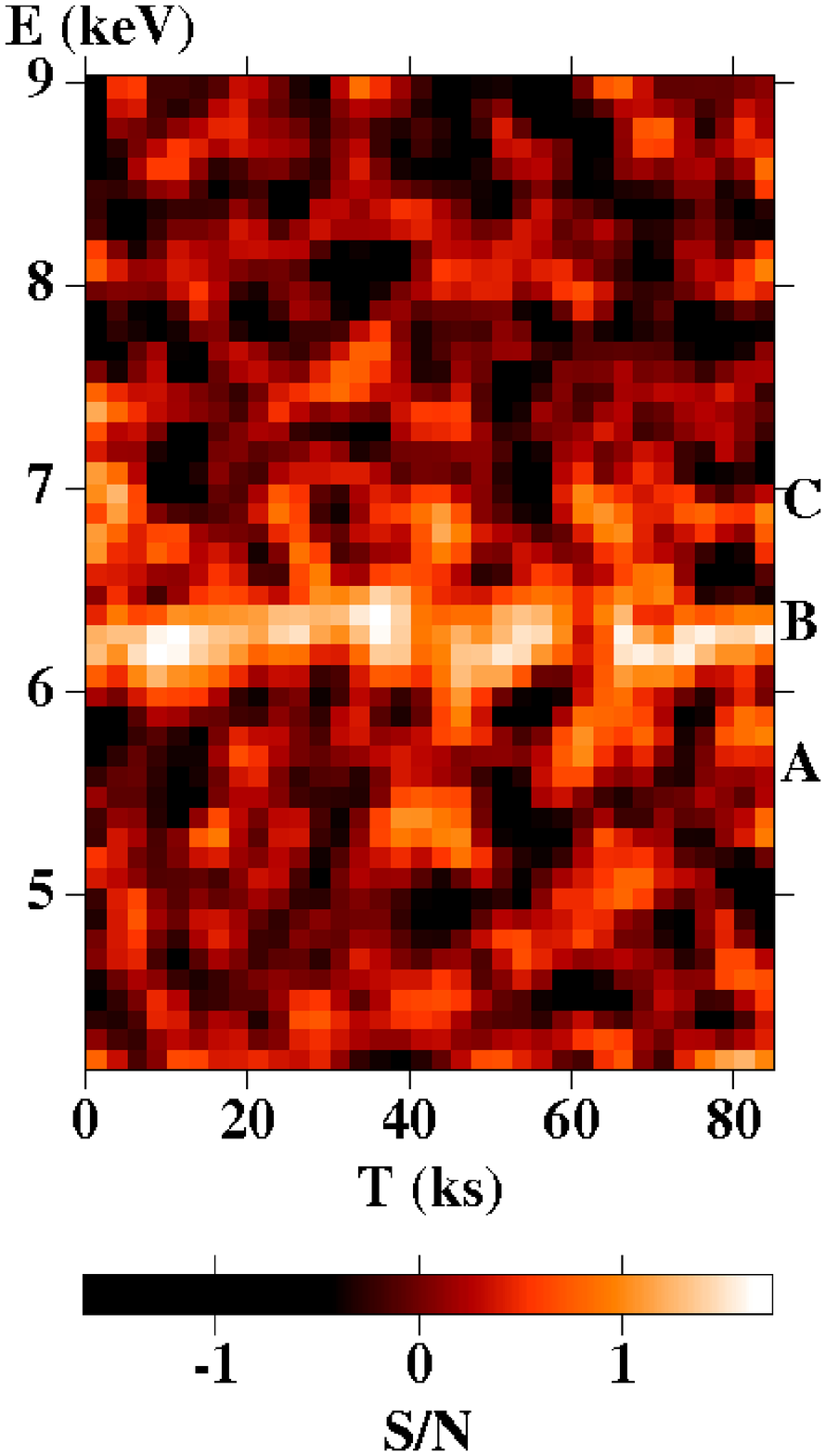} & \includegraphics[width=4.8cm,height=4.0cm]{figure/light_curves/MRK509_86ks_lc.ps}\\
 & & \\
\end{tabular}
\end{figure*}

\begin{figure*}
\caption{\emph{Left panels}: data-to-model (power law plus cold absorption) ratios of the 4--9 keV time-averaged spectra; \emph{middle panels}: S/N map of excess residuals in the time-energy plane (at a time resolution of 2500 s); \emph{right panels}: 0.3--10 keV background subtracted continuum and residuals light curves (in bands A, B and C). The light curves are renormalized for the corresponding average flux.}
\label{fig:maps3}
\centering
\vspace{1.5cm}
\begin{tabular}{p{4cm}p{4cm}p{4cm}}
\includegraphics[height=4.0cm,width=4.0cm]{figure/spectra/MRK509_70ks_spec.ps} & \includegraphics[height=4.2cm,width=4.0cm]{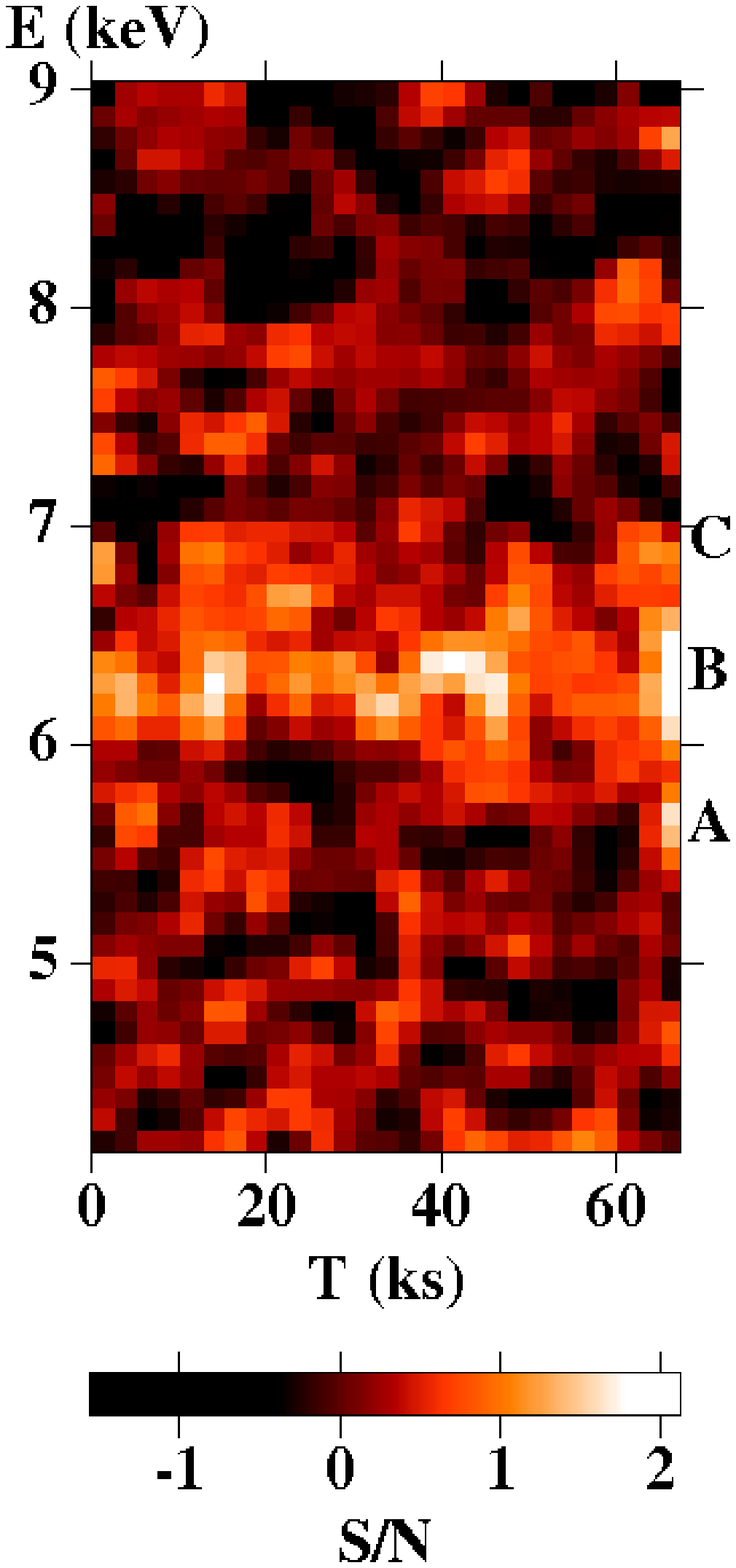} & \includegraphics[width=4.8cm,height=4.0cm]{figure/light_curves/MRK509_70ks_lc.ps}\\
 & & \\
\includegraphics[height=4.0cm,width=4.0cm]{figure/spectra/MCG-6-30-15_89ks_spec.ps} & \includegraphics[height=4.2cm,width=4.0cm]{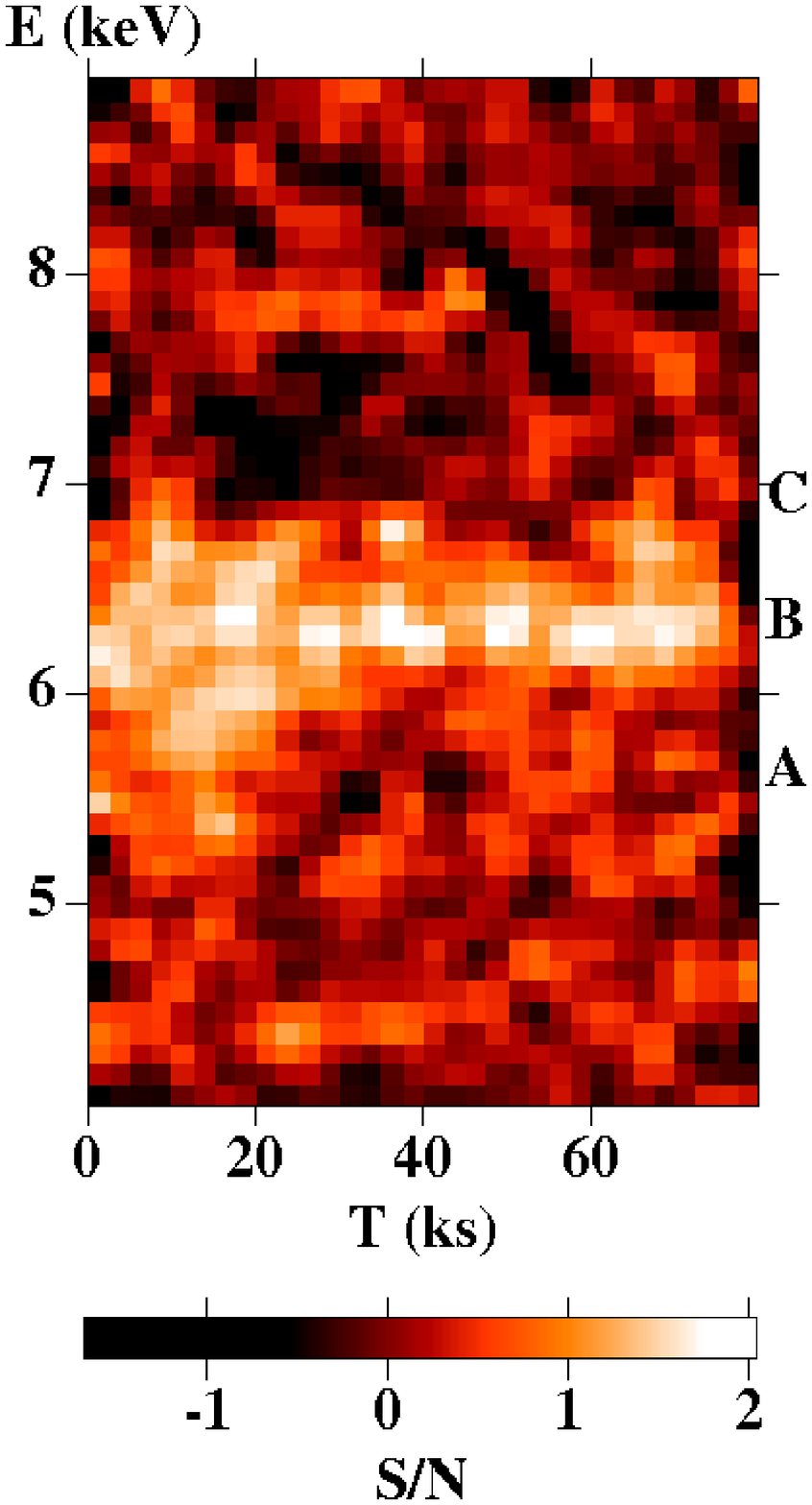} & \includegraphics[width=4.8cm,height=4.0cm]{figure/light_curves/MCG-6-30-15_89ks_lc.ps}\\
 & & \\
\includegraphics[height=4.0cm,width=4.0cm]{figure/spectra/MCG-6-30-15_129ks_spec.ps} & \includegraphics[height=4.2cm,width=4.0cm]{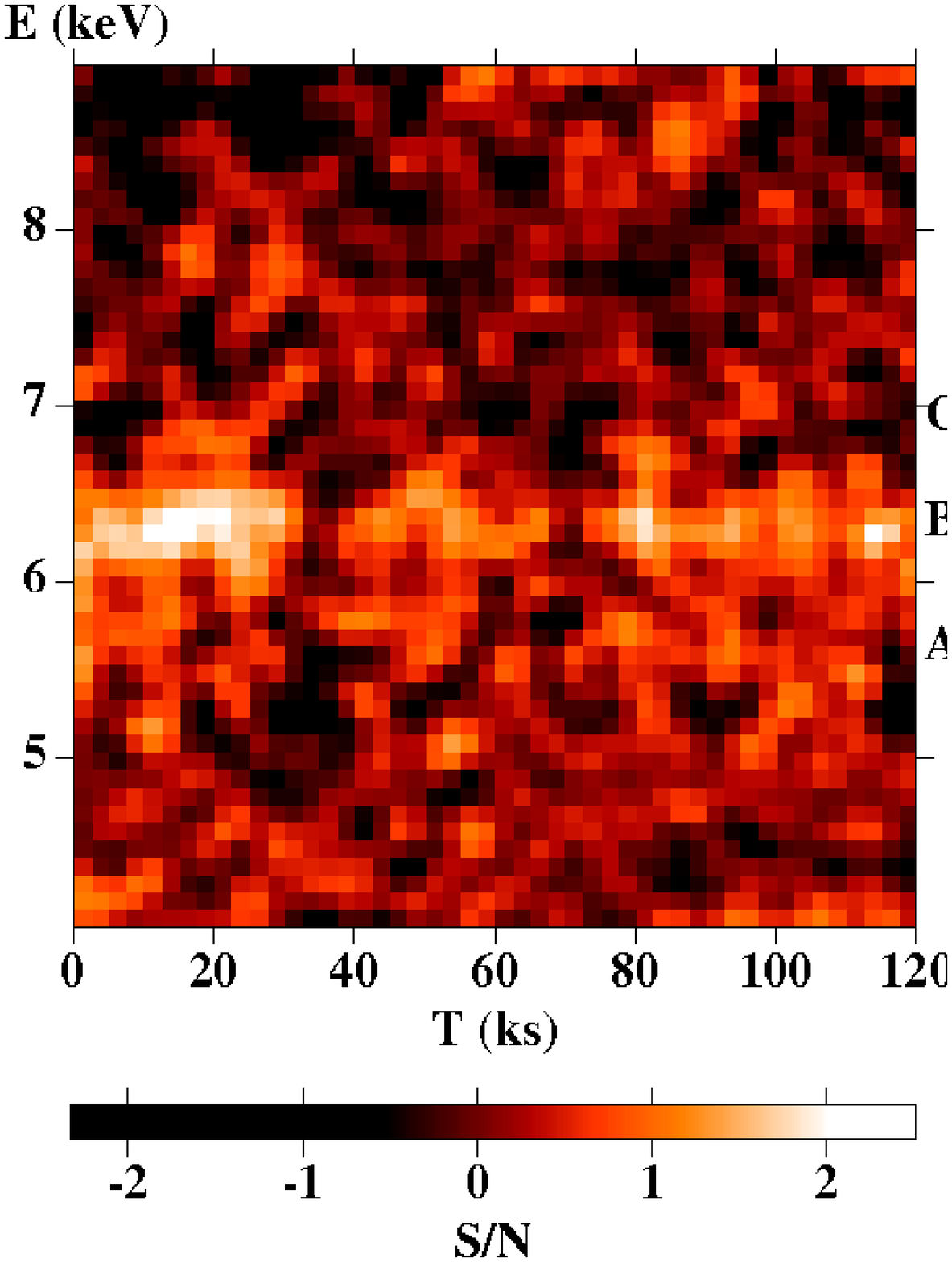} & \includegraphics[width=4.8cm,height=4.0cm]{figure/light_curves/MCG-6-30-15_129ks_lc.ps}\\
 & & \\
\includegraphics[height=4.0cm,width=4.0cm]{figure/spectra/MCG-6-30-15_66ks_spec.ps} & \includegraphics[height=4.2cm,width=4.0cm]{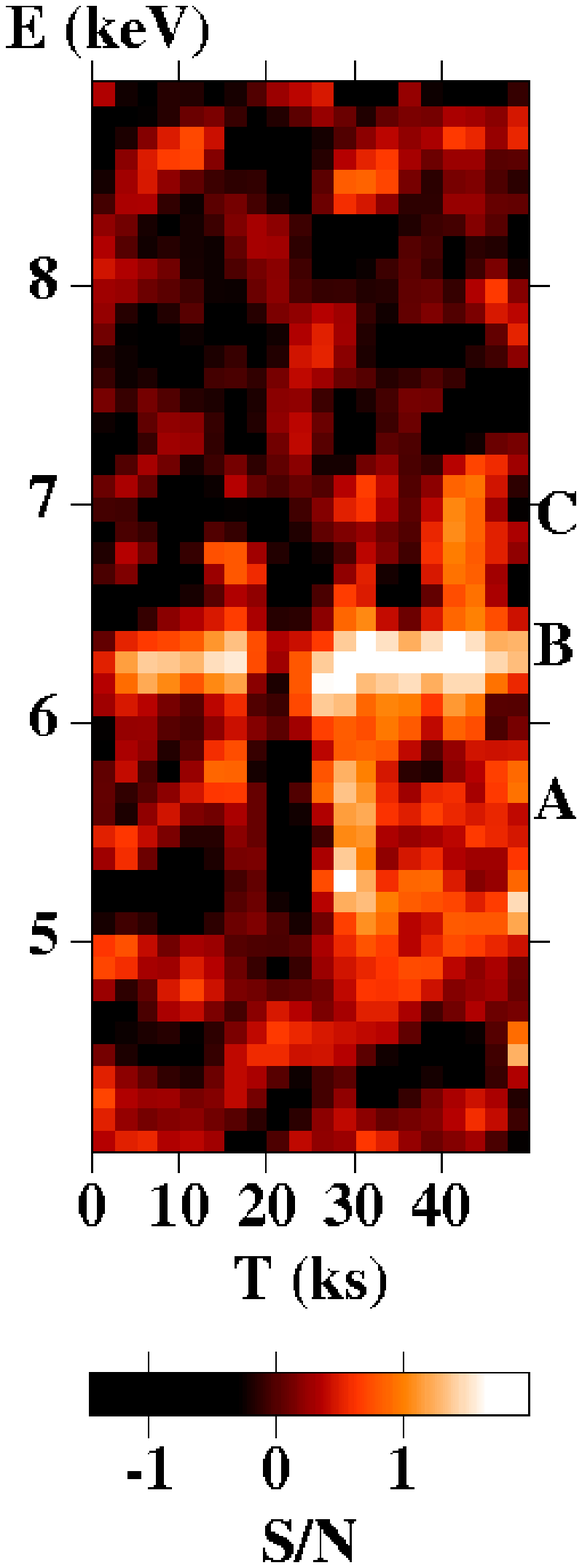} & \includegraphics[width=4.8cm,height=4.0cm]{figure/light_curves/MCG-6-30-15_66ks_lc.ps}\\
 & & \\
\includegraphics[height=4.0cm,width=4.0cm]{figure/spectra/NGC7314_44ks_spec.ps} & \includegraphics[height=4.2cm,width=4.0cm]{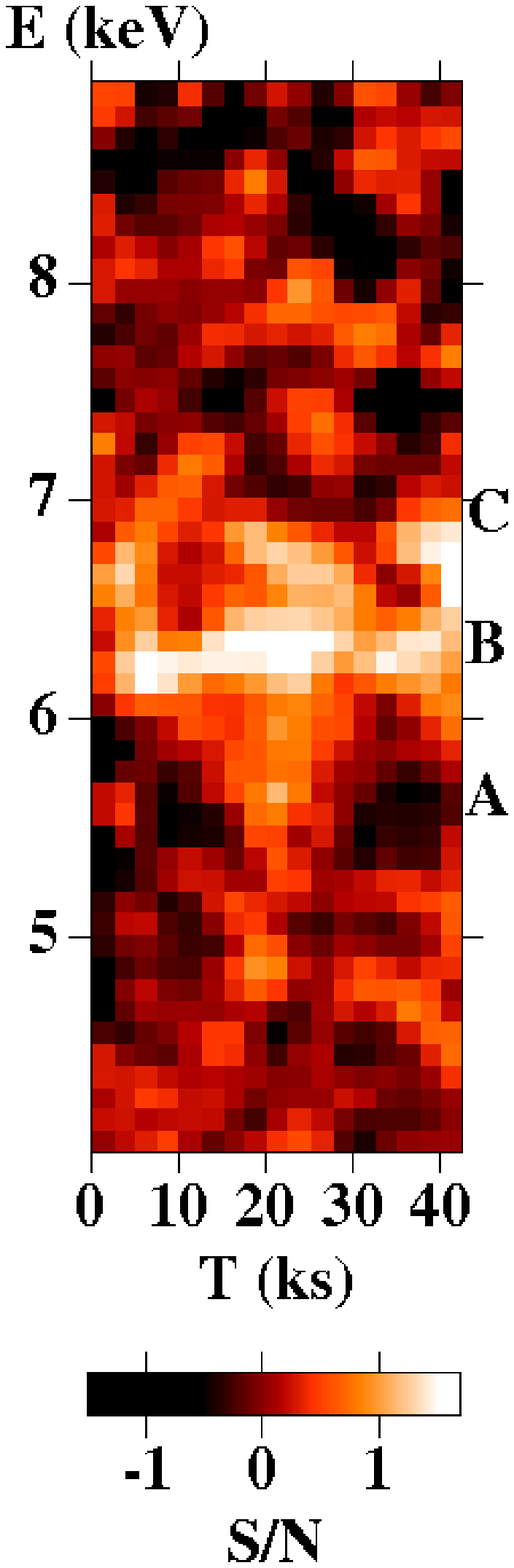} & \includegraphics[width=4.8cm,height=4.0cm]{figure/light_curves/NGC7314_44ks_lc.ps}\\
 & & \\

\end{tabular}
\end{figure*}

\begin{figure*}
\caption{\emph{Left panels}: data-to-model (power law plus cold absorption) ratios of the 4--9 keV time-averaged spectra; \emph{middle panels}: S/N map of excess residuals in the time-energy plane (at a time resolution of 2500 s); \emph{right panels}: 0.3--10 keV background subtracted continuum and residuals light curves (in bands A, B and C). The light curves are renormalized for the corresponding average flux.}
\label{fig:maps4}
\centering
\vspace{1.5cm}
\begin{tabular}{p{4cm}p{4cm}p{4cm}}
\includegraphics[height=4.0cm,width=4.0cm]{figure/spectra/MKN279_38ks_spec.ps} & \includegraphics[height=4.2cm,width=4.0cm]{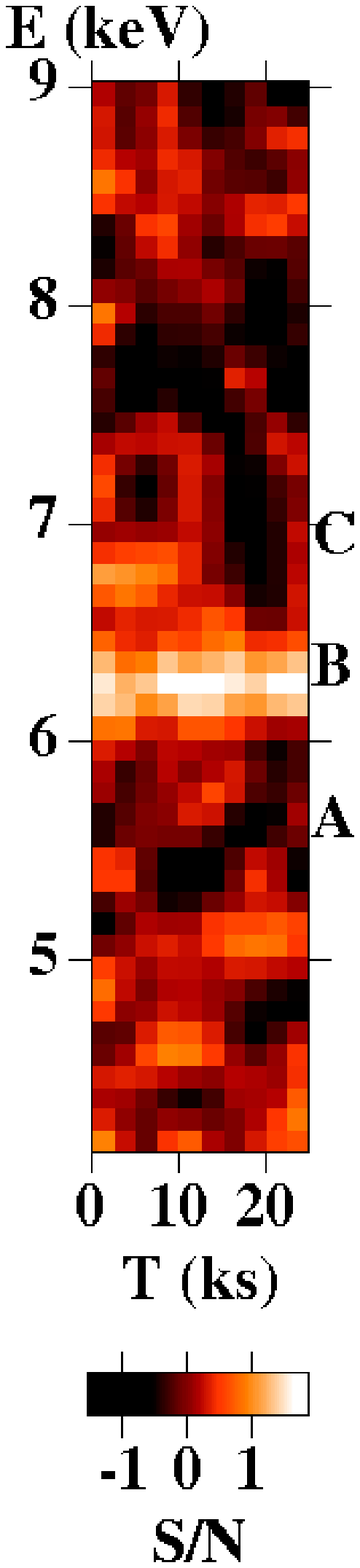} & \includegraphics[width=4.8cm,height=4.0cm]{figure/light_curves/MKN279_38ks_lc.ps}\\
 & & \\
\includegraphics[height=4.0cm,width=4.0cm]{figure/spectra/MRK766_99ksC_spec.ps} & \includegraphics[height=4.2cm,width=4.0cm]{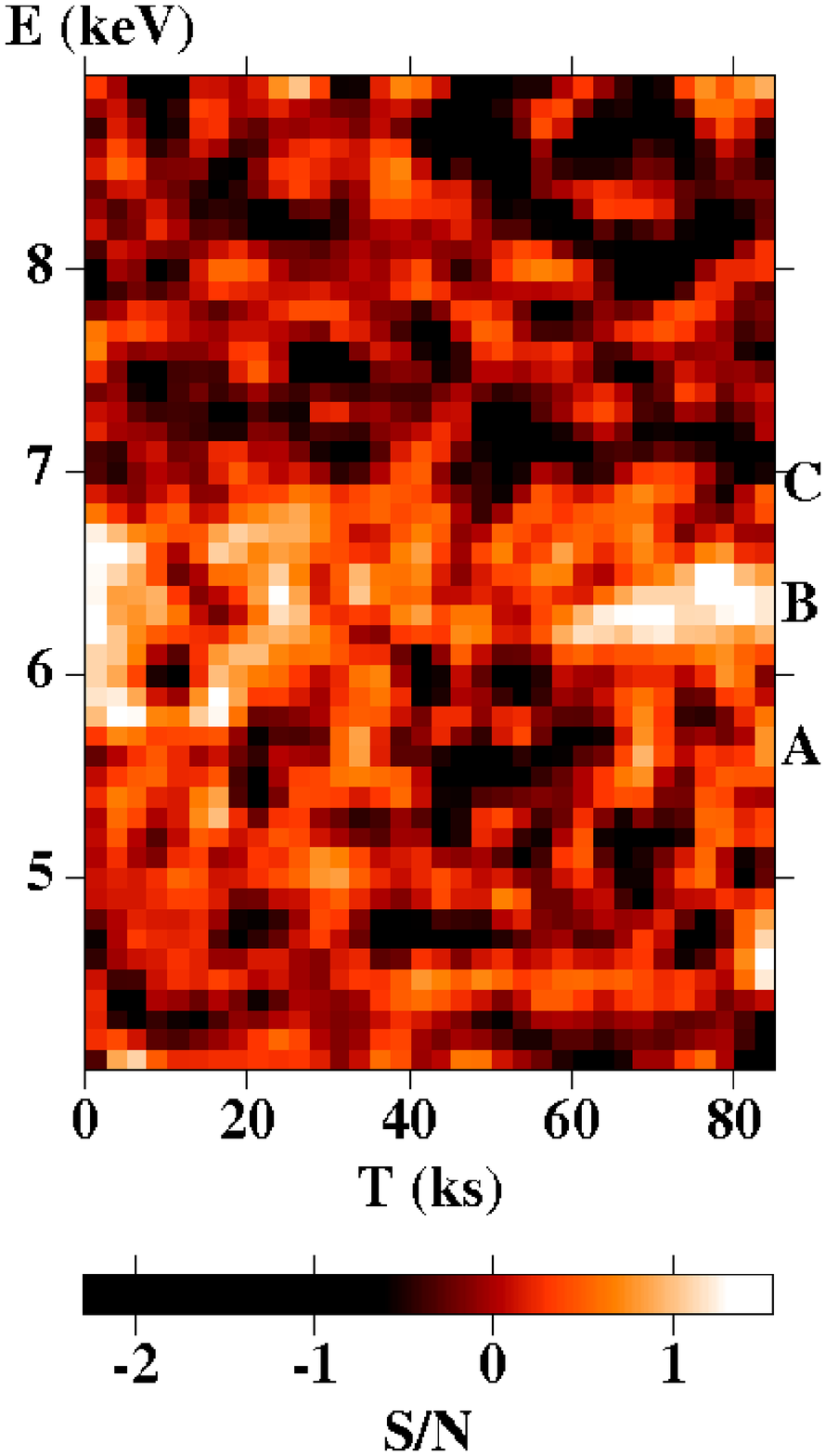} & \includegraphics[width=4.8cm,height=4.0cm]{figure/light_curves/MRK766_99ksC_lc.ps}\\
 & & \\
\includegraphics[height=4.0cm,width=4.0cm]{figure/spectra/MRK766_35ks_spec.ps} & \includegraphics[height=4.2cm,width=4.0cm]{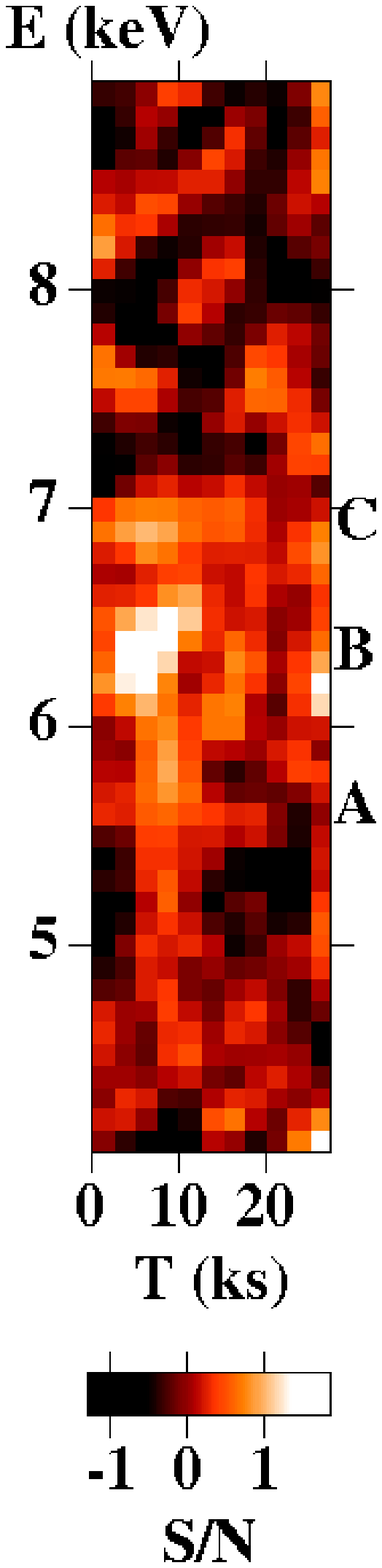} & \includegraphics[width=4.8cm,height=4.0cm]{figure/light_curves/MRK766_35ks_lc.ps}\\

\includegraphics[height=4.0cm,width=4.0cm]{figure/spectra/NGC7469_19ks_spec.ps} & \includegraphics[height=4.2cm,width=4.0cm]{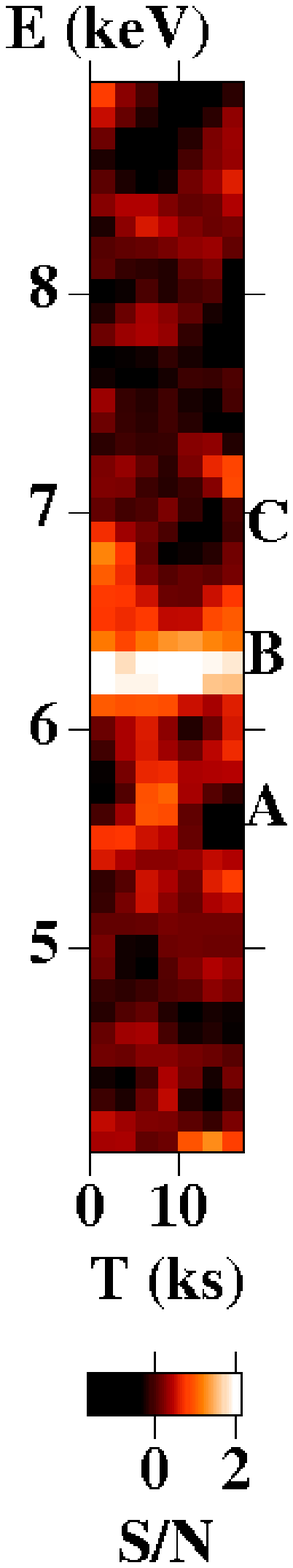} & \includegraphics[width=4.8cm,height=4.0cm]{figure/light_curves/NGC7469_19ks_lc.ps}\\
 & & \\
\includegraphics[height=4.0cm,width=4.0cm]{figure/spectra/NGC7469_25ks_spec.ps} & \includegraphics[height=4.2cm,width=4.0cm]{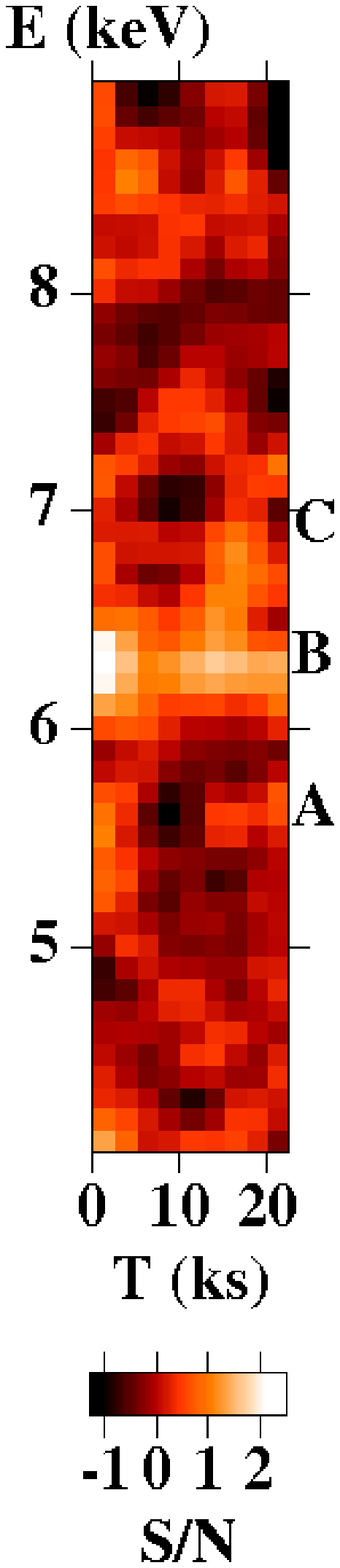} & \includegraphics[width=4.8cm,height=4.0cm]{figure/light_curves/NGC7469_25ks_lc.ps}\\
 & & \\
\end{tabular}
\end{figure*}

\begin{figure*}
\caption{\emph{Left panels}: data-to-model (power law plus cold absorption) ratios of the 4--9 keV time-averaged spectra; \emph{middle panels}: S/N map of excess residuals in the time-energy plane (at a time resolution of 2500 s); \emph{right panels}: 0.3--10 keV background subtracted continuum and residuals light curves (in bands A, B and C). The light curves are renormalized for the corresponding average flux.}
\label{fig:maps5}
\centering
\vspace{1.5cm}
\begin{tabular}{p{4cm}p{4cm}p{4cm}}
\includegraphics[height=4.0cm,width=4.0cm]{figure/spectra/ESO141-G055_31ks_spec.ps} & \includegraphics[height=4.2cm,width=4.0cm]{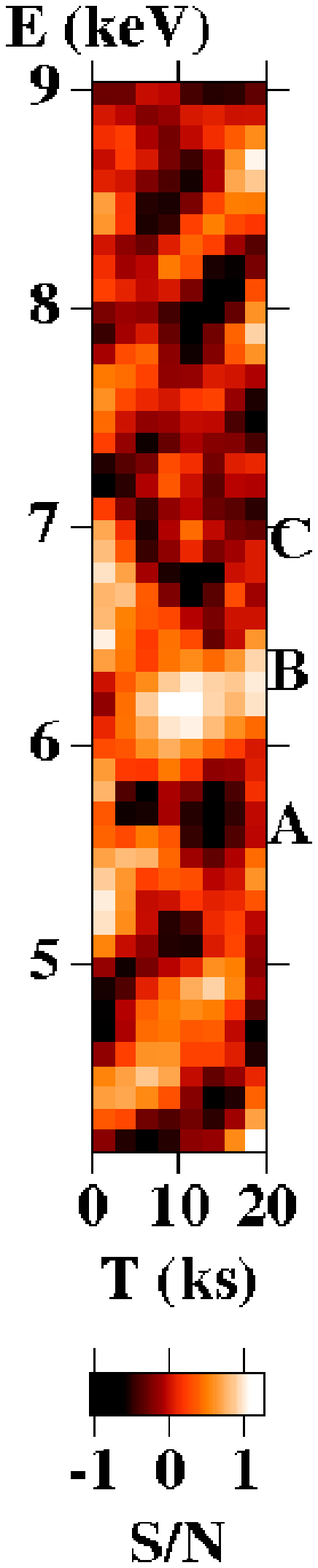} & \includegraphics[width=4.8cm,height=4.0cm]{figure/light_curves/ESO141-G055_31ks_lc.ps}\\
 & & \\ 
\includegraphics[height=4.0cm,width=4.0cm]{figure/spectra/MRK79_20ksA_spec.ps} & \includegraphics[height=4.2cm,width=4.0cm]{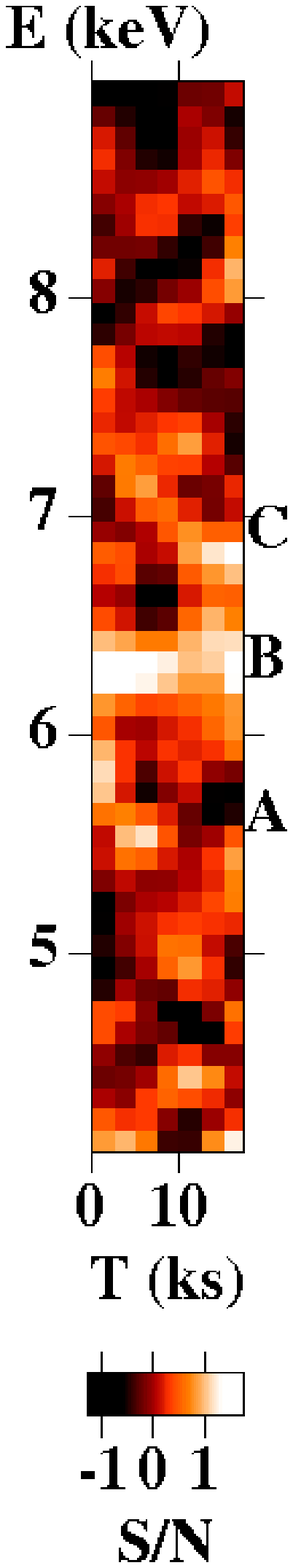} & \includegraphics[width=4.8cm,height=4.0cm]{figure/light_curves/MRK79_20ksA_lc.ps}\\
 & & \\
\includegraphics[height=4.0cm,width=4.0cm]{figure/spectra/MRK79_89ks_spec.ps} & \includegraphics[height=4.2cm,width=4.0cm]{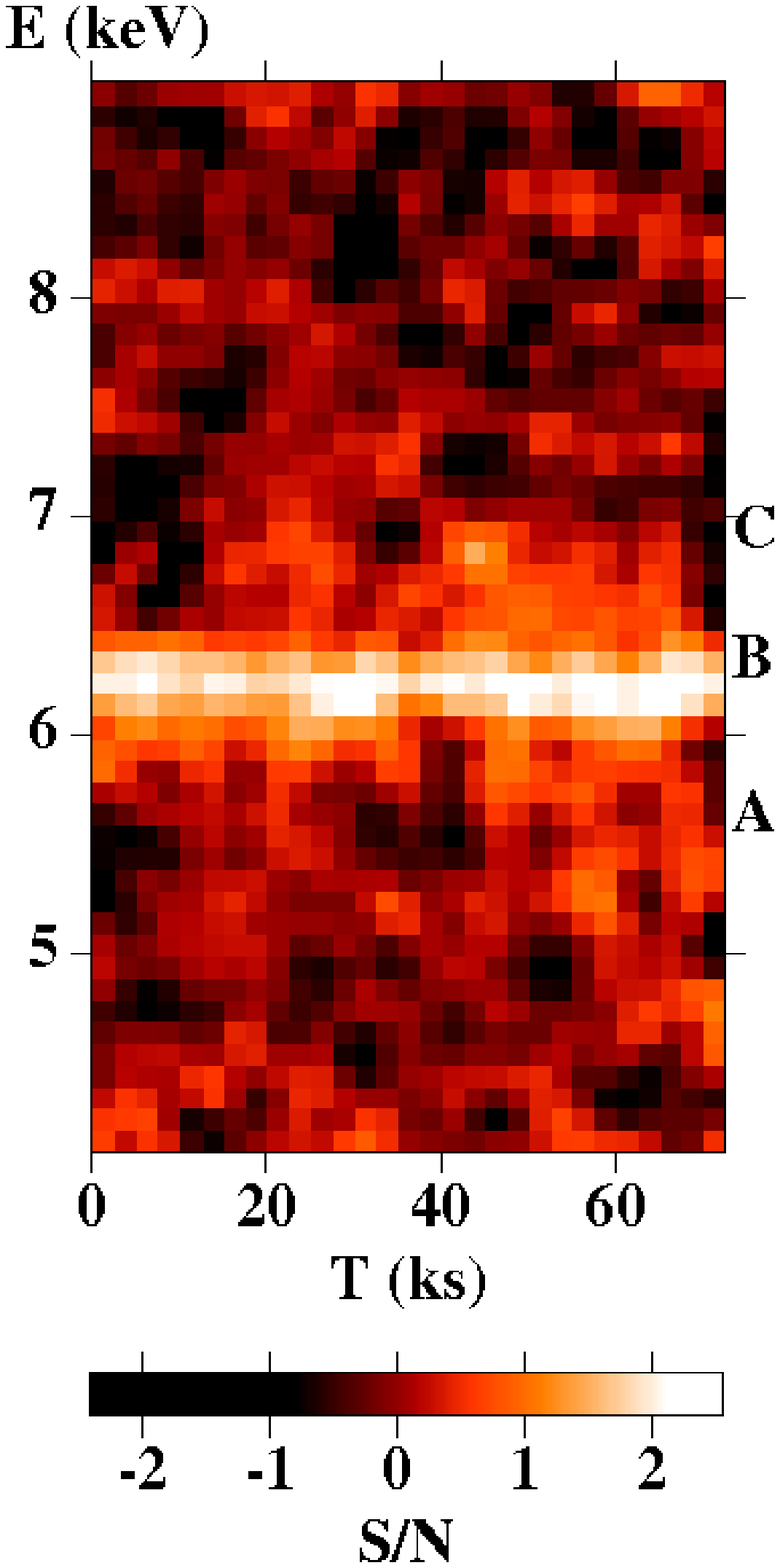} & \includegraphics[width=4.8cm,height=4.0cm]{figure/light_curves/MRK79_89ks_lc.ps}\\
 & & \\
\includegraphics[height=4.0cm,width=4.0cm]{figure/spectra/ESO198-G024_34ks_spec.ps} & \includegraphics[height=4.2cm,width=4.0cm]{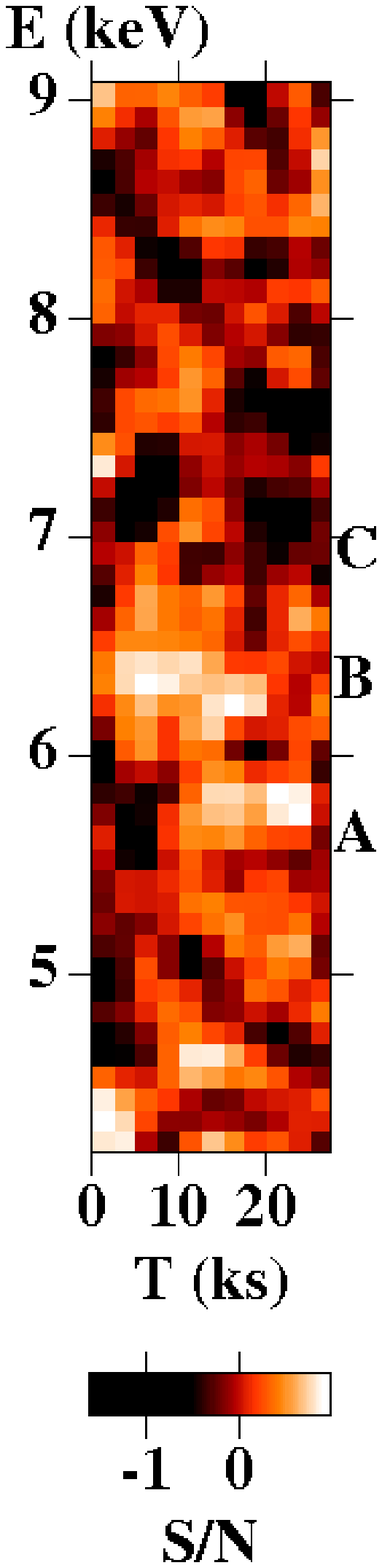} & \includegraphics[width=4.8cm,height=4.0cm]{figure/light_curves/ESO198-G024_34ks_lc.ps}\\
 & & \\
\includegraphics[height=4.0cm,width=4.0cm]{figure/spectra/AKN564_102ks_spec.ps} & \includegraphics[height=4.2cm,width=4.0cm]{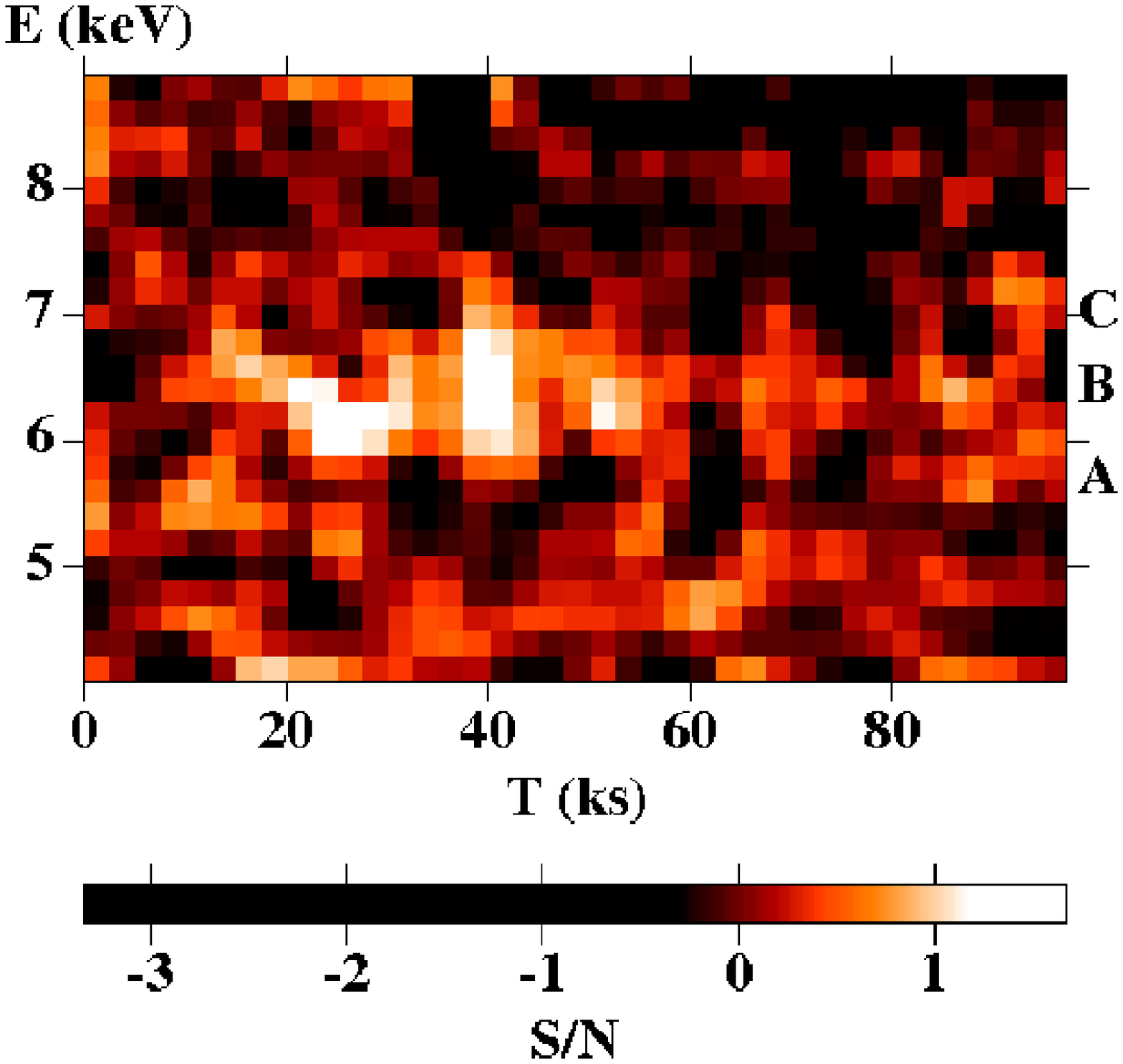} & \includegraphics[width=4.8cm,height=4.0cm]{figure/light_curves/AKN564_102ks_lc.ps}\\
\end{tabular}
\end{figure*}

\begin{figure*}
\caption{\emph{Left panels}: data-to-model (power law plus cold absorption) ratios of the 4--9 keV time-averaged spectra; \emph{middle panels}: S/N map of excess residuals in the time-energy plane (at a time resolution of 2500 s); \emph{right panels}: 0.3--10 keV background subtracted continuum and residuals light curves (in bands A, B and C). The light curves are renormalized for the corresponding average flux.}
\label{fig:maps6}
\centering
\vspace{1.5cm}
\begin{tabular}{p{4cm}p{4cm}p{4cm}}
\includegraphics[height=4.0cm,width=4.0cm]{figure/spectra/MRK704_22ks_spec.ps} & \includegraphics[height=4.2cm,width=4.0cm]{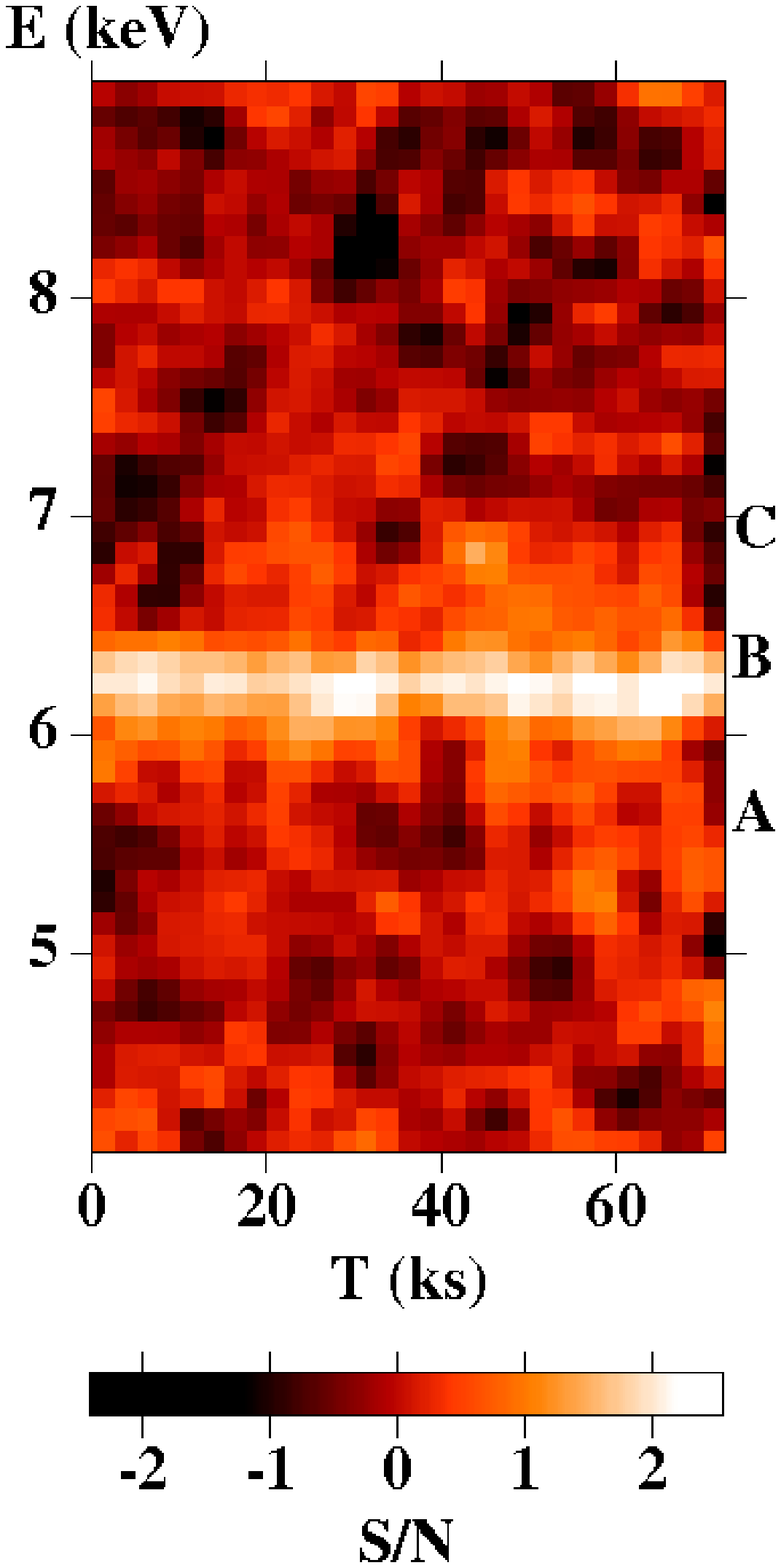} & \includegraphics[width=4.8cm,height=4.0cm]{figure/light_curves/MRK704_22ks_lc.ps}\\

\end{tabular}
\end{figure*}

The method used to produce excess time-energy maps is extensively described in
IMF04 and Tombesi et al. (2007) and briefly summarized in the following:
\begin{description}
\item[i)] each observation was divided into slices according to the chosen
  time resolution;
\item[ii)] for each slice, the continuum was determined by excluding the Fe K energy band (i.e. typically at E$\sim$5--7 keV), and by rebinning the spectrum to
  have at least 50 counts per energy bin in order to garantee a reliable continuum determination in the high energy part of the spectrum (E$>$7.5 keV);
\item[iii)] the baseline spectral model for the continuum is a power law plus
  cold absorption (with the column density N$_{H}$ parameter fixed at the value obtained from the average spectrum);
\item[iv)] data were then rebinned at the energy resolution of the
  map. Residuals to the best fit continuum model were computed in the
  4--9 keV energy band and visualized on a time vs energy plane;
\item[v)] a circular Gaussian low-pass filter with $\sigma_{lp}$ = 0.85 pixel is used to smooth the image in the time-energy plane so as to reduce random noise between adjacent pixels.
\end{description}
In the maps, signal--to--noise (S/N)
ratios of residual flux are displayed (see Figs. \ref{fig:maps1}--\ref{fig:maps6}). The noise includes contributions from uncertainties in the subtracted background and power law model components,
and in the excess residual counts.\\
The excess map technique is useful to identify spectrally and temporally transient, narrow features and trace their energy/intensity temporal evolution.
 As these features could appear at an energy which is not known apriori we defined three physically motivated energy bands for the extraction of residuals light curves, which allowed to infer the statistical significance of such structures in a uniform and systematic way:
\begin{description}
\item[A:] 5.4--6.1 keV, red shifted Fe K$\alpha$ line band;
\item[B:] 6.1--6.8 keV, neutral and/or mildly ionized Fe K$\alpha$ line band;
\item[C:] 6.8--7.2 keV, Fe K$\beta$ and/or highly ionized Fe K$\alpha$ band.
\end{description}
 Despite the widths of the chosen energy bands are greater than the spectral binning of the maps, only relatively narrow features are picked up in the residuals analysis. Any other broad feature spreading across the entire width of each band would be accounted for by the continuum modelling for each time resolved spectrum. Moreover, adopting wider energy intervals allows to increase the statistics in the integrated residual counts.\\
 Light curves of the resulting excess flux and relative excess maps are reported for the observations where significant variability has been detected (see Figs. \ref{fig:maps1}--\ref{fig:maps6}). The light curves are renormalized for their corresponding average value. In the figures the maps are all shown at a time resolution of 2.5 ks, to better display excess residuals.
The procedure used to calculate the light curves errors is outlined in Sect. \ref{errors}.

\subsection{Light curves error estimation and variability significance}
\label{errors}

The excess map technique is useful to reveal the presence of narrow transient
features (IMF04) and is not affected by the presence of broad line components (as the adopted continuum model would mimic the shape of such lines). The smoothing process applied
(see previous section) prevents, however, the direct determination of the lines
significance. For instance, the errors on the flux measurements in the time
series cannot be estimated using a counting statistics as, after the smoothing,
adjacent pixels in the map are no more independent. Hence we followed the
procedure described in IMF04 to assess the errors
via simulations. We simulated N$_{sim}=$1000 time-energy maps, assuming
 constant spectral components (whose parameters are those of the average spectrum best-fit model, as reported in Table \ref{tab:models}) and a power law normalization
varying as the 0.3--10 keV light curve\footnote{In the case of MCG-6-30-15 the curvature induced by the broad Fe line is treated as due to continuum emission (see Appendix). Since the broad line (E$\sim$3--7 keV) is less variable than the broad band 0.3--10 keV continuum (e.g. Vaughan \& Fabian 2004), the power law normalization is assumed to follow the 4--9 keV light curve in the simulated maps.}. These simulations are carried out using the \emph{fakeit} Xspec task, which allows to create fake source and background spectra with a Poisson noise distribution, which approaches the normal distribution whenever the number of counts per bin is $>$20.\\
The square root of the
  simulated light curves mean variance (in the band of interest)
  is thus taken as error of the real light curve. This is justified
  by the fact that in the simulations every Gaussian component flux is assumed
  constant (and consequently equal to zero in case of no line detection), hence a measure of the variance is a good estimate of the real time series errors.\\
The variability significance has been calculated by comparing the real light curve 
variance ($\sigma_{r}^{2}$) with the simulated ones ($\sigma_{s}^{2}$). If N
is the number of simulated light curves for which $\sigma_{s}^{2}$$>$$\sigma_{r}^{2}$, the variability significance is given by (1-N/N$_{sim}$). Table \ref{tab:varsig} contains a list of the computed variability significances in the considered energy bands.

\begin{figure}[hbtp!]
\centering

\includegraphics[scale=0.35,angle=0]{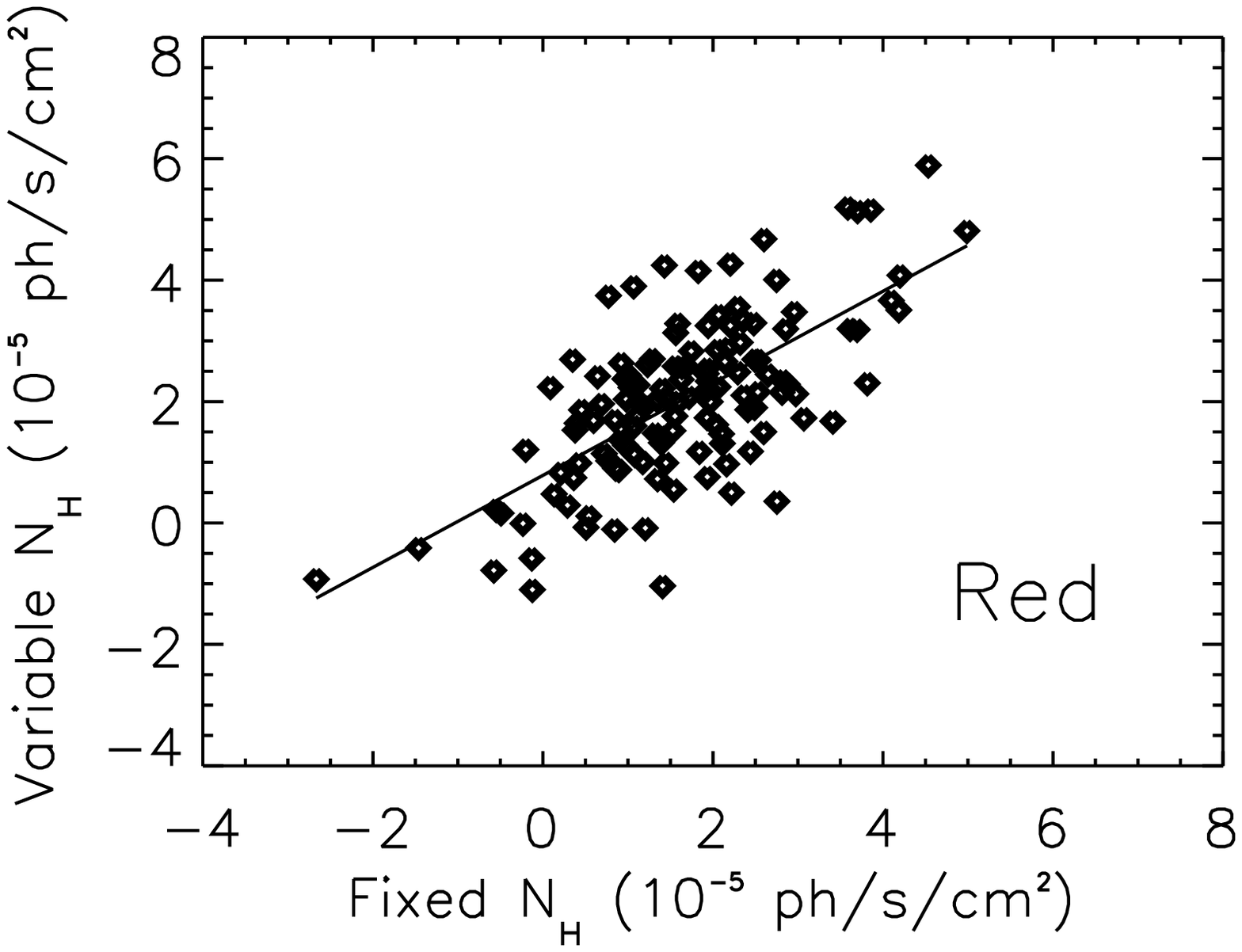}\hspace{0.5cm}
\includegraphics[scale=0.35,angle=0]{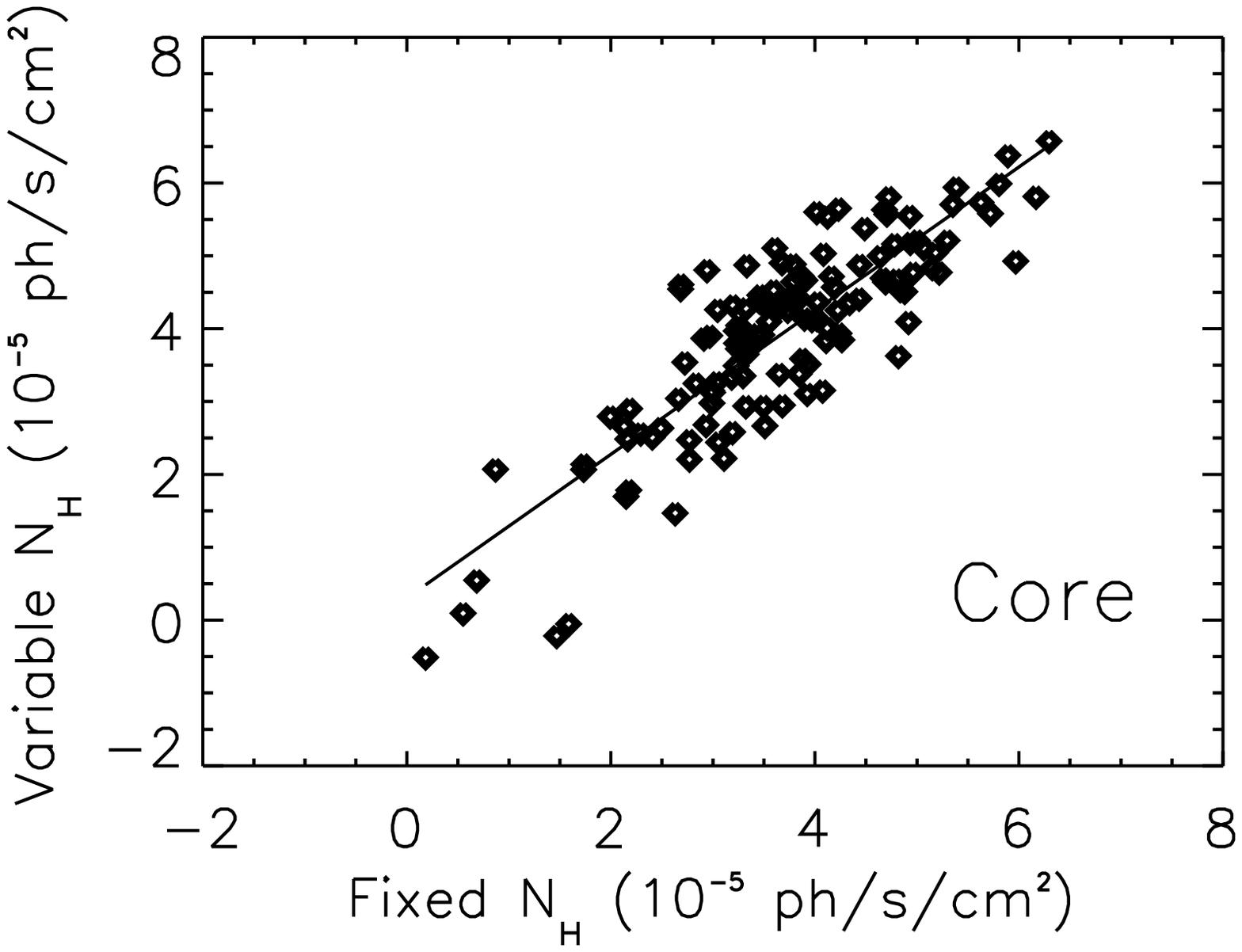}\hspace{0.5cm}
\includegraphics[scale=0.35,angle=0]{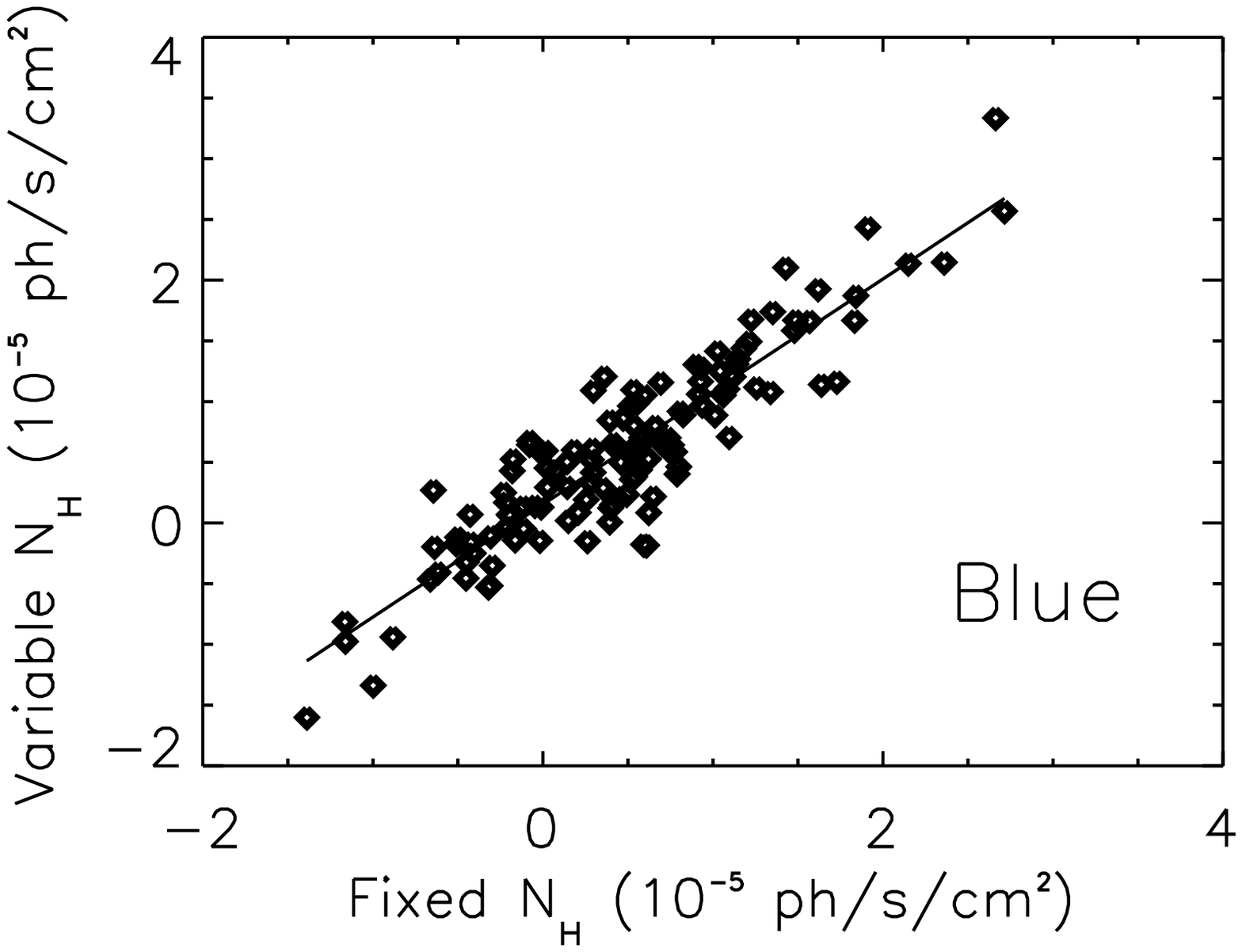}

  \caption{Comparison between line flux measurements for subtracted continuum model including a power law plus either a fixed or variable cold absorption N$_{H}$ parameter. The fluxes are obtained from residuals in the A, B and C energy bands (from the top to the bottom panel, respectively).}
\label{corr}
    \end{figure}

\subsection{Robustness against continuum modelling and background filtering}

The technique just summarized assumes a simple modelization for
  the underlying continuum (i.e. a power law plus cold absorption). As this
  assumption may affect the flux estimation of the variable features, we
  tested its robustness by considering a more complex continuum model, comprising an
  ionized absorber component (\emph{absori} model in Xspec, Done et
  al. 1992) rather than the cold absorption one. The test was performed for IC
  4329a, the brightest source of the sample, with the highest statistics to
  provide the most accurate estimate of the continuum. For each energy band, we found that the calculated residuals are within the flux errors derived by the simple power law plus cold absorption continuum fit. On average, the difference between residuals calculated with the two continuum models is around 30\% of the light curve error. This is due to several factors: the narrow band considered for the analysis (E$=$4--9 keV), the high energy resolution of the maps
  (suitable for the study of narrow spectral features) and the lower variability
  level of secondary components (i.e. reflection
  continuum and ionized absorber systems) with respect to the primary power
  law. Hence, we conclude that the underlying continuum in the chosen bandpass is well represented by a simple power law, with cold absorption being a good enough approximation for the modelling of
  the residual curvature.\\
It is worth noting that the reflection continuum associated to the narrow and constant Fe K$\alpha$ line dominates over that associated to any transient feature. This dominant reflection component will provide a constant contribution in the flux measurements. Hence it will not introduce spurious variability in the residuals maps.\\
As a consequence of the adopted technique and high energy resolution (i.e. $\Delta E=$0.1-0.2 keV), the maps will be sensitive only to narrow spectral features (i.e. both spectral lines and narrow structures of the relativistic line profile.)
 We further checked if the assumption of a fixed value for the cold absorption N$_{H}$ parameter in the continuum subtraction process (see Sect. \ref{sec:proced}) can produce spurious variable features (most of all in the redshifted energy band). This might be the case, for instance, in the presence of any residual curvature induced by variable and/or complex absorbers (i.e. warm absorber) and/or more complex continuum modelling.
The test is exploited for the long-look (rev. 301,302 and 303) of MCG -6-30-15, this source being characterized by a strongly variable X-ray flux and a complex spectrum (including both a warm absorber and a relativistic Fe line). In Figs.\ref{corr} we show a comparison between the residual line flux measurements (as obtained from the excess map analysis in band A, B and C) for the subtracted continuum model including (1) a variable or (2) a fixed N$_{H}$. As expected, the line fluxes are well correlated (the linear Pearson correlation coefficients are 0.68, 0.85 and 0.91 respectively for the three bands), although the scattering from the best fit linear model increases for lower energy bands, where the effects of continuum modelling are stronger. However the relative error due to the deviation in the line flux (1)-flux (2) plot from the linear model is smaller than the statistical error in the line flux measurement as obtained from simulations (see Sect. \ref{errors}). It is worth noticing that the scattering depends also on the energy and time resolutions adopted, decreasing for lower resolutions.\\
We conclude that our assumption of constant N$_{H}$ does not invalidate our results. Being the case of MCG -6-30-15 one of the most representative of a strongly variable bright source with broad relativistic Fe line emission, we infer that this assumption can be considered valid for all the other sources of the sample.

As some AGN show a spectral steepening when they get brighter, we also tested whether significant variations in the power law $\Gamma$ parameter, which might influence the line flux measurements, are registered during each observation where significant variability in at least one of the three energy bands has been observed. The fractional deviation of the power law index from a constant model is always $\leq$20\%. We rule out the possibility that such small deviations can produce spurious variations in the line fluxes, being the statistical errors on the fluxes generally larger. This result validates again our assumption of a constant $\Gamma$ parameter in the simulations (see Sect. \ref{errors}).
 Finally we exploited an a-posteriori check on the filtered background in order to rule out this can lead to fake results. Indeed energy-dependence of soft-p$^{+}$ flares might in principle produce spurious variable features in the spectrum of the source. To check this we inspected the background 4-9 keV light curves during observations where significant variability in the Fe K band was found. After soft-p$^{+}$ flares removal, the background contributes less than 8\% to the 4-9 keV count rate in most observations. Such a low background fraction is not expected to have any effect on our results.
However during 7 of these observations the fraction of background count rate, even after soft-p$^{+}$ flares removal, remains high, falling between 11\% and 38\% of the source one. To check if this may induce spurious energy dependent variability in the source spectrum we inspected the background spectra during periods of maximum count rate and compared them to the time averaged one. We fitted these spectra using a broken power law model, obtaining in all the cases a good fit. We found that the background spectral shape does not change, apart from a significant increase in normalization. This means that no energy dependence is present in flares.

\begin{figure}[hbtp!]
\centering

\includegraphics[scale=0.35,angle=270]{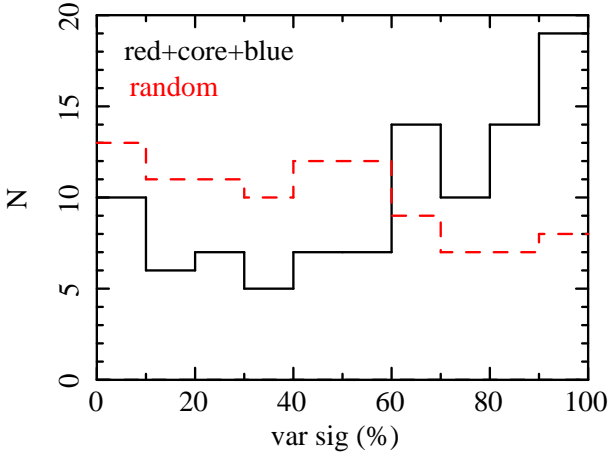}\hspace{0.5cm}

  \caption{Distribution of variability significances in the three analysed energy bands of the $FOM$-selected sub-sample (continuous curve). The dashed curve represents the distribution obtained from simulations whose variability is produced by random fluctuations.}
\label{histo1}
    \end{figure}

\begin{figure}[hbtp!]
\centering

\includegraphics[scale=0.35,angle=270]{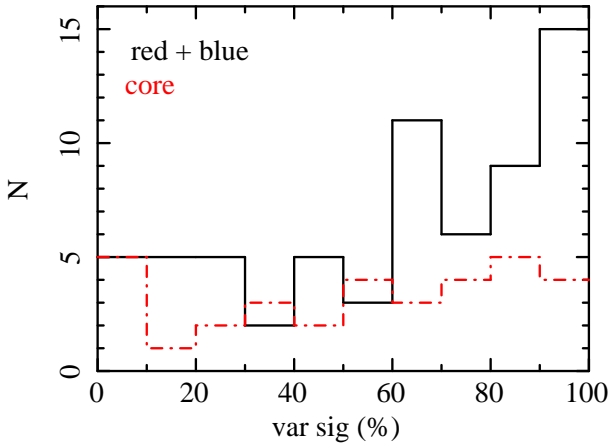}

  \caption{Distribution of variability significances of the $FOM$-selected sub-sample: comparison between the ``core'' (B band, dashed line) and the ``red+blue'' (bands A and C, continuous line) distributions.}
\label{histo2}
    \end{figure}

\section{Results from excess maps}
\label{sec:results}

The excess map technique revealed variability at $\geq$90\% confidence level (against random noise) \emph{within at least one} of the three (red, rest-frame and blue) energy bands in 26 out of 72 observations (see Table \ref{tab:varsig}). 
Whereas, the overall number of energy bands showing excess variability in the sample is 34 with significances in the range 90--99.7\% each. 
In order to assess the statistical significance of this result we carried out a binomial test.
As we looked for variability within three distinguished energy bands 
(see Sect. \ref{sec:proced}) the total number of trials that must be taken into account in the computation is 216 (i.e. number of observations times number of analysed energy bands). Furthermore, we make the conservative assumption that all the observed variable features are detected at the lower limit confidence threshold, i.e. 90\%. 
The test yielded 99.5\%, i.e. about 3$\sigma$, as the probability for rejection of the null hypothesis. In other words, the number of variable features detections in the sample is significantly greater than the one expected from a random distribution.
In the former calculation we conservatively assumed the three energy bands to be independent. This is justified by the fact that, during the same observation, significant variability is rarely observed in more than one band (i.e. only in 7 out of 34 cases). Although from a physical point of view we would expect the three bands to be somewhat linked in their variability properties, we must indeed take into account several observational effects which allow to consider the three bands independently. For example, the presence of blended constant components (this effect should be dominant in Band B, due to the presence of the ``core'' of the Fe K$\alpha$ line, which plausibly arises in the outer regions of the disk or even farther), and the fact that S/N ratio decreases at higher energies, may result in an underestimate of the variability significance in both B and C bands.\\
As next step we tried to estimate and quantify potential observational biases and/or limits that may hamper the detection of Fe K variability.
One of the main observational issue in our analysis is the duration of the exposure. Physical considerations about the choice of the maps temporal sampling have been done referring to the orbital time scale at 10 r$_{g}$ (See Sect. \ref{sec:proced}).
This is an arbitrary choice, and does not address the possibility for variability phenomena to occur at shorter time scales. However, it represents an observational bias which cannot be easily removed, as the quality of current data does not allow to sample shorter orbital periods. In fact, in our analysis, sampling time scales must be chosen in order to preserve good statistics into each time resolved spectrum, and should be able, at the same time, to probe short-term variability of the features in the Fe K band, even of relativistic ones. As relativistic effects are expected to be still observable at r$\sim$ 10 r$_{g}$, the choice of considering the estimated orbital time scale at such distance as a characteristic time scale of the system is well justified. Unfortunately, for some of the analysed observations the total exposure is too short compared to the orbital time scale at 10 r$_{g}$.
This in principle forbids detection of features varying on time scales either of this order or longer.\\
At the same time, the other important issue is the brightness of the source, as the possibility of obtaining significant detections depends on the total number of counts collected during the characteristic time scale of variations.\\
In order to account for these two biases, we made a further selection on the sample. We defined the figure of merit ($FOM$):
$$
FOM= \frac{F_{2-10\ keV} \times T}{T_{orb}}\ \ (erg\ s^{-1}\ cm^{-2}),
$$
as the total 2-10 keV flux per unit area during the whole observation of duration $T$, normalized for the estimated orbital time scale $T_{orb}$. This is equivalent to the total number of counts rescaled for the orbital time-scale.

We arbitrarily sampled all the observations having $FOM$$\geq$1.4$\times$10$^{-10}$ erg s$^{-1}$ cm$^{-2}$, which corresponds to the inclusion of those with an exposure 10 times longer than the orbital time scale and with a flux lower limit of $F_{2-10 keV}$$\geq$1.4$\times$10$^{-11}$ erg s$^{-1}$ cm$^{-2}$. This selection yields a total number of 33 ``good'' observations\footnote{IC 4329a (rev. 670), MCG -5-23-16 (rev. 1099), NGC 3783 (rev. 193, 371, 372), NGC 3516 (rev. 245, 1250, 1251, 1252, 1253), MCG-6-30-15 (rev. 108A, 108B, 301, 302, 303), NGC 7314 (rev. 256, 1172), NGC 3227 (rev. 1279), MRK 766 (rev. 82, 265, 999, 1000, 1001, 1002, 1003, 1004), NGC 7469 (rev. 912, 913), NGC 4051 (rev. 263, 541), MRK 110 (rev. 904), AKN 564 (rev. 930), NGC 4593 (rev. 465).} out of 72. Carrying out a binomial test on this sub-sample (which includes a total of 19 energy bands showing excess variability) we obtain a slightly higher probability, i.e. P=99.6\%, that the recorded significant detections are not random. In the computation of $FOM$ the main source of uncertainty is the BH mass, which affects the orbital time scale estimate. 
However, in most cases we used masses derived from optical/UV reverberation mapping measurements (see Ferrarese \& Ford 2005 for a discussion on the reliability of reverberation mapping technique and a comparison with other methods).
We collected different mass estimates (see references in Table \ref{tab:models}), obtained either via direct reverberation measurements or through secondary mass estimators based on reverberation mapping, from recent literature (Wang \& Zhang 2007, Peterson et al. 2004, Bian \& Zhao 2003, Kaspi et al. 2000 and Ho 1998). 
For each source, we used the average mass value in the calculation of $T_{orb}$ (see Table \ref{tab:models}).
In four cases (MCG -5-23-16, NGC 7314, MR 2251-178 and ESO 511-G030) reverberation mapping measurements were not available. Hence mass estimates from other methods were adopted (e.g. the photoionization method, Wandel \& Mushotsky 1986, Padovani \& Rafanelli 1988, or accretion disk models spectral fitting,  Brunner et al. 2007). 
In those sources (MCG -6-30-15, NGC 526A, ESO 198-G024, AKN 564 and NGC 7213) where only one reverberation mass value was found in literature, available estimates from different methods were added in the computation of the average mass.
Fig. \ref{histo1} shows the distribution of variability significances in the three bands from this sub-sample. Assuming that statistical fluctuations dominate the observed variability in the data, we would expect the distribution of significances to be flat. Indeed, in this case, variances in the real data greater than those observed in the simulated light curves would be recorded 10\% of the times with significances below 10\%, 10\% of the times with significances between 10\%-20\%, and so on, leading to a flat proability density distribution. The dashed histogram in Fig. \ref{histo1} illustrates the distribution of significances derived from simulated data whose variability is due to random fluctuations (assuming the same total number of observations as the real ones).
We observe that the var-sig distribution is significantly
peaked at high values, with a deficit at low significances,
thus indicating the effective presence of the
features we are catching.
 In particular, a K-S test strongly disfavours the observed distribution to be consistent with the random one (P$_{KS}=$0.13\%). 
Moreover, if we only focus on Band A and C where relativistic, energy shifted (either redwards or bluewards) features are expected to be detectable (Fig. \ref{histo2}, solid line), the clustering towards high significances results even more prominent (P$_{KS_{red+blue}}$=0.06\%). Conversely, the distribution is smoother and similar to the random one (Fig. \ref{histo2}, dashed line) when the energy band containing the narrow ``core'' of the E$=$6.4 keV Fe K$\alpha$ component is isolated (P$_{KS_{core}}$=62\%). This effect is indeed expected, since the narrow component contributes with an almost steady flux to the total band counts being most plausibly produced in regions far from the nucleus.

\section{Summary and discussion}

The main goal of this paper is to find and monitor transient emission and absorption features in the Fe K line energy band of radio-quiet and bright AGNs. In this respect, as the detection of transient features is seriously biased by the lack of good statistics (time resolved spectra are usually characterized by low S/N, which prevents, in most cases to obtain high significance detections), the best method to derive strong conclusions on their reliability is by analysing a statistically complete sample. This issue is for the first time addressed in the present work, where we present both null and positive results from the uniform analysis of an almost (94\%) complete sample of sources, allowing to draw statistically meaningful conclusions.\\
In this respect we studied a flux selected sub-sample of the FERO sample, defined by GBD06. The sample includes the 30 brightest (in the RXTE Slew Survey, XSS, with a flux lower limit of F$_{2-10 keV}\geq 1.5\times 10^{-11}$erg cm$^{-2}$s$^{-1}$) radio-quiet AGNs observed by XMM-{\it Newton} as of 1 February 2009. The observations we analysed (for a total of 72) are chosen to have a duration $\geq$10 ks, in order to carry out a meaningful temporal analysis.\\
Adopting the method of searching for excess variability (IMF04) in the energy bands of the redshifted (Band A), neutral-mildly ionized (Band B) and highly ionized Fe K$\alpha$ and/or neutral Fe K$\beta$ (Band C) lines, we revealed significant (with a confidence level $\geq$90\%) signatures of variable features in 26 out of 72 observations, with probabilities in the range 90\%--99.7\%. Considering the total number of energy bands showing excess variability, this translates into 34 detections out 216 monitored energy bands. A binomial test for the significance of this result yielded a probability of 99.5\% (i.e. $\sim$3$\sigma$) that the overall detections are not by chance.\\
We further tested the detection frequency of variable features, after taking into account several observational biases (i.e. short duration of the exposure as compared with a typical dynamical time-scale of the system, as well as low flux observations) which might seriously prevent positive detections. We tried to remove these biases by defining a $FOM$ ratio (see Sect. \ref{sec:results}).
 As these observational limits forbid us to draw conclusions about Fe K variability properties in the faintest sources of the sample, we will focus the following discussion on the $FOM$-selected dataset.
 Once the selection is accomplished, the distribution of variability significances in the overall three energy bands (with significant detections ranging from 91.7\% to 99\%), results peaked towards high values (i.e. $\geq$90\%). Moreover, subtracting the contribution of the ``core'' energy band (band B), which is not expected to show signatures of strong variability due to the dominance of the constant Fe K$\alpha$ emission, the skewness of the distribution towards high significances increases. All these results have been statistically checked through a K-S test, using a uniform distribution as the reference one.\\
From the significances distribution analysis, it is reliable to assert that variability is commonly observed in the Fe K complex energy band of bright sources (see Fig. \ref{histo1}), provided the typical variability time scale is well sampled by the duration of the observation and the source is sufficiently bright. Infact, our estimated detection frequency is of 13 observations out of 33, showing variability in at least one of the three sampled energy bands, i.e. $\sim$39\%.
 Moreover, from Fig. \ref{histo2} it is clear that the bulk of variability comes from the red- and blue- shifted (with respect to the neutral ``core'') energy bands. This is most probably due to the fact that, if variability phenomena are still taking place in the neutral Fe K$\alpha$ band, they result overwhelmed by the constancy (on short time scales) of the Fe line ``core''.
Indeed, the detection frequency of variable features in the red- plus blue- shifted energy bands is 12 out of 33 observations, i.e. $\sim$36\%. It is worth noting that this is very similar to the fraction of observations (i.e. 12/37$\sim$32\%) for which the characteristic emission radius of the relativistic component is constrained to be $<$50 r$_{g}$ in the NOGR07 sample.\\
Understanding the origin of the variable features in our sample from current data is not easy.
Any excess of variability recorded in band A is most probably due to redshifted Fe K emission. Indeed this energy band, apart from the relativistic Fe K line components, is expected to include only the Fe K$\alpha$ high order Compton shoulders, which, however, are predicted to be very faint and not observable with current instruments (e.g. George \& Fabian 1991, Matt 2002).
It has been claimed that partial covering can induce spectral curvature in this band (e.g. Miller et al. 2008). Indeed, an ionized complex absorbing system might introduce further complexity and be responsible for variability as a consequence of variations in its properties (e.g. ionization state, covering factor, etc.). Although hard to predict, we expect that, in general, the partial covering gas would imprint its signature in the B band as well as in the A and C bands. The plot of Fig. \ref{histo2} conversely, clearly shows a lack of correlated variations between the ``core'' and the ``red+blue'' energy bands, disfavouring the partial covering interpretation.\\
On the other hand the other two bands, in addition to the constant (on short time scales) Fe K$\alpha$ and Fe K$\beta$ emission components, can also include both the blue peak of a relativistic Fe line and ionized Fe K absorption features, which may be responsible for the detected variability as well. 
Disentangling between the two cases would require a deeper analysis on any single case. However, residuals observed in band B are always positive, which is proper of variable features in emission. On the contrary, there are three cases (NGC 3516 rev. 1250, 1251 and 1252) where variable residuals in band C are negative, hence, most probably due to variable absorption features. This view is confirmed by the detection of two absorption lines in the time-averaged spectra of the three observations at these energies. In all the other cases the average residuals flux in band C is positive, pointing to an origin in emitting material, most probably the peak of a relativistic Fe line. Assuming this is the viable interpretation, we can give 10/33, i.e. $\sim$30\%, as an estimate for the detection frequency of relativistic variable features in bright and radio-quiet type 1 AGN, again of the same of order of the fraction deduced by NOGR07.\\
Inspection of residuals light curves of the entire sample allows us to roughly compute the time scale at which the observed variations take place. In almost all the cases these variations have not a regular shape. Only in three cases (IC 4329a rev. 670, NGC 3783 rev. 372 and MRK 509 rev. 1073) they are characterized by a repeated, flares-like pattern.
For these and other observations, a deeper study can be addressed (i.e. NGC 3516, IMF04; MKN 766, Turner et al. 2006; NGC 3783, Tombesi et al. 2007; IC 4329a, De Marco et al. 2009), and we refer to the relative papers for an exhaustive discussion.
In general, we decided to assume as the temporal time-scale of the variations, the time interval between the maximum peak in the light curve and either the following or the preceding flux minimum. The range of estimated time-scales is $\Delta t$$=$6--30 ks. This allowing to compute an upper limit on the extension of the region responsible for the observed variations. As variability from a source cannot be observed on time scales shorter than the light crossing time of the source itself, the upper limit on the size of the region is given by $R<c\Delta t$ (e.g. Mushotzky et al. 1993). If we consider the average $\langle\Delta t\rangle \sim$15.5 ks (which is also the average value of the $FOM$-selected sub-sample) as the reference one, the derived upper limit on the region extension is $R\leq 4.6\times 10^{14}$cm$\sim$140 r$_{g}$ for a black hole mass equal to the average value of the masses in the entire sample, i.e. $\langle M_{BH}\rangle \sim 2.24 \times 10^{7}$M$_{\odot}$. These calculations point to an origin into relatively small regions.\\
Whether these regions are flares illuminated zones of the accretion disk or blobs of outflowing gas is not clear from our data, as overall the sensitivity of the observations does not allow to trace a clear pattern of variability. If associated to orbital motions, the observed variability would imply an average emitting radius of $\langle r\rangle\sim$7.9 r$_{g}$ (assuming the average values of $\langle\Delta t\rangle \sim$15.5 ks and $\langle M_{BH}\rangle \sim 2.24 \times 10^{7}$M$_{\odot}$ for the tim-scale and the black hole mass respectively).
However, in most of the cases detection of significant variability does not come with clear intensity modulations in the observed light curves, disfavouring the ``hot spot'' interpretation (e.g. Nayakshin \& Kazanas 2001, Dov\v{c}iak et al. 2004, Goosmann et al. 2007). Nevertheless, this does not completely rule out the model as the estimate of spot-induced modulations significance strongly depends on the possibility of observing the spot emission for a sufficiently long time (e.g. Vaughan 2005). This ultimately relies on the length of the observation and the permanence of the primary source flare. Moreover, the occurrence of several flare events illuminating the disk is expected to generate more complex line patterns.\\
 While a discrete spot-like region of the disk would produce a smooth intensity/energy modulation pattern as a consequence of its orbital motion, this would not be the case if the emitting region has a ring-like geometry. In that instance, the energy modulations are expected to be weak, in agreement with results from extensive analysis of the variability patterns of some observations of our sample (NGC 3783, Tombesi et al. 2007; IC 4329a, De Marco et al. 2009). Many authors pointed out the possibility to produce spiral density waves in non-self-gravitating models of accretion disks (e.g. Caunt \& Tagger 2001; Hawley 2001). The effects of such structures on the Fe emission line would essentially be the production of several sub-peaks into the line profile, characterized by quasi-periodic variability as a consequence of the motion, and independent on continuum variations (e.g. Karas, Martocchia \& Subr 2001, Hartnoll \& Blackman 2002).\\
Another suggested scenario is the production of line emission (via recombination of highly ionized Fe), during the ejection phase of blobs of gas originating from disk instabilities (Wang et al. 2000). In this case, detection of both red- and blue- shifted lines from both sides of the jet are expected. Some of the observed variable features in our sample are consistent with this picture, which is however not able to explain the modulations registered in some sources (e.g. NGC 3516, IMF04; MKN 766, Turner et al. 2006; NGC 3783, Tombesi et al. 2007; IC 4329a, De Marco et al. 2009).\\
The present work is the first attempt to search for such patterns of variability. It makes use of the best quality data available up to now for this kind of studies. However, disentangling among possible interpretations is still quite difficult. A qualitative step forward into the understanding of these issues will be provided by forthcoming missions, as the \emph{International X-ray Observatory} (IXO).

\begin{acknowledgements}
     This paper is based on observations obtained with the XMM--{\it Newton}
satellite, an ESA funded mission with contributions by ESA Member States and USA.
     BDM, MC, MD, GP and FT acknowledge financial support from ASI under contracts
     ASI/INAF I/023/05/0 and I/088/06/0. BDM and AC acknowledge MIUR for financial support. GP aknowledges ANR
for support under grant number ANR-06-JCJC-0047. GM acknowledges the Ministerio de Ciencia e Innovaci\'on and CSIC for support through a Ram\'on y Cajal contract. BDM aknowledges C. Evoli for helpful discussions. The authors thank the anonymous referee for suggestions that led to significant improvements in the paper.

\end{acknowledgements}

\onecolumn

\onecolumn
\begin{scriptsize}

\longtab{2}{
\begin{landscape}
\begin{longtable}{l c c c c c c c c c c c c c}
\caption{Sources list: (1) name; (2) RXTE count rate; (3) redshift; (4)
  estimated SMBH mass; (5) estimated Keplerian orbital period at r$=$10 r$_g$; (6) references for the mass estimates; (7) revolution number;
  (8) 2--10 keV flux; (9) duration of the exposure;
  (10) chosen time resolution for the excess map analysis; (11) chosen
  energy resolution for the excess map analysis; (12) $FOM$ value (see Sect. \ref{sec:results}); (13) best fit value of the power law index; (14) best fit value of the cold absorption column density. The $\Gamma$ and N$_{H}$ values are obtained assuming a simple ``power law + cold absorption'' for the fit of the time-averaged 4--9 keV continuum. The errors are computed at 90\% confidence level.}\label{tab:info}\\
\hline\hline      
Sources & \multicolumn{1}{c}{RXTE} &  \multicolumn{1}{c}{z} &  \multicolumn{1}{c}{$\langle$Log(M$_{BH}$/M$_{\odot}$)$\rangle$} &  \multicolumn{1}{c}{$\langle$T$_{orb}$(10r$_{g}$)$\rangle$} &  \multicolumn{1}{c}{Ref.} &  \multicolumn{1}{c}{Rev. Num.} &  \multicolumn{1}{c}{F$_{2-10 keV}$} &  \multicolumn{1}{c}{Obs. Duration} &  \multicolumn{1}{c}{$\Delta$t} &  \multicolumn{1}{c}{$\Delta$E} &  \multicolumn{1}{c}{$FOM$} &  \multicolumn{1}{c}{$\Gamma$} &  \multicolumn{1}{c}{N$_{H}$}\\
       &  \multicolumn{1}{c}{(cts/s)}  &  &  &  \multicolumn{1}{c}{(ks)} &   &      &     \multicolumn{1}{c}{($10^{-11}$ erg cm$^{-2}$ s$^{-1}$)} &  \multicolumn{1}{c}{(ks)}  &   \multicolumn{1}{c}{(ks)} &  \multicolumn{1}{c}{(keV)} &  \multicolumn{1}{c}{($10^{-11}$ erg cm$^{-2}$ s$^{-1}$)} &   &   \multicolumn{1}{c}{($10^{22}$ cm$^{-2}$)}\\
(1)  &  \multicolumn{1}{c}{(2)} &  \multicolumn{1}{c}{(3)} &  \multicolumn{1}{c}{(4)}  &   \multicolumn{1}{c}{(5)}  &  \multicolumn{1}{c}{(6)} &  \multicolumn{1}{c}{(7)}  &  \multicolumn{1}{c}{(8)} &  \multicolumn{1}{c}{(9)} &  \multicolumn{1}{c}{(10)} &   \multicolumn{1}{c}{(11)} &  \multicolumn{1}{c}{(12)} &  \multicolumn{1}{c}{(13)} &  \multicolumn{1}{c}{(14)} \\
\hline
\hline
\endfirsthead
\caption{continued.}\\
\hline\hline
Sources & \multicolumn{1}{c}{RXTE} &  \multicolumn{1}{c}{z} &  \multicolumn{1}{c}{$\langle$Log(M$_{BH}$/M$_{\odot}$)$\rangle$} &  \multicolumn{1}{c}{$\langle$T$_{orb}$(10r$_{g}$)$\rangle$} &  \multicolumn{1}{c}{Ref.} &  \multicolumn{1}{c}{Rev. Num.} &  \multicolumn{1}{c}{F$_{2-10 keV}$} &  \multicolumn{1}{c}{Obs. Duration} &  \multicolumn{1}{c}{$\Delta$t} &  \multicolumn{1}{c}{$\Delta$E} &  \multicolumn{1}{c}{$FOM$} &  \multicolumn{1}{c}{$\Gamma$} &  \multicolumn{1}{c}{N$_{H}$}\\
       &  \multicolumn{1}{c}{(cts/s)}  &  &  &  \multicolumn{1}{c}{(ks)} &   &      &     \multicolumn{1}{c}{($10^{-11}$ erg cm$^{-2}$ s$^{-1}$)} &  \multicolumn{1}{c}{(ks)}  &   \multicolumn{1}{c}{(ks)} &  \multicolumn{1}{c}{(keV)} &  \multicolumn{1}{c}{($10^{-11}$ erg cm$^{-2}$ s$^{-1}$)} &   &   \multicolumn{1}{c}{($10^{22}$ cm$^{-2}$)}\\
(1)  &  \multicolumn{1}{c}{(2)} &  \multicolumn{1}{c}{(3)} &  \multicolumn{1}{c}{(4)}  &   \multicolumn{1}{c}{(5)}  &  \multicolumn{1}{c}{(6)} &  \multicolumn{1}{c}{(7)}  &  \multicolumn{1}{c}{(8)} &  \multicolumn{1}{c}{(9)} &  \multicolumn{1}{c}{(10)} &   \multicolumn{1}{c}{(11)} &  \multicolumn{1}{c}{(12)} &  \multicolumn{1}{c}{(13)} &  \multicolumn{1}{c}{(14)} \\
\hline
\hline
\endhead
\hline
\endfoot
IC 4329a & 7.29 & 0.0161 & 7.29 & 19 & 1,2,3,4 & 210 & 15.3 & 10 & 2.5 & 0.1 & 8.0 & 2.04$^{+0.16}_{-0.08}$  & 2.66$^{+2.13}_{-2.27}$ \\
         &      &        &      &    &         &  670 & 9.48 & 133 & 2.5 & 0.1 & 65.8 & 1.73$^{+0.03}_{-0.04}$  & 0.35$^{+0.67}_{-0.16}$ \\
\hline
MCG -5-23-16 & 6.46 & 0.0085 &  7.60  & 39  & 13 & 363  & 7.43 & 22 & 2.5 & 0.1 & 4.2 & 1.69$^{+0.10}_{-0.13}$ & $<$3.50 \\   
             &      &        &        &     &    & 1099 & 8.29 & 122 & 2.5 & 0.1 & 25.9 & 1.79$^{+0.03}_{-0.06}$ & 2.67$^{+0.50}_{-0.94}$ \\
\hline
NGC 3783 & 4.90 & 0.0097 & 7.01 & 10 & 1,2,3,4,5 & 193 & 5.27 & 37 & 2.5 & 0.1 & 19.7 & 1.86$^{+0.05}_{-0.06}$ & 2.80$^{+0.57}_{-0.95}$ \\
         &      &        &      &   &            & 371 & 3.96 & 131 & 2.5 & 0.1 & 52.3 & 1.67$\pm$0.20 & 1.61$\pm 1.10$ \\
         &      &        &      &   &            & 372 & 5.56 & 136 & 2.5 & 0.1 & 76.3 & 1.82$^{+0.03}_{-0.07}$ & 2.56$^{+0.50}_{-0.70}$ \\
\hline
NGC 5548 & 3.58 & 0.0170 & 7.71 & 50 & 1,2,3,4,5 & 191 & 3.14 & 23 & 2.5 & 0.1 & 1.4 & 1.67$^{+0.11}_{-0.06}$ & $<$1.54\\
         &      &        &      &    &           & 290 & 3.86 & 93 & 2.5 & 0.1 & 7.1 & 1.77$\pm$0.07 & $<$1.83 \\
         &      &        &      &    &           & 291 & 4.89 & 37 & 2.5 & 0.1 & 3.6 & 1.89$\pm$0.10 & 2.13$^{+1.36}_{-1.30}$ \\
\hline
NGC 3516 & 3.21 & 0.0088 & 7.21 & 16 & 1,2,3,5 & 245 & 2.34 & 129 & 5.0 & 0.1 & 19.2 & 1.81$\pm$0.09 & 5.32$^{+1.16}_{-1.17}$ \\
         &      &        &      &    &         & 352 & 1.53 & 128 & 5.0 & 0.1 & 12.4 & 1.82$^{+0.11}_{-0.17}$ & 7.82$^{+1.24}_{-1.39}$ \\
         &      &        &      &    &         & 1250 & 5.24 & 52 & 2.5 & 0.1 & 17.3 & 2.15$^{+0.06}_{-0.09}$ & 4.76$^{+0.75}_{-1.23}$ \\
         &      &        &      &    &         & 1251 & 4.60 & 69 & 2.5 & 0.1 & 20.1 & 2.21$^{+0.06}_{-0.07}$ &  6.07$^{+0.96}_{-0.93}$ \\
         &      &        &      &    &         & 1252 & 3.64 & 68 & 2.5 & 0.1 & 15.7 & 2.29$^{+0.09}_{-0.08}$ &  8.62$^{+1.12}_{-0.95}$ \\
         &      &        &      &    &         & 1253 & 4.48 & 68 & 2.5 & 0.1 & 19.3 & 2.26$^{+0.07}_{-0.15}$ & 6.41$^{+1.14}_{-1.15}$ \\
\hline
MRK 509 & 3.12 & 0.0344 & 7.84 & 68  & 1,2,3,4,5 & 161  & 2.97 & 30 & 2.5 & 0.1 & 1.3 & 1.63$^{+0.13}_{-0.11}$ & $<$3.04 \\
        &      &        &      &     &           & 250  & 3.84 & 43 & 2.5 & 0.1 & 2.4 & 1.72$^{+0.10}_{-0.15}$ & $<$2.73 \\
        &      &        &      &     &           & 1073 & 3.67 & 85 & 2.5 & 0.1 & 4.6 & 1.69$\pm$0.03 & $<$0.59 \\
        &      &        &      &     &           & 1074 & 3.70 & 46 & 2.5 & 0.1 & 2.5 & 1.74$^{+0.09}_{-0.02}$ & $<$0.72 \\
        &      &        &      &     &           & 1168 & 4.26 & 69 & 2.5 & 0.1 & 4.4 & 1.71$^{+0.08}_{-0.05}$ &  $<$0.69 \\
\hline
MCG -6-30-15 & 3.08 & 0.0077 & 5.84 & 0.67 & 3,8 & 108A & 2.62 & 43 & 2.5 & 0.1 & 168.0 &  2.01$^{+0.19}_{-0.32}$ & $<$1.56 \\
             &      &        &      &      &     & 108B & 3.60 & 55 & 2.5 & 0.1 & 295.3 & 2.50$\pm$0.08 & 7.30$\pm$1.14 \\
             &      &        &      &      &     & 301 & 4.11 & 84 & 2.5 & 0.1 & 514.9 & 2.47$\pm$0.05 & 6.77$\pm$0.90 \\
             &      &        &      &      &     & 302 & 4.53 & 127 & 2.5 & 0.1 & 858.1 & 2.28$^{+0.06}_{-0.04}$ & $\pm$0.70 \\
             &      &        &      &      &     & 303 & 4.01 & 125 & 2.5 & 0.1 & 747.6 & 2.17$^{+0.29}_{-0.19}$ & 1.37$^{+0.80}_{-0.87}$ \\
\hline
MCG +8-11-11 & 2.36 & 0.0205 & 7.18 & 15 & 3 & 794 & 4.52 & 38 & 2.5 & 0.1 & 11.6 & 1.64$^{+0.09}_{-0.07}$ & $<$1.84 \\
\hline
NGC 7314 & 2.16 & 0.0048 & 6.67 & 4.6 & 12 & 256  & 3.93 & 43 & 2.5 & 0.1 & 36.9 & 2.12$\pm$0.09 & 2.68$^{+1.25}_{-1.27}$ \\
         &      &        &      &     &    & 1172 & 1.54 & 82 & 5.0 & 0.1 & 27.5 & 1.94$^{+0.14}_{-0.11}$ & 3.84$^{+1.85}_{-1.44}$ \\
\hline
AKN 120 & 2.14 & 0.0323 & 8.14 & 134 & 1,2,3,4,5 & 679 & 3.76 & 112 & 2.5 & 0.1 & 3.1 & 1.97$^{+0.06}_{-0.07}$ & 1.03$^{+0.73}_{-0.99}$ \\
\hline
MRK 279 & 2.13 & 0.0305 & 7.79 & 60 & 1,2,3,5 & 1087 & 2.60 & 59 & 5.0 & 0.1 & 2.6 & 1.91$^{+0.06}_{-0.09}$ & $<$2.38 \\
        &      &        &      &    &         & 1088 & 2.43 & 59 & 5.0 & 0.1 & 2.4 & 1.75$^{+0.05}_{-0.10}$ & $<$2.07 \\
        &      &        &      &    &         & 1089 & 2.38 & 38 & 5.0 & 0.1 & 1.5 & 2.03$^{+0.08}_{-0.15}$ & 4.20$^{+2.06}_{-2.00}$ \\
\hline
MR 2251-178 & 2.10 & 0.0640 & 8.40 & 246  & 10  & 447 & 2.03 & 64 & 5.0 & 0.1 & 0.5 & 1.61$\pm$0.09 & $<$2.14 \\
\hline
NGC 3227 & 2.10 & 0.0039 & 7.39 & 24 & 1,2,3,4,5 & 178 & 0.85 & 35 & 5.0 & 0.1 & 1.2 & 1.63$^{+0.20}_{-0.13}$ & 7.84$^{+2.75}_{-2.04}$ \\
         &      &        &      &    &           & 1279 & 3.64 & 106 & 2.5 & 0.1 & 16.1 & 1.92$\pm$0.08 & 2.89$^{+1.08}_{-1.15}$ \\
\hline
IRAS 05078+1626 & 2.08 & 0.0179 & 7.04 & 11 & 1 & 1410 & 2.41 & 58 & 5.0 & 0.1 & 13.0 & 1.78$^{+0.10}_{-0.09}$ & 2.53$^{+1.46}_{-1.25}$ \\
\hline
MRK 590 & 1.95 & 0.0264 & 7.37 & 23 & 1,2,4,5 & 837 & 0.56 & 108 & 9.0 & 0.1 & 2.6 & 2.30$^{+0.37}_{-0.15}$ & 5.21$^{+4.66}_{-1.97}$ \\
\hline
MRK 766 & 1.77 & 0.0129 & 5.92 & 0.82 & 3 & 82 & 1.52 & 39 & 5.0 & 0.1 & 72.7 & 2.25$^{+0.17}_{-0.16}$ & 3.46$\pm$2.29 \\
        &      &        &      &      &   & 265 & 2.44 & 129 & 5.0 & 0.1 & 386.0 & 2.10$^{+0.08}_{-0.04}$ & $<$1.20 \\
        &      &        &      &      &   & 999 & 0.72 & 95 & 8.0 & 0.1 & 83.9 & 1.85$^{+0.23}_{-0.32}$ & 5.13$^{+2.54}_{-2.30}$ \\
        &      &        &      &      &   & 1000 & 1.13 & 98 & 6.5 & 0.1 & 135.8 & 1.99$^{+0.20}_{-0.07}$ & $<$2.77 \\
        &      &        &      &      &   & 1001 & 1.40 & 98 & 6.0 & 0.1 & 168.3 & 2.13$^{+0.11}_{-0.12}$ & $<$2.92 \\
        &      &        &      &      &   & 1002 & 1.74 & 95 & 5.0 & 0.1 & 202.7 & 2.27$^{+0.10}_{-0.11}$ & 3.59$^{+1.23}_{-1.46}$ \\
        &      &        &      &      &   & 1003 & 1.48 & 98 & 5.5 & 0.1 & 177.9 & 2.01$^{+0.17}_{-0.04}$ & $<$2.23 \\
        &      &        &      &      &   & 1004 & 1.30 & 35 & 5.0 & 0.1 & 55.8 & 2.10$^{+0.15}_{-0.08}$ & $<$5.06 \\
\hline
NGC 7469 & 1.69 & 0.0164 & 7.14 & 13 & 1,2,3,4,5 & 192A & 2.49 & 18 & 5.0 & 0.1 & 3.4 & 1.88$^{+0.17}_{-0.15}$ & $<$3.97 \\
         &      &        &      &    &           & 192B & 2.66 & 23 & 5.0 & 0.1 & 4.6 & 1.79$^{+0.11}_{-0.05}$ & $<$1.18 \\
         &      &        &      &    &           & 912 & 2.83 & 85 & 2.5 & 0.1 & 18.0 & 1.92$^{+0.08}_{-0.07}$ & 1.55$^{+1.18}_{-0.93}$ \\
         &      &        &      &    &           & 913 & 2.93 & 79 & 2.5 & 0.1 & 17.3 & 1.97$^{+0.07}_{-0.08}$ & 2.24$\pm$1.09 \\
\hline
MCG -2-58-22 & 1.61 & 0.0469 & 8.00 & 99 & 1,3 & 180 & 3.02 & 10 & 2.5 & 0.1 & 0.3 & 1.68$^{+0.22}_{-0.15}$ & $<$4.04 \\
\hline
NGC 526A & 1.61 & 0.0191 & 6.98 & 9 & 1,13 & 647 & 2.20 & 44 & 2.5 & 0.1 & 10.5 & 1.61$^{+0.05}_{-0.12}$ & 2.56$^{+1.30}_{-1.44}$ \\
\hline
NGC 4051 & 1.49 & 0.0023 & 5.90 & 0.77 & 2,3,4 & 263 & 2.35 & 117 & 5.5 & 0.1 & 355.8 & 2.20$^{+0.09}_{-0.07}$ & 2.86$^{+1.07}_{-1.10}$ \\
         &      &        &      &      &       & 541 & 0.56 & 50 & 6.0 & 0.2 & 36.2 & 1.56$^{+0.23}_{-0.31}$ & $<$4.35 \\
\hline
ESO 141-G055 & 1.37 & 0.0360  & 7.51 & 31 & 1 & 1435 A & 2.82 & 20 & 5.0 & 0.1 & 1.8 & 2.06$^{+0.21}_{-0.15}$ & $<$3.88 \\
             &      &         &      &    &   & 1435 B & 2.54 & 21 & 5.0 & 0.1 & 1.7 & 2.15$^{+0.23}_{-0.21}$ & $<$5.42 \\
             &      &         &      &    &   & 1436 & 2.31 & 41 & 5.0 & 0.1 & 3.0 & 2.10$^{+0.19}_{-0.13}$ & $<$3.31 \\
             &      &         &      &    &   & 1445 & 2.88 & 78 & 5.0 & 0.1 & 7.2 & 1.94$^{+0.06}_{-0.08}$ & $<$0.78 \\
\hline
UGC 3973 & 1.32 & 0.0222 & 7.74 & 54 & 1,2,4,5 & 1247 & 2.33 & 21 & 5.0 & 0.1 & 0.9 & 2.02$^{+0.18}_{-0.17}$ & $<$4.65 \\
         &      &        &      &    &         & 1263 & 2.01 & 20 & 5.0 & 0.1 & 0.7 & 1.64$^{+0.16}_{-0.09}$ & $<$3.17 \\
         &      &        &      &    &         & 1332 & 1.84 & 20 & 5.0 & 0.1 & 0.7 & 1.63$^{+0.23}_{-0.13}$ & $<$4.25 \\
         &      &        &      &    &         & 1535 & 0.77 & 87 & 6.0 & 0.1 & 1.2 & 1.70$^{+0.11}_{-0.20}$ & 5.87$^{+3.32}_{-1.37}$ \\
\hline
Fairall 9 & 1.25 & 0.0470 & 8.00 & 98 & 1,2,3,4,5 & 105 & 1.16 & 29 & 5.0 & 0.1 & 0.3 & 1.73$^{+0.19}_{-0.16}$ & $<$5.00 \\
\hline
ESO 198-G024 & 1.21 & 0.0455 & 8.38 & 235 & 1,9 & 207 & 1.30 & 31 & 5.0 & 0.1 & 0.2 & 1.84$^{+0.16}_{-0.15}$ & $<$3.43 \\
             &      &        &      &     &     & 1128 & 1.02 & 122 & 6.0 & 0.1 & 0.5 & 1.75$^{+0.19}_{-0.09}$ & 1.83$^{+1.78}_{-1.22}$ \\
\hline
MRK 110 & 1.17 & 0.0353 & 6.99 & 9.5 & 2,3,4,5 & 904 & 2.80 & 47 & 2.5 & 0.1 & 13.8 & 1.87$\pm$0.10 & 1.79$^{+1.38}_{-1.36}$ \\
\hline
AKN 564 & 1.13 & 0.0247 & 6.24 & 1.7 & 3,7 &  930 & 1.60 & 99 & 5.0 & 0.2 & 94.1 & 2.50$^{+0.11}_{-0.04}$ & $<$1.28 \\
\hline
NGC 7213 & 1.12 & 0.0058 & 8.04 & 107 & 1,11 & 269 & 2.16 & 42 & 5.0 & 0.1 & 0.9 & 1.69$^{+0.10}_{-0.03}$ & $<$0.85 \\
\hline
ESO 511-G030 & 1.10 & 0.0224 & 8.66 & 448 & 6 & 1402 & 1.93 & 109 & 5.0 & 0.1 & 0.5 & 1.89$\pm$0.08 & 2.03$^{+1.15}_{-1.07}$ \\
\hline
MRK 704 & 1.06 & 0.0292 & 7.55 & 35 & 1 & 1074 & 0.99 & 21 & 5.0 & 0.2 & 0.6 & 1.88$^{+0.40}_{-0.17}$ & 6.58$^{+5.44}_{-2.47}$ \\
        &      &        &      &    &   & 1630 & 1.07 & 98 & 5.5 & 0.1 & 3.0 & 1.87$^{+0.19}_{-0.08}$ & $<$3.57 \\
\hline
NGC 4593 & 1.05 & 0.0047 & 6.91 & 7.9 & 1,2,3,5 & 465 & 4.00 & 76 & 2.5 & 0.1 & 38.5 & 1.84$^{+0.06}_{-0.03}$ & 1.21$^{+0.91}_{-0.93}$ \\
\hline
\hline   
\end{longtable}
References: (1) Wang \& Zhang 2007; (2) Peterson et al. 2004; (3) Bian \& Zhao 2003; (4) Kaspi et al. 2000; (5) Ho 1998; (6) Winter et al. 2008; (7) Botte et al. 2004; (8) Vaughan et al. 2003; (9) Rokaki \& Boisson 1999; (10) Brunner et al. 1997; (11) Nelson \& Whittle 1995; (12) Padovani \& Rafanelli 1988; (13) Wandel \& Mushotzky 1986.
\end{landscape}
}

\end{scriptsize}


\onecolumn
\begin{scriptsize}

\longtab{3}{
\begin{landscape}
\begin{longtable}{l c c c c c c c c c c c c c c c c}
\caption{Gaussian components included in the best-fit models to the 4--9 keV time-averaged spectra, obtained assuming a ``power law + cold absorption'' as the continuum model. 
 For each Gaussian component the line rest-frame centroid energy E (keV), width $\sigma$ (keV), equivalent width EW (eV) and flux $I$ ($10^{-5}$ ph s$^{-1}$ cm$^{-2}$) are reported. All the Gaussians are detected at $\geq$99\% ($\Delta \chi^{2} \geq$9.21) from confidence contours for line intensity vs energy. The errors are computed at 90\% confidence level.}\label{tab:models}\\
\hline\hline             
\multicolumn{1}{l}{Source}   &  \multicolumn{4}{c}{Red}  &  \multicolumn{4}{c}{Core} & \multicolumn{4}{c}{Blue} & \multicolumn{4}{c}{Absorption} \\
\cline{2-17}
\multicolumn{1}{l}{(rev. number)}  &  \multicolumn{1}{c}{E}   &  \multicolumn{1}{c}{$\sigma$}   & \multicolumn{1}{c}{EW} & \multicolumn{1}{c |}{$I$}  &   \multicolumn{1}{c}{E}   &  \multicolumn{1}{c}{$\sigma$}   & \multicolumn{1}{c}{EW} & \multicolumn{1}{c |}{$I$}  &   \multicolumn{1}{c}{E}   &  \multicolumn{1}{c}{$\sigma$}   & \multicolumn{1}{c}{EW} & \multicolumn{1}{c |}{$I$}  &   \multicolumn{1}{c}{E}   &  \multicolumn{1}{c}{$\sigma$}   & \multicolumn{1}{c}{EW} & \multicolumn{1}{c}{$I$} \\ 
 \multicolumn{1}{l}{(1)}  &  \multicolumn{1}{c}{(2)}   &  \multicolumn{1}{c}{(3)}   & \multicolumn{1}{c}{(4)} & \multicolumn{1}{c |}{(5)}  &   \multicolumn{1}{c}{(6)}   &  \multicolumn{1}{c}{(7)}   & \multicolumn{1}{c}{(8)} & \multicolumn{1}{c |}{(9)}  &   \multicolumn{1}{c}{(10)}   &  \multicolumn{1}{c}{(11)}   & \multicolumn{1}{c}{(12)} & \multicolumn{1}{c |}{(13)}  &   \multicolumn{1}{c}{(14)}   &  \multicolumn{1}{c}{(15)}   & \multicolumn{1}{c}{(16)} & \multicolumn{1}{c}{(17)} \\ 
\hline
\hline
\endfirsthead
\caption{continued.}\\
\hline\hline
\multicolumn{1}{l}{Source}   &  \multicolumn{4}{c}{Red}  &  \multicolumn{4}{c}{Core} & \multicolumn{4}{c}{Blue} & \multicolumn{4}{c}{Absorption} \\
\cline{2-17}
\multicolumn{1}{l}{(rev. number)}  &  \multicolumn{1}{c}{E}   &  \multicolumn{1}{c}{$\sigma$}   & \multicolumn{1}{c}{EW} & \multicolumn{1}{c |}{$I$}  &   \multicolumn{1}{c}{E}   &  \multicolumn{1}{c}{$\sigma$}   & \multicolumn{1}{c}{EW} & \multicolumn{1}{c |}{$I$}  &   \multicolumn{1}{c}{E}   &  \multicolumn{1}{c}{$\sigma$}   & \multicolumn{1}{c}{EW} & \multicolumn{1}{c |}{$I$}  &   \multicolumn{1}{c}{E}   &  \multicolumn{1}{c}{$\sigma$}   & \multicolumn{1}{c}{EW} & \multicolumn{1}{c}{$I$} \\ 
 \multicolumn{1}{l}{(1)}  &  \multicolumn{1}{c}{(2)}   &  \multicolumn{1}{c}{(3)}   & \multicolumn{1}{c}{(4)} & \multicolumn{1}{c |}{(5)}  &   \multicolumn{1}{c}{(6)}   &  \multicolumn{1}{c}{(7)}   & \multicolumn{1}{c}{(8)} & \multicolumn{1}{c |}{(9)}  &   \multicolumn{1}{c}{(10)}   &  \multicolumn{1}{c}{(11)}   & \multicolumn{1}{c}{(12)} & \multicolumn{1}{c |}{(13)}  &   \multicolumn{1}{c}{(14)}   &  \multicolumn{1}{c}{(15)}   & \multicolumn{1}{c}{(16)} & \multicolumn{1}{c}{(17)} \\ 
\hline
\hline
\endhead
\hline
\endfoot
IC 4329a &
  &  &  &  &
 6.43$^{+0.04}_{-0.03}$ &  $<0.1$ &  59$^{+24}_{-19}$ &  9.9 &
  &  &  &  &
  &  &  &  \\
(rev. 210)   &  &  &  &  &  &  &  &  &  &  &  &  &  &  &  &  \\
\hline

IC 4329a &
 6.34$^{+0.10}_{-0.31}$  &  0.34$^{+0.17}_{-0.12}$ &  52$^{+19}_{-30}$ &  5.9 &
 6.40$\pm$0.01  & 0.06$\pm$0.03 & 52$^{+21}_{-15}$ &  5.7 &
 6.97$^{+0.06}_{-0.08}$ & 0.10$^{+0.08}_{-0.07}$ & 17$^{+17}_{-9}$ & 1.5 &
 7.67$^{+0.05}_{-0.02}$ & $<0.11$ & -11$^{+4}_{-6}$ &  -0.8 \\
 (rev. 670)  &  &  &  &  &  &  &  &  &  &  &  &  &  &  &  &  \\
\hline

MCG-5-23-16 &
 6.19$^{+0.16}_{-0.20}$ & 0.37$^{+0.20}_{-0.24}$ & 74$^{+33}_{-37}$ & 7.6 &
 6.42$^{+0.02}_{-0.01}$ & $<0.1$ & 66$^{+36}_{-15}$ & 6.2 &
  &  &  &  &
  &  &  &  \\
 (rev. 363)  &  &  &  &  &  &  &  &  &  &  &  &  &  &  &  &  \\
\hline

MCG-5-23-16 & 
 6.35$^{+0.05}_{-0.06}$ & 0.34$^{+0.07}_{-0.06}$ & 90$^{+15}_{-17}$ & 9.6 &
 6.41$\pm$0.01 & $<0.06$ & 41$^{+11}_{-9}$ & 4.4 &
 7.02$\pm$0.03 & $<0.09$  &  13$^{+6}_{-5}$ & 1.1 &
  &  &  &  \\
(rev. 1099)   &  &  &  &  &  &  &  &  &  &  &  &  &  &  &  &  \\
\hline

NGC 3783 &
 6.21$\pm$0.05 & $<0.13$ & $<28$ & 1.0 &
 6.38$\pm$0.01 & 0.05$^{+0.15}_{-0.04}$ & 109$^{+10}_{-11}$ & 6.8 &
 7.00$\pm$0.04 & $<0.02$ & 26$\pm$9 & 1.4 &
 6.60$\pm$0.04 & $<0.16$ & -16$^{+7}_{-39}$ & -1.1 \\
(rev. 193)   &  &  &  &  &  &  &  &  &  &  &  &  &
 7.68$\pm$0.10  & 0.20$^{0.18}_{-0.09}$  & -57$\pm$18 & -2.5 \\
\hline

NGC 3783 &
 5.80$^{+0.27}_{-0.20}$ & 0.45$^{+0.27}_{-0.18}$ & 66$^{+45}_{-20}$ & 3.5 &
 6.39$\pm 0.02$ & 0.06$\pm 0.02$ & 137$^{+12}_{-13}$ & 6.4 &
 6.98$\pm$0.04 & 0.10$\pm$0.04 &  59$\pm$14 & 2.3 &
 &  &  &  \\
(rev. 371)  &  &  &  &  &  &  &  &  &  &  &  &  &  &  &  &  \\
\hline

NGC 3783 & 
 6.09$^{+0.32}_{-0.06}$ & 0.46$^{+0.65}_{-0.05}$ & 45$^{+22}_{-18}$ & 3.4 & 
 6.40$^{+0.02}_{-0.02}$ & 0.06$^{+0.04}_{-0.03}$ & 95$^{+14}_{-9}$ & 6.2 &
 7.01$^{+0.03}_{-0.15}$ & $<0.16$ & 26$^{+8}_{-9}$ & 1.4 &
 6.67$^{+0.25}_{-0.19}$ & $<0.47$ & -20$\pm$7 & -1.4 \\
 (rev. 372)  &  &  &  &  &  &  &  &  &  &  &  &  &  &  &  &  \\
\hline

NGC 5548 &
  &  &  &  &
 6.40$\pm$0.02 & $<0.06$ & 74$^{+15}_{-16}$ & 2.5 &
  &  &  &  &
  &  &  &  \\
 (rev. 191)  &  &  &  &  &  &  &  &  &  &  &  &  &  &  &  &  \\
\hline

NGC5548 &
  &  &  &  &
 6.39$\pm$0.02 & 0.09$^{+0.03}_{-0.02}$ & 79$^{+9}_{-10}$ & 3.3 &
  &  &  &  &
  &  &  &  \\
 (rev. 290) &  &  &  &  &  &  &  &  &  &  &  &  &  &  &  &  \\
\hline

NGC 5548 &
  &  &  &  &
 6.37$\pm$0.03 & 0.06$<0.11$ & 57$^{+11}_{-13}$ & 3.1 &
  &  &  &  &
  &  &  &  \\
 (rev. 291) &  &  &  &  &  &  &  &  &  &  &  &  &  &  &  &  \\
\hline

NGC 3516 & 
 6.29$^{+0.08}_{-0.22}$ & 0.22$^{+0.11}_{-0.15}$ & 63$^{+21}_{-30}$ & 2.3 &
 6.42$\pm$0.01 & $<0.05$ & 88$^{+15}_{-14}$ & 2.9 &
  &  &  &  &
  &  &  &  \\
 (rev. 245)  &  &  &  &  &  &  &  &  &  &  &  &  &  &  &  &  \\
\hline

NGC 3516 &
 6.40$^{+0.11}_{-0.12}$ & 0.62$^{+0.08}_{-0.13}$ & 214$^{+42}_{-41}$ & 6.3 &
 6.41$\pm$0.01 & 0.07$\pm$0.01 & 207$\pm$15 & 5.2 &
  &  &  &  &
  &  &  &  \\
 (rev. 352)  &  &  &  &  &  &  &  &  &  &  &  &  &  &  &  &  \\
\hline

NGC 3516 &
 5.98$^{+0.07}_{-0.04}$ & $<0.17$ & 19$\pm$7 & 1.4 &
 6.39$\pm 0.02$ & 0.09$^{+0.03}_{-0.02}$ & 88$\pm$10 & 5.5 &
  &  &  &  &
  6.72$\pm$0.03 & $<0.09$ & -22$^{+7}_{-8}$ & -1.3 \\
(rev. 1250) &  &  &  &  &  &  &  &  &  &  &  &  &
 7.03$^{+0.04}_{-0.03}$ & $<0.09$ & -27$\pm$8 & -1.3 \\
\hline

NGC 3516 &
  &  &  &  &
 6.47$^{+0.07}_{-0.03}$ & 0.18$^{+0.07}_{-0.04}$ & 221$^{+126}_{-56}$ & 10.9 &
  &  &  &  &
 6.66$^{+0.01}_{-0.05}$ & 0.08$\pm$0.04 & -69$^{+51}_{-54}$ & -4.5 \\
 (rev. 1251)  &  &  &  &  &  &  &  &  &  &  &  &  &
 7.04$\pm 0.06$ & $<0.08$ & -28$^{+15}_{-82}$ & -1.3 \\
 \hline

NGC 3516 &
  &  &  &  &
 6.39$\pm 0.01$ & 0.10$^{+0.02}_{-0.03}$ & 134$^{+8}_{-15}$ & 6.8 &
  &  &  &  &
 6.77$^{+0.04}_{-0.02}$ & $<0.12$ & -23$^{+7}_{-9}$ & -1.1 \\
(rev. 1252)  &  &  &  &  &  &  &  &  &  &  &  &  &
 7.07$^{+0.02}_{-0.01}$ & $<0.08$ & -28$\pm$8 & -1.1 \\
\hline

NGC 3516 &
  &  &  &  &
 6.46$^{+0.04}_{-0.13}$ & 0.14$^{+0.07}_{-0.02}$ & 184$^{+221}_{-36}$ & 9.1 &
  &  &  &  &
 6.66$^{+0.09}_{-0.15}$ & 0.09$\pm$0.08 & -57$^{+47}_{-107}$ & -3.4 \\
 (rev. 1253)  &  &  &  &  &  &  &  &  &  &  &  &  &
 7.08$^{+0.07}_{-0.08}$ & 0.01$^{+0.09}_{-0.05}$ & -27$\pm$10 & -1.2 \\
\hline

MRK 509 &
  &  &  &  &
 6.38$\pm$0.03 & 0.08$\pm$0.05 & 76$^{+16}_{-18}$ & 2.6 &
  &  &  &  &
  &  &  &  \\
 (rev. 161)  &  &  &  &  &  &  &  &  &  &  &  &  &  &  &  & \\
\hline

MRK 509 &
  &  &  &  &
 6.42$\pm$0.03 & 0.01$^{+0.36}_{-0.01}$ & 30$^{+17}_{-13}$ & 1.4 &
 6.64$^{+0.26}_{-0.14}$ & 0.23$^{+0.13}_{-0.15}$ & 48$\pm$24 & 2.1 &
  &  &  &  \\
(rev. 250)  &  &  &  &  &  &  &  &  &  &  &  &  &  &  &  & \\
\hline

MRK 509 &
  &  &  &  &
 6.43$\pm$0.02 & 0.12$\pm$0.03 & 74$\pm$10 eV & 2.8 &
 7.02$^{+0.06}_{-0.07}$ & $<0.17$ & 24$^{+10}_{-9}$ & 0.8 &
 8.00$^{+0.04}_{-0.02}$ & $<0.06$ & -20$^{+8}_{-11}$ & -0.5 \\
(rev. 1073)  &  &  &  &  &  &  &  &  &  &  &  &  &  &  &  & \\
\hline

MRK 509 &
  &  &  &  &
 6.44$^{+0.03}_{-0.04}$ & 0.12$^{+0.03}_{-0.05}$ & 76$^{+12}_{-15}$ & 2.8 &
 6.90$^{+0.03}_{-0.05}$ & $<0.08$ & 27$\pm$10 & 0.9 &
  &  &  &  \\
 (rev. 1074)  &  &  &  &  &  &  &  &  &  &  &  &  &  &  &  & \\
\hline

MRK 509 &
  &  &  &  &
 6.44$\pm$0.04 & 0.10$^{+0.07}_{-0.06}$ & 39$\pm$12 & 1.8 &
 6.91$^{+0.34}_{-0.24}$ & 0.66$^{+0.36}_{-0.39}$ & 120$^{+44}_{-54}$ & 4.7 &
 7.33$^{+0.04}_{-0.02}$ & $<0.13$ & -26$^{+11}_{-10}$ & -1.0 \\
(rev. 1168)  &  &  &  &  &  &  &  &  &  &  &  &  &
 8.50$\pm$0.05 & $<0.12$ & -22$^{+10}_{-13}$ & -0.6 \\
\hline

MCG-6-30-15 &
 5.84$^{+0.53}_{-1.37}$ & 0.51$^{+1.09}_{-0.47}$ & 98$^{+639}_{-72}$ & 3.2 &
 6.45$^{+0.04}_{-0.10}$ & 0.28$^{+1.11}_{-0.10}$ & 162$^{+73}_{-84}$ & 4.1 & 
  &  &  &  &
 6.67$\pm$0.04 & $<0.11$ & -38$^{+8}_{-205}$ & -1.1 \\
(rev. 108A)  &  &  &  &  &  &  &  &  &  &  &  &  &  &  &  & \\
\hline

MCG-6-30-15 &
  &  &  &  &
 6.38$\pm$0.02 & $<0.09$ & 57$\pm$10 & 2.5 &
  &  &  &  &
  &  &  &  \\
 (rev. 108B)  &  &  &  &  &  &  &  &  &  &  &  &  &  &  &  & \\
\hline

MCG-6-30-15 &
  &  &  &  &
 6.43$\pm$0.04 & 0.25$\pm$0.05 & 112$^{+15}_{-16}$ & 5.3 & 
  &  &  &  &
  &  &  &  \\
 (rev. 301)  &  &  &  &  &  &  &  &  &  &  &  &  &  &  &  & \\
\hline

MCG-6-30-15 &
 5.84$^{+0.12}_{-0.48}$ & 0.21$^{+0.28}_{-0.61}$ & 25$\pm$10 & 1.5 &
 6.42$\pm$0.03 & 0.13$^{+0.08}_{-0.03}$ & 67$^{+8}_{-24}$ & 3.3 &
  &  &  &  &
  &  &  &  \\
 (rev. 302)  &  &  &  &  &  &  &  &  &  &  &  &  &  &  &  & \\
\hline

MCG-6-30-15 &
 5.37$^{+0.25}_{-0.26}$ & 0.71$^{+0.22}_{-0.16}$ & 146$^{+366}_{-12}$ & 8.5 &
 6.40$\pm$0.03 & 0.21$\pm$0.04 & 130$^{+16}_{-50}$ & 5.3 &
 6.90$\pm$0.04 & $<0.07$ keV & 13$^{+8}_{-6}$ & 0.5 &
  &  &  &  \\
 (rev. 303)  &  &  &  &  &  &  &  &  &  &  &  &  &  &  &  & \\
\hline

MCG +8-11-11 &
  &  &  &  &
 6.42$\pm$0.01 & 0.06$^{+0.02}_{-0.05}$ & 119$^{+11}_{-25}$ & 5.9 &
 6.90$^{+0.58}_{-0.51}$ & 0.15$^{+0.85}_{-0.73}$ & 43$^{+18}_{-17}$ & 1.9 &
  &  &  &  \\
 (rev. 794)  &  &  &  &  &  &  &  &  &  &  &  &  &  &  &  & \\
\hline

NGC 7314 &
  &  &  &  &
 6.40$\pm$0.03 & 0.10$^{+0.05}_{-0.04}$ & 72$\pm$14 & 3.1 & 
 6.86$^{+0.06}_{-0.07}$ & 0.12$^{+0.05}_{-0.06}$ &  52$\pm$16 & 1.9 & 
  &  &  &  \\
 (rev. 256)  &  &  &  &  &  &  &  &  &  &  &  &  &  &  &  & \\
\hline

NGC 7314 & 
  &  &  &  &
 6.42$\pm$0.02 & 0.10$^{+0.02}_{-0.03}$ & 134$^{+18}_{-19}$ & 2.4 &
  &  &  &  &
  &  &  &  \\
 (rev. 1172)  &  &  &  &  &  &  &  &  &  &  &  &  &  &  &  & \\
\hline

ARK 120 & 
  &  &  &  &
 6.41$^{+0.02}_{-0.07}$ & 0.15$^{+0.04}_{-0.07}$ & 127$^{+9}_{-14}$ & 4.8 & 
 6.94$^{+0.14}_{-0.31}$ &  0.12$^{+0.26}_{-0.05}$ & 42$^{+9}_{-10}$ & 1.4 &
  &  &  &  \\
 (rev. 679)  &  &  &  &  &  &  &  &  &  &  &  &  &  &  &  & \\ 
\hline

MRK 279 &
  &  &  &  &
 6.43$\pm$0.02 & 0.08$\pm$0.03 & 96$^{+15}_{-16}$ & 2.6 & 
  &  &  &  & 
  &  &  &  \\
 (rev. 1087)  &  &  &  &  &  &  &  &  &  &  &  &  &  &  &  & \\ 
\hline

MRK 279 & 
  &  &  &  &
 6.41$^{+0.06}_{-0.07}$ & 0.08$^{+0.07}_{-0.02}$ & 99$^{+23}_{-16}$ & 2.5 & 
  &  &  &  &
 6.69$^{+0.04}_{-0.05}$ & $<0.08$ & -22$^{+8}_{-52}$ & -0.6 \\ 
 (rev. 1088)  &  &  &  &  &  &  &  &  &  &  &  &  &
 7.76$^{+0.07}_{-0.08}$ & 0.12$^{+0.15}_{-0.07}$ & -55$^{+18}_{-21}$ & -1.0 \\  
\hline

MRK 279 &
  &  &  &  &
 6.41$\pm$0.03 & 0.08$\pm$0.04 & 113$\pm$21 & 3.1 &
  &  &  &  & 
  &  &  &  \\
 (rev. 1089)  &  &  &  &  &  &  &  &  &  &  &  &  &  &  &  & \\ 
\hline

MR 2251-178 & 
  &  &  &  &
 6.42$^{+0.05}_{-0.06}$ & 0.10$\pm$0.08 & 51$^{+14}_{-15}$ & 1.2 &
  &  &  &  &
  &  &  &  \\
 (rev. 447)  &  &  &  &  &  &  &  &  &  &  &  &  &  &  &  & \\ 
\hline

NGC 3227 &
  &  &  &  &
 6.39$\pm$0.01 &  0.05$\pm$0.02 & 202$^{+26}_{-22}$ & 2.7 & 
  &  &  &  &
  &  &  &  \\
(rev. 178)  &  &  &  &  &  &  &  &  &  &  &  &  &  &  &  & \\ 
\hline

NGC 3227 & 
 6.02$^{+0.08}_{-0.05}$ & $<0.18$ & 27$\pm$9 & 1.3 & 
 6.40$\pm 0.01$ & 0.06$\pm$0.02 & 102$^{+11}_{-10}$ & 4.2 & 
  &  &  &  &
  &  &  &  \\
 (rev. 1279)  &  &  &  &  &  &  &  &  &  &  &  &  &  &  &  & \\ 
\hline

IRAS 05078+1626 &
  &  &  &  &
 6.38$^{+0.02}_{-0.03}$ & $<0.10$ & 94$^{+13}_{-33}$ & 2.7 &
 6.66$^{+0.09}_{-0.32}$ & $<0.35$ & 17$\pm$10 & 0.5 & 
  &  &  &  \\
 (rev. 1410)  &  &  &  &  &  &  &  &  &  &  &  &  &  &  &  & \\ 
\hline

MKN 590 &
  &  &  &  &
 6.44$\pm$0.02 & 0.07$^{+0.03}_{-0.04}$ & 172$^{+29}_{-31}$ & 1.1 & 
 7.02$^{+0.09}_{-0.11}$ & 0.07$^{+0.2}_{-0.02}$ & 96$^{+31}_{-37}$ & 0.5 &
  &  &  &  \\
 (rev. 837)  &  &  &  &  &  &  &  &  &  &  &  &  &  &  &  & \\ 
\hline

MKN 766 & 
  &  &  &  &
 6.42$^{+0.04}_{-0.03}$ & $<0.13$ & 50$^{+22}_{-29}$ & 0.8 &
 6.80$^{+0.12}_{-0.17}$ & 0.22$^{+0.17}_{-0.12}$ &  114$^{+40}_{-39}$ & 1.6 &
  &  &  &  \\
 (rev. 82)  &  &  &  &  &  &  &  &  &  &  &  &  &  &  &  & \\ 
\hline

MKN 766 &
 5.62$^{+0.04}_{-0.03}$ & $<0.71$ & 15$^{+7}_{-6}$ & 0.4 & 
 6.38$^{+0.05}_{-0.04}$ & $<0.10$ & 15$^{+20}_{-7}$ & 0.4 &
 6.62$^{+0.31}_{-0.05}$ & 0.32$^{+0.06}_{-0.14}$ & 159$^{+17}_{-31}$ & 3.4 &
  &  &  &  \\
 (rev. 265)  &  &  &  &  &  &  &  &  &  &  &  &  &  &  &  & \\
\hline

MKN 766 &
 5.98$^{+0.19}_{-0.28}$ & 0.42$^{+0.15}_{-0.20}$ & 185$^{+57}_{-58}$ & 1.9 &
 6.46$\pm$0.03 & $<0.13$ & 62$^{+19}_{-55}$ & 0.7 &
 6.70$^{+0.10}_{-0.13}$ & $<0.40$ & 43$^{+20}_{-39}$ & 0.4 &
  &  &  &  \\
 (rev. 999)  &  &  &  &  &  &  &  &  &  &  &  &  &  &  &  & \\
\hline

MKN 766 &
 5.87$^{+0.24}_{-0.29}$ & 0.58$^{+0.48}_{-0.09}$ & 165$^{+49}_{-45}$ & 2.2 &
 6.40$\pm$0.03 & $<0.28$ & 47$^{+21}_{-24}$ & 0.6 &
 6.62$^{+0.06}_{-0.09}$ & 0.09$^{+0.10}_{-0.05}$ & 56$^{+17}_{-24}$ & 0.7 &
  &  &  &  \\
 (rev. 1000)  &  &  &  &  &  &  &  &  &  &  &  &  &  &  &  & \\
\hline

MKN 766 &
 5.86$^{+0.05}_{-0.06}$ & $<0.12$ & 19$\pm$10 & 0.3 &
 6.41$^{+0.15}_{-0.08}$ & $<0.28$ & 64$^{+18}_{-38}$ & 0.9 &
 6.83$^{+0.17}_{-0.19}$ & 0.17$^{+0.19}_{-0.11}$ & 99$^{+23}_{-25}$ & 1.2 &
  &  &  &  \\
 (rev. 1001)  &  &  &  &  &  &  &  &  &  &  &  &  &  &  &  & \\
\hline

MKN 766 &
 5.78$^{+0.06}_{-0.05}$ & $<0.12$ & 18$\pm$9 &  0.4 &
 6.40$^{+0.02}_{-0.03}$ & $<0.09$ & 39$^{+14}_{-13}$ & 0.7 &
 6.77$^{+0.06}_{-0.07}$ & 0.20$^{+0.07}_{-0.06}$ & 106$^{+20}_{-22}$ & 1.7 &
  &  &  &  \\
 (rev. 1002)  &  &  &  &  &  &  &  &  &  &  &  &  &  &  &  & \\
\hline

MKN 766 &
 5.93$^{+0.03}_{-0.02}$ & $<0.20$ & 20$^{+12}_{-8}$ & 0.3 &
 6.54$^{+0.07}_{-0.06}$ & 0.26$\pm$0.06 & 148$\pm$25 & 2.0 &
  &  &  &  &
  &  &  &  \\
 (rev. 1003)  &  &  &  &  &  &  &  &  &  &  &  &  &  &  &  & \\
\hline

MKN 766 &
  &  &  &  &
 6.44$^{+0.20}_{-0.12}$ & 0.27$^{+0.12}_{-0.20}$ & 139$^{+115}_{-80}$ & 1.9 &
  &  &  &  &
  &  &  &  \\
 (rev. 1004)  &  &  &  &  &  &  &  &  &  &  &  &  &  &  &  & \\
\hline

NGC 7469 & 
  &  &  &  &
 6.39$\pm$0.02 & $<0.07$ & 108$^{+21}_{-22}$ & 2.8 &
  &  &  &  &
  &  &  &  \\
 (rev. 192A)  &  &  &  &  &  &  &  &  &  &  &  &  &  &  &  & \\
\hline

NGC 7469 &
  &  &  &  &
 6.41$\pm$0.03 & $<0.14$ & 90$^{+17}_{-22}$ & 2.4 &
  &  &  &  &
  &  &  &  \\
 (rev. 192B)  &  &  &  &  &  &  &  &  &  &  &  &  &  &  &  & \\
\hline

NGC 7469 &
  &  &  &  &
 6.40$\pm$0.01 & 0.06$\pm$0.02 & 95$^{+10}_{-9}$ & 2.8 & 
 7.00$\pm$0.05 & $<0.14$ & 21$\pm$9 & 0.5 &
  &  &  &  \\
 (rev. 912)  &  &  &  &  &  &  &  &  &  &  &  &  &  &  &  & \\
\hline

NGC 7469 &
  &  &  &  &
 6.42$\pm$0.02 & 0.08$\pm$0.02 & 94$^{+12}_{-11}$ & 2.9 &
  &  &  &  &
  &  &  &  \\
 (rev. 913)  &  &  &  &  &  &  &  &  &  &  &  &  &  &  &  & \\
\hline

MCG-2-58-22 &
  &  &  &  &
 6.39$^{+0.08}_{-0.09}$  & 0.17$^{+0.16}_{-0.08}$ & 107$^{+40}_{-37}$ & 3.6 &
  &  &  &  &
  &  &  &  \\
 (rev. 180)  &  &  &  &  &  &  &  &  &  &  &  &  &  &  &  & \\
\hline

NGC 526A &
  &  &  &  &
 6.38$^{+0.03}_{-0.02}$ & $<0.09$ & 34$^{+14}_{-12}$ & 1.0 &
 6.59$^{+0.23}_{-0.14}$ & 0.32$^{+0.22}_{-0.13}$ & 82$^{+30}_{-32}$ & 2.2 &
 7.22$^{+0.04}_{-0.05}$ & $<0.12$ & -19$^{+10}_{-15}$ & -0.5 \\
 (rev. 647)  &  &  &  &  &  &  &  &  &  &  &  &  &  &  &  & \\
\hline

NGC 4051 &
 6.27$^{+0.24}_{-0.20}$ & 0.41$^{+0.43}_{-0.22}$ & 68$^{+26}_{-32}$ & 1.9 &
 6.39$\pm 0.01$ & $<0.07$ &  68$^{+11}_{-13}$ & 1.8 &
  &  &  &  &
 7.08$^{+0.06}_{-0.05}$ & $<0.18$ & -24$^{+11}_{-12}$ & -0.5 \\
 (rev. 263)  &  &  &  &  &  &  &  &  &  &  &  &  &  &  &  & \\
\hline

NGC 4051 &
 6.13$^{+0.11}_{-0.18}$ & 0.65$^{+0.16}_{-0.28}$ & 485$^{+98}_{-95}$ & 3.6 &
 6.44$\pm$0.02 & 0.07$^{+0.03}_{-0.04}$ & 196$^{+28}_{-36}$ & 1.5 &
 7.08$^{+0.03}_{-0.04}$ & $<0.21$ & 49$^{+31}_{-27}$ & 0.3 &
  &  &  &  \\
 (rev. 541)  &  &  &  &  &  &  &  &  &  &  &  &  &  &  &  & \\
\hline

ESO 141-G055 &
  &  &  &  &
 6.35$^{+0.11}_{-0.09}$ & $<0.27$ & 70$^{+31}_{-32}$ & 2.0 &
  &  &  &  &
  &  &  &  \\ 
 (rev. 1435 A)  &  &  &  &  &  &  &  &  &  &  &  &  &  &  &  & \\
\hline

ESO 141-G055 &
  &  &  &  &
 6.54$^{+0.19}_{-0.20}$ & 0.39$^{+0.20}_{-0.31}$ & 202$^{+67}_{-70}$ & 5.1 &
  &  &  &  &
  &  &  &  \\
 (rev. 1435 B)  &  &  &  &  &  &  &  &  &  &  &  &  &  &  &  & \\
\hline

ESO 141-G055 &
  &  &  &  &
 6.35$\pm$0.07 & 0.17$^{+0.08}_{-0.07}$ & 117$^{+36}_{-33}$ & 2.7 &
 6.90$\pm$0.03 & $<0.11$ & 51$^{+24}_{-22}$ & 1.0 &
  &  &  &  \\
 (rev. 1436)  &  &  &  &  &  &  &  &  &  &  &  &  &  &  &  & \\
\hline

ESO 141-G055 &
  &  &  &  &
 6.42$^{+0.03}_{-0.02}$ & $<0.09$ & 30$^{+17}_{-13}$ & 0.9 &
 6.62$^{+0.20}_{-0.17}$ & 0.45$^{+0.26}_{-0.33}$ & 126$^{+35}_{-59}$ & 3.5 &
 8.31$^{+0.03}_{-0.04}$ & $<0.11$ & -75$\pm 20$ & -1.3 \\
 (rev. 1445)  &  &  &  &  &  &  &  &  &  &  &  &  &  &  &  & \\
\hline

UGC 3973 &
  &  &  &  &
 6.44$\pm$0.03 & $<0.08$ & 69$\pm$20 & 1.7 &
  &  &  &  &
  &  &  &  \\
 (rev. 1247)  &  &  &  &  &  &  &  &  &  &  &  &  &  &  &  & \\
\hline

UGC 3973 & 
  &  &  &  &
 6.42$\pm$0.03 & 0.04$^{+0.08}_{-0.02}$ & 84$^{+21}_{-23}$ & 1.8 &
  &  &  &  &
  &  &  &  \\
 (rev. 1263)  &  &  &  &  &  &  &  &  &  &  &  &  &  &  &  & \\
\hline

UGC 3973 &
  &  &  &  &
 6.38$\pm$0.04 & 0.11$\pm$0.04 & 122$^{+31}_{-30}$ & 2.5 &
  &  &  &  & 
  &  &  &  \\
 (rev. 1332)  &  &  &  &  &  &  &  &  &  &  &  &  &  &  &  & \\
\hline

UGC 3973 & 
  &  &  &  &
 6.35$\pm$0.01 & 0.11$^{+0.03}_{-0.02}$ & 284$^{+26}_{-29}$ & 3.1 &
  &  &  &  &
  &  &  &  \\
 (rev. 1535)  &  &  &  &  &  &  &  &  &  &  &  &  &  &  &  & \\
\hline

Fairall 9 &
  &  &  &  &
 6.40$\pm$0.03 & $<0.11$ & 126$^{+24}_{-26}$ & 1.7 &
  &  &  &  &
  &  &  &  \\
 (rev. 105)  &  &  &  &  &  &  &  &  &  &  &  &  &  &  &  & \\
\hline

ESO 198-G24 & 
  &  &  &  &
 6.51$\pm$0.11 & $<0.36$  & 60$\pm$29 & 0.8 &
  &  &  &  &
  &  &  &  \\
 (rev. 207)  &  &  &  &  &  &  &  &  &  &  &  &  &  &  &  & \\
\hline

ESO 198-G024 &
  &  &  &  &
 6.41$\pm$0.02 & 0.07$^{+0.02}_{-0.05}$ & 77$^{+13}_{-14}$ & 0.9 &
  &  &  &  &
 7.56$\pm$0.13 & $<0.33$ & -49$^{+23}_{-25}$ & -0.4 \\
 (rev. 1128)  &  &  &  &  &  &  &  &  &  &  &  &  &  &  &  & \\
\hline

MKN 110 &
  &  &  &  &
 6.41$\pm$0.04 & $<0.13$ & 46$\pm$13 & 1.4 &
  &  &  &  &
  &  &  &  \\
 (rev. 904)  &  &  &  &  &  &  &  &  &  &  &  &  &  &  &  & \\
\hline

AKN 564 & 
  &  &  &  &
 6.54$\pm$0.10 & 0.28$^{+0.11}_{-0.08}$ & 94$^{+24}_{-25}$ & 1.2 &
  &  &  &  &
  &  &  &  \\
 (rev. 930)  &  &  &  &  &  &  &  &  &  &  &  &  &  &  &  & \\
\hline

NGC 7213 &
  &  &  &  &
 6.40$\pm$0.02 & $<0.08$ & 79$^{+17}_{-21}$ & 1.8 & 
 6.74$^{+0.13}_{-0.16}$ & 0.18$^{+0.12}_{-0.10}$ & 59$\pm$24 & 1.3 &
  &  &  &  \\
 (rev. 269)  &  &  &  &  &  &  &  &  &  &  &  &  &  &  &  & \\
\hline

ESO 511-G030 &
  &  &  &  &
 6.39$\pm$0.02 & 0.12$^{+0.04}_{-0.03}$ & 95$\pm$13 & 2.0 &
  &  &  &  &
  &  &  &  \\
 (rev. 1402)  &  &  &  &  &  &  &  &  &  &  &  &  &  &  &  & \\
\hline

MRK 704 &
  &  &  &  &
 6.37$^{+0.06}_{-0.07}$ & 0.13$^{+0.07}_{-0.06}$ & 159$\pm$43 & 2.3 &
  &  &  &  &
  &  &  &  \\
 (rev. 1074)  &  &  &  &  &  &  &  &  &  &  &  &  &  &  &  & \\
\hline

MRK 704 &
  &  &  &  &
 6.38$\pm$0.02 & 0.11$\pm$0.04 & 140$^{+16}_{-22}$ & 1.6 &
  &  &  &  &
  &  &  &  \\
 (rev. 1630)  &  &  &  &  &  &  &  &  &  &  &  &  &  &  &  & \\
\hline

NGC 4593 &
  &  &  &  &
 6.36$\pm$0.01 & $<0.11$ & 121$\pm$10 & 5.2 &
 6.93$^{+0.05}_{-0.06}$ & 0.09$^{+0.12}_{-0.08}$ & 37$^{+10}_{-11}$ & 1.4 &
  &  &  &  \\
 (rev. 465)  &  &  &  &  &  &  &  &  &  &  &  &  &  &  &  & \\
\hline
\hline   
\end{longtable}
\end{landscape}
}

\end{scriptsize}


\begin{table*}
\caption{Variability significances obtained from excess map analysis in the three energy bands: 5.4--6.1 keV (band A), 6.1--6.8 keV (band B) and 6.8--7.2 keV (band C). Confidence levels $\geq$90\% are marked in boldface.}
\label{tab:varsig}
\centering
\vspace{0.2cm}
\begin{scriptsize}
\begin{tabular}{l c c c|l c c c}
\hline\hline             
Sources & Band A & Band B & Band C & Sources & Band A & Band B & Band C \\
       & (\%)  & (\%) & (\%) &   &  (\%)    &    (\%) & (\%) \\
\hline
IC 4329a (rev. 210) & 35.8 & {\bf 99.7} & 71.2 &  MRK 590 (rev. 837) & 71.3 & 11.6 & 6.8 \\
IC 4329a (rev. 670) & {\bf 98.5}  & 74.5  & 11.2 & & & & \\
 & & & & MRK 766 (rev. 82) & 20.3 & 32.2 & 45.0 \\
MCG -5-23-16 (rev. 363) & 31.0  & 77.3 &  66.9 & MRK 766 (rev. 265) & 1.4  & 80.2  & 49.5  \\
MCG -5-23-16 (rev. 1099) & 63.2 & 79.8  & 89.5  & MRK 766 (rev. 999) & 70.8 & 38.8  & 71.0  \\
 & & & & MRK 766 (rev. 1000) & 45.6 & 60.7  & 3.1 \\
NGC 3783 (rev. 193) & 36.7 & 25.6 & 82.7 & MRK 766 (rev. 1001) & 47.6 & 59.8  & 70.8 \\
NGC 3783 (rev. 371) & 68.1  & 5.3  & 58.3 & MRK 766 (rev. 1002) & 84.7 & 70.0  & 66.1  \\
NGC 3783 (rev. 372) & {\bf 96.7}  & 5.3  & 65.2  & MRK 766 (rev. 1003) & {\bf 97.2}  & 84.6  & 29.9  \\
 & & & &  MRK 766 (rev. 1004) & {\bf 95.7}  & {\bf 95.6}  & 24.3  \\
NGC 5548 (rev. 191) & 61.9 & 6.8 & {\bf 94.3} & & & & \\
NGC 5548 (rev. 290) & 23.0   & 82.9 & 74.7 & NGC 7469 (rev. 192A) & 42.9 & 4.6  & {\bf 96.6}\\
NGC 5548 (rev. 291) & 77.3  & 51.1  & 14.9 & NGC 7469 (rev. 192B) & 81.6 & 40.5 & {\bf 94.5}  \\
 & & & & NGC 7469 (rev. 912) & 84.5  & 86.9  & 86.7 \\
NGC 3516 (rev. 245) & 30.2 & 28.6 & {\bf 98.2} &  NGC 7469 (rev. 913) & 66.6  & 43.3  & 3.4 \\
NGC 3516 (rev. 352) & 88.6  & 29.6 & 6.5 & & & & \\
NGC 3516 (rev. 1250) & 14.4  & 88.7  & {\bf 95.7} &  MCG -2-58-22 (rev. 180) & 54.4 & 44.9  & 49.4  \\
NGC 3516 (rev. 1251) & {\bf 91.7}  & 62.8 & {\bf 93.7} & & & & \\
NGC 3516 (rev. 1252) & 67.9  & 39.8 & {\bf 99} & NGC 526A (rev. 647) & 78.7 & 58.7  & 52.3  \\
NGC 3516 (rev. 1253) & 77.0  & 4.2  & 66.6 & & & & \\
 & & & & NGC 4051 (rev. 263) & 65.4 & 57.9  & 78.3  \\
MRK 509 (rev. 161) & {\bf 95.3} & {\bf 90.4} & 42.3 & NGC 4051 (rev. 541) & 20.3 & 53.4  & 82.3  \\
MRK 509 (rev. 250) & 66.1  & 3.1  & 72.2  & & & & \\
MRK 509 (rev. 1073) & {\bf 92.7}  & 13.7  & 88.1 & ESO 141-G055 (rev. 1435 A) & {\bf 93.4} & 9.0  & {\bf 93.0} \\ 
MRK 509 (rev. 1074) & 83.7  & 30.0 & 25.1 & ESO 141-G055 (rev. 1435 B) & 13.0  & 12.8 & 20.8 \\
MRK 509 (rev. 1168) & 27.1  & {\bf 91.2}  & 65.8 & ESO 141-G055 (rev. 1436) & 19.1  & 20.8 & 15.8  \\
 & & & &  ESO 141-G055 (rev. 1445) & 60.7 & 59.8 & 10.7  \\
MCG -6-30-15 (rev. 301) & {\bf 97.0} & 82.5 & {\bf 93.7} & & & & \\
MCG -6-30-15 (rev. 302) & 63.7  & {\bf 96.5}  & 82.8  & UGC 3973 (rev. 1247) & 70.6 & 37.8 & 52.3 \\
MCG -6-30-15 (rev. 303) & 83.1  & 67.4 & 4.7 & UGC 3973 (rev. 1263) & {\bf 94.6}  & 41.5 & 74.7 \\
MCG -6-30-15 (rev. 108A) & 81.6  & 79.6  & 65.4 & UGC 3973 (rev. 1332) & 37.3 & 48.4 & 16.2 \\
MCG -6-30-15 (rev. 108B) & {\bf 97.0}  & {\bf 96.4}  & {\bf 94.5} & UGC 3973 (rev. 1535) & 55.6 & 24.7 & {\bf 95.9} \\
 & & & & & & & \\
MCG +8-11-11 (rev. 794) & 14.1 & 29.9 & 20.3 & Fairall 9 (rev. 105) & 71.8 & 25.6 & 82.9 \\
 & & & & & & & \\
NGC 7314 (rev. 256) & {\bf 96.2} & 43.3 & 67.2 & ESO 198-G024 (rev. 207) & {\bf 92.4} & 54.7 & 35.4 \\
NGC 7314 (rev. 1172) & 70.9  & 11.5  & 17.1 & ESO 198-G024 (rev. 1128) & 86.7 & 27.6 & 2.7 \\
 & & & & & & & \\
AKN 120 (rev. 679) & 62.5 & 44.0 & 45.9 & MRK 110 (rev. 904) & 27.7 & 5.6 & 56.8 \\
 & & & & & & & \\
MRK 279 (rev. 1087) & 64.1 & 44.0 & 66.0 & AKN 564 (rev. 930) & 12.1 & {\bf 94.0} & {\bf 95.5} \\ 
MRK 279 (rev. 1088) & 17.5  & 69.1  & 31.1 & & & & \\
MRK 279 (rev. 1089) & 14.1  & 3.6  & {\bf 94.5} & NGC 7213 (rev. 269) & 64.7 & 79.2 & 67.1 \\
 & & & & & & & \\
MR 2251-178 (rev. 447) & 58.2  & 72.3 & 83.4 & ESO 511-G030 (rev. 1042) & 44.3 & 87.5 & 11.8 \\
 & & & & & & & \\
NGC 3227 (rev. 178) & 72.4 & 26.4 & 32.2 & MRK 704 (rev. 1074) & 85.7 & {\bf 91.3} & 15.5 \\
NGC 3227 (rev. 1279) & 40.9  & 53.4  & 17.4 & MRK 704 (rev. 1630) & 36.8 & 58.2 & 78.0 \\
 & & & &    &   &   &  \\
IRAS 05078+1626 (rev. 1410) & 24.3 & 21.7 & 9.8 & NGC 4593 (rev. 465) & 9.2 & 9.9 & 54.0 \\
& & & &   &   &   &  \\
\hline
\hline
\end{tabular}
\end{scriptsize}
\end{table*}

\appendix
\section{Notes on single sources}

The aim of this section is to discuss and compare our results with previous studies, as many of the sources of our sample have been extensively investigated in literature. 
In general, the best fit models we used as templates for the simulations (see Table \ref{tab:models}) are in good agreement with the ones derived by other authors from deeper broad X-ray band analysis. We mostly refer to the NOGR07 survey for broad Fe lines, as they analysed many of the observations treated in this paper. 
In the following we will focus on the most relevant sources.

\emph{IC 4329a}: The addition of a broad and redshifted emission component to the best fit model is required at 99\% confidence level (from intensity vs energy contour plots) during rev. 670. The detection of this line is confirmed by previous studies (Markowitz et al. 2006; NOGR07) on the same data set. However, its origin is controversial, being, most probably produced at r$>$50 r$_{g}$, where relativistic effects cannot be robustly measured (see NOGR07). We registered significant flux variations within the energy band comprising this component, on time scales of $\sim$32 ks. We refer to De Marco et al. 2009 for a more comprehensive study of such emission feature.\\
 On the contrary the redshifted component appears to be absent (in agreement with NOGR07 results) during the preceding observation (rev. 210). In this case, the non-detection of the line may be a consequence of the short exposure ($\sim$10 ks). However, significant variability is detected in band B, pointing to the plausible presence of some kind of persisting, short-time scale variability phenomenon taking place in this source.\\
Detection of a blueshifted ($\sim 0.1c$) and constant flux absorption line is registered during rev. 670, in agreement with Markowitz et al. 2006 and NOGR07, possibly ascribable to FeXXVI. 

\emph{MCG -5-23-16}: This source has been deeply monitored via simultaneous observations with XMM-{\it Newton}, \emph{Chandra} and Suzaku. This allowed accurate determination of the underlying continuum and detection of a relativistic Fe K$\alpha$ line during rev. 1099. We confirm detection of a broad component during both rev. 363 and 1099, after restricting the analysis energy band to E$=$4--9 keV. This line does not show significant variability, in good agreement with previous studies (Dewangan et al. 2003, Braito et al. 2007, NOGR07). 

\emph{NGC 3783}: We revealed a redshifted emission line in all the three observations (rev. 193, 371, 372), in agreement with published results (Blustin et al. 2002; Reeves et al. 2004; NOGR07). However, NOGR07 associate the broad component observed during rev. 193 to emission originated at r$>$50r$_{g}$. We also found signatures of a narrow absorption line at E$\sim$6.6 keV during rev. 103 and 372, ascribed by Reeves et al. 2004 to a blend of FeXXIII, FeXXIV and FeXXV lines at the systemic velocity of the source. Reeves et al. observed that the line decreased its intensity between rev. 371 and rev. 372, indicating a variability time scale of $\sim$10$^{5}$ s for the highly ionized absorber. Although our technique does not test variability between different revolutions, we can nonetheless confirm these findings, as we do not have strong evidence for an absorption line detected during rev. 371. However, it is worth noting that the line is present during rev. 193, in good agreement with results of NOGR07.\\
An exhaustive temporal study of the spectral features in the 3 XMM-{\it Newton} observations of NGC 3783 has been published by Tombesi et al. 2007. Our analysis resambled the overall results of this paper. 

\emph{NGC 3516}: Narrow, energy shifted emission components in the Fe K line profile were for the first time detected during a simultaneous \emph{Chandra} and XMM-{\it Newton} (rev. 352) observation of NGC 3516 (Turner et al. 2002). 
Variability of the narrow redshifted emission feature at E$\sim$6.08 keV (Bianchi et al. 2004) has been widely investigated by IMF04 for rev. 245 using the excess map technique, finding a variability confidence level of 97\%. The pre-defined redshifted energy band we used for our analysis (band A, see Sect. \ref{sec:proced}) is broader than the energy band monitored by IMF04 (i.e. 5.8-6.2 keV), hence the light curve variations are smoothed out. However, restricting the energy band to the one considered in IMF04 we achieve the same variability confidence level (i.e. 97.1\%).\\
In addition, different layers of absorbing gas were revealed during rev. 245 and 352 (Turner et al. 2005, NOGR07). Here we did not not account for these components as the narrow energy band in which we focus our analysis is not strongly affected by them. However, the technique we used to trace variability is nonetheless able to reveal high energy transient components of the absorbing gas.
On the contrary, we observed two deep absorption lines in the average spectra of NGC 3516 during four consecutive revolutions (i.e. rev. 1250, 1251, 1252, 1253, covering the period 2006 Oct. 6--13), in the 6.7--7.1 keV regime (coincident with the C band), which can be associated to H-like and He-like species of Fe. These lines were ascribed by Turner et al. 2008 to a disk wind with outflowing velocities in the range 1000--2000 km s$^{-1}$.
All these absorbing structures seem responsible for the significant variability revealed in band C during rev. 245, 1250, 1251 and 1252. We point out that the lack of variability during rev. 1253 is not due to statistical problems as the observation has approximately the same duration and flux as the previous three. 

\emph{MRK 509}: As pointed out by NOGR07, the statistical significance of the relativistic Fe K$\alpha$ line during rev. 161 is comparable to the one obtained using a distant reflector model. Hence we adopted a simple, relatively narrow Gaussian template to fit the line.
We further observed two absorption features during rev. 1073 and 1168. One of these (E$\sim$7.33 keV) is also revealed in the synthesis spectrum (averaged out over all the five exposures) of Ponti et al. 2009, and its origin is ascribed to highly ionized outflowing (v$\sim$14000 km s$^{-1}$) gas. On the contrary the high energy one (E$\sim$8.0 keV) is also detected by Cappi et al. 2009 and Tombesi et al. 2009.
We observed significant variability in band B during rev. 161 and 1168, in agreement with results obtained from the total RMS spectrum analysis (Ponti et al. 2009). Moreover we found signatures of significant variability in Band A during rev. 161 and 1073.

\emph{MCG -6-30-15}: In this particular case the relativistic Fe K$\alpha$ line (e.g. Wilms et al. 2001) extends over the entire energy band where we focused our analysis. As the band is too narrow to obtain a good fit of the broad Fe line, the induced residual spectral bending can be modelled as due to continuum emission. Hence, as a consequence of the adopted technique and energy resolution (i.e. $\Delta E=$0.1 keV), the mapping procedure is sensitive only to narrow spectral features above the broad line, and the corresponding flux measurements are robust against continuum modelling. 
The RMS spectrum (with a time resolution of 6 ks) of the two observations carried out during rev. 108 shows a steep rise of variability at E$\sim$4.5 keV with a peak at E$\sim$5.0 keV (Ponti et al. 2004). We analysed the 2 observations with a finest time resolution (i.e. 2.5 ks) and found excess variability in all the three bands (E$\sim$5.4-7.2 keV) during rev. 108B. The energy bands we analysed exclude the strong variability event observed in the RMS spectrum.\\
The variations registered in the entire Fe K band during rev. 108B are most probably driven by the strong flare observed in the continuum light curve (Ponti et al. 2004; Iwasawa \& Miniutti 2004).
Papadakis et al. 2005 studied the frequency-resolved spectra (Revnivtsev et al. 1999) of MCG -6-30-15 during rev. 108-303 and found that, although the broad Fe K$\alpha$ line shows no significant variations on time scales less than $\sim$1-2 days, some residuals between 5--7 keV are suggestive of small amplitude variations in a line-like feature. Moreover the drop observed in the spectra in the Fe K$\alpha$ regime is less than the one expected from a constant line.
Vaughan \& Fabian 2004, using the flux-flux linear correlation (between soft band, 1--2 keV, and hard band, 3--10 keV) technique, deduced that flux variations are dominated by changes in the normalization of the power law component. This is in agreement with our assumption in the Monte Carlo simulations, where we assume that only the power law normalization varies, while the other components are constant. In addition there must be an additional component that varies little, and contributes more in the 3--10 keV band than in the 1--2 keV. The largest fractional contribution from this component occurs in the Fe K band, where it accounts for $\sim$40\% of the total flux.
Once the variable continuum is subtracted from the total spectrum (see Sect. \ref{sec:proced}), the residuals should include only the nearly constant reflection component. Hence, the significant variations we observed in these residuals (during rev. 108B, 301 and 302) must be due to additional line-like features overlapping the broad Fe line profile, in agreement with deductions of Papadakis et al. 2005. 

\emph{NGC 7314}: a relativistic line or a distant reflector model can fit equally well the spectrum of this source during rev. 256 (NOGR07). For our analysis we assumed a simple model which accounts for the neutral and highly ionized Fe K$\alpha$ emission lines. We found significant variations redward the neutral component during rev. 256. The detection of a narrow emission feature at E$\sim$5.8 keV was also claimed by Yaqoob et al. (2003), in a \emph{Chandra} observation taken 19--20 July 2002, and interpreted as redshifted Fe K$\alpha$ emission. A hot spot origin of the line was ruled out by the authors because the estimated spot location resulted too large (i.e. r$>$132 r$_{g}$) to cause sufficient gravitational and Doppler redshift. In our spectrum, however, the feature is observed for about 20 ks (i.e. about half the duration of the observation). If produced in a spot -- assuming a SMBH mass of 5$\times 10^{6}$M$_{\odot}$ (Padovani \& Rafanelli 1988) -- the emitting radius should be r$>$26 r$_{g}$ in order for the emission to correspond to less than one orbit (i.e. a spot and not an annulus). This estimated lower limit on the radius is still within the upper limit of 132 r$_{g}$.\\
It is worth noting that the target of the second XMM-{\it Newton} observation analysed in this paper (rev. 1172) is the cluster XMMUJ2235.3-2557, while NGC 7314 is observed off-axis, reducing the X-ray telescope's effective area.

\emph{NGC 3227}: We detected a redshifted narrow emission feature during the longest observation (rev. 1279). Estimated parameters of this feature are in agreement with those derived in Markowitz et al. 2009 and double checked with MOS data. The energy of the feature results inconsistent with a Compton shoulder minimum energy (i.e. E$>$6.24 keV). We did not find any significant evidence for variability in the Fe K band, in agreement with results of time resolved analysis in Markowitz et al. 2009.

\emph{MKN 766}: A relativistic Fe K line was detected by NGO07 during rev. 82 and 265. To fit the curvature induced by this line we adopted a broad Gaussian component whose centroid energy falls in the energy range typical of a highly ionized Fe K$\alpha$ and/or Fe K$\beta$ line.
Turner et al. 2006 reported on significant energy modulation (detected via excess map technique) during rev. 265, within E$\sim$6.1-6.8 keV (band B), tracing a sinusoid of period $\sim$150 ks. They associated this emission to orbiting material at an estimated distance of r=115 r$_{g}$ from the central BH. At such distance the increase in flux during the approaching phase of the orbital motion is not expected to be strong (e.g. Dovciak et al. 2004), in agreement with the results of our temporal analysis, which rules out the presence of any significant intensity modulation in the data.
An extensive analysis of the entire set of observations (rev. 82, 265, 999-1004) available for this source, was carried out by Miller et al. 2006. They found variability in the low-ionization band (E$=$6.08-6.48 keV), significant at 95\%. This value is in agreement with the variability significance we estimated in Band B (which includes the low-ionization band of Miller et al.) during rev. 1004.

\emph{NGC 4051}: We found evidence for broad emission at Fe K redshifted energies in both the observations (rev. 263 and 541). For what concerns rev. 263, our result is consistent with NOGR07. We did not detect significant variability in this source, in agreement with the RMS spectrum (time resolution 2 ks) analysis carried out by Ponti et al. 2006.

\emph{AKN 564}: \emph{ASCA} observations of AKN 564 revealed an Fe K$\alpha$ line origin in highly ionized material (dominated by H-like ions). The line results significantly broad (Turner et al. 2001). Variability measurements down to time scales of approximately a week constrained the bulk of the line to originate at small distances from the nucleus.
Previous studies (Papadakis et al. 2007) on the $\sim$100 ks XMM-{\it Newton} observation (rev. 930) revealed a Fe K$\alpha$ line component slightly broader than the energy resolution of the EPIC-pn detector and rather weak (i.e. EW$\sim$70 eV), in agreement with our best fit model (Table \ref{tab:models}). However the line cannot be unambiguously ascribed to relativistic emission (see Papadakis et al. 2007; NOGR07). We detected significant variability in both band B and C where the only residuals visible in the map coincide with the Fe line. They seem to be absent at the very beginning of the observation; we rule out this to be a background subtraction effect, because the source signal results very high with respect to the background level (during the first 10 ks the background count rate in the 4--9 keV energy band is just 5\% the source one).

\end{document}